\documentclass[prd,preprint,eqsecnum,nofootinbib,amsmath,amssymb,tightenlines,superscriptaddress]{revtex4}
\pdfoutput=1
\usepackage[pdftex]{graphicx}
\usepackage{epstopdf}
\usepackage{bm}
\usepackage{amsfonts}
\usepackage{amssymb}
\usepackage{slashbox}

\usepackage{dsfont}

\def\b{{\bm b}}
\def\x{{\bm x}}
\def\y{{\bm y}}
\def\z{{\bm z}}
\def\p{{\bm p}}
\def\q{{\bm q}}
\def\k{{\bm k}}
\def\l{{\bm l}}

\def\B{{\bm B}}
\def\C{{\bm C}}

\def\bcalP{{\bm{\mathcal P}}}

\def\uOmega{\underline{\Omega}}
\def\Ybar{\overline{Y}}

\def\Nc{N_{\rm c}}
\def\CA{C_{\rm A}}

\def\alphas{\alpha_{\rm s}}
\def\Re{\operatorname{Re}}

\def\tr{\operatorname{tr}}
\def\sgn{\operatorname{sgn}}
\def\grad{{\bm\nabla}}

\def\ix{{\rm i}}
\def\fx{{\rm f}}
\def\xx{{\rm x}}
\def\xbx{{\bar{\rm x}}}
\def\yx{{\rm y}}
\def\ybx{{\bar{\rm y}}}
\def\zx{{\rm z}}

\def\yfrak{{\mathfrak y}}
\def\Bfrak{{\mathfrak B}}
\def\Time{\mathbb{T}}
\def\Fint{\Phi}
\def\BnBint{\Psi}
\def\seq{{\rm seq}}
\def\tform{t_{\rm form}}
\def\tformx{t_{{\rm form},x}}
\def\tformy{t_{{\rm form},y}}

\def\gammaE{\gamma_{\rm\scriptscriptstyle E}}

\begin {document}



\title
    {
      The LPM effect in sequential bremsstrahlung 2:
      factorization
    }

\author{Peter Arnold}
\affiliation
    {%
    Department of Physics,
    University of Virginia,
    Charlottesville, Virginia 22904-4714, USA
    \medskip
    }%
\author{Han-Chih Chang}
\affiliation
    {%
    Department of Physics,
    University of Virginia,
    Charlottesville, Virginia 22904-4714, USA
    \medskip
    }%
\author{Shahin Iqbal}
\affiliation
    {%
    Department of Physics,
    University of Virginia,
    Charlottesville, Virginia 22904-4714, USA
    \medskip
    }%
\affiliation
    {%
    National Centre for Physics, \\
    Quaid-i-Azam University Campus,
    Islamabad, 45320 Pakistan
    \medskip
    }%

\date {\today}

\begin {abstract}%
{%
   The splitting processes of bremsstrahlung and pair production in a medium
   are coherent over large distances in the very high energy limit,
   which leads to a suppression known as the Landau-Pomeranchuk-Migdal
   (LPM) effect.  In this paper, we continue analysis of
   the case when the coherence
   lengths of two consecutive splitting processes overlap (which is
   important for understanding corrections to standard treatments
   of the LPM effect in QCD), avoiding soft-gluon approximations.
   In particular, this paper analyzes the subtle problem of how to precisely
   separate overlapping double splitting (e.g.\ overlapping
   double bremsstrahlung)
   from the case of
   consecutive, independent bremsstrahlung (which is the case that would
   be implemented in a Monte Carlo simulation based solely on
   single splitting rates).  As an example of the method,
   we consider the rate of real double gluon bremsstrahlung
   from an initial gluon
   with various simplifying assumptions
   (thick media; $\hat q$ approximation; large $\Nc$; and
   neglect for the moment of processes involving 4-gluon vertices)
   and explicitly compute
   the correction $\Delta\,d\Gamma/dx\,dy$ due
   to overlapping formation times.
}%
\end {abstract}

\maketitle
\thispagestyle {empty}

{\def\boldmath{}\tableofcontents}
\newpage


\section{Introduction and Results}
\label{sec:intro}

When passing through matter, high energy particles lose energy
by showering, via the splitting processes of hard bremsstrahlung and
pair production.  At very high energy, the quantum mechanical duration
of each splitting process, known as the formation time, exceeds
the mean free time for
collisions with the medium, leading to a significant reduction in
the splitting rate known as the Landau-Pomeranchuk-Migdal (LPM)
effect \cite{LP,Migdal}.%
\footnote{
  English translations of ref.\ \cite{LP} may be found
  in ref.\ \cite{LPtranslate}.
}
A long-standing problem in field theory has
been to understand how to implement this effect in cases where
the formation times of two consecutive splittings overlap.

Let $x$ and $y$ be the longitudinal
momentum fractions of two consecutive bremsstrahlung gauge bosons.
In the limit $y \ll x \ll 1$, the problem of overlapping formation
times has been analyzed at leading logarithm order in
refs.\ \cite{Blaizot,Iancu,Wu}
in the context of
energy loss of high-momentum partons traversing
a QCD medium (such as a quark-gluon plasma).
Two of us \cite{2brem} subsequently developed and implemented field theory
formalism needed for the more general case where $x$ and $y$ are
arbitrary.  However, we only computed a subset of the
interference effects.  Most significantly, we deferred the analysis
of effects that require carefully disentangling (i) the computation of
double bremsstrahlung with overlapping formation
times from (ii) the naive approximation of double bremsstrahlung
as two, consecutive, quantum-mechanically independent
single-bremsstrahlung processes.
In this paper, we compute the effects that require this
careful disentanglement.

In the remainder of this introduction, we will first qualitatively
discuss what effect overlapping formation times have on a simplified
Monte Carlo picture of shower development, which will help us
later set up the technical details of how to approach
explicit calculations.
We then give a more precise description of exactly what effects
we calculate in this paper versus which are further deferred to later work.
With those caveats, we present interim results for the example of
real, double gluon bremsstrahlung $g \to ggg$ in the
case of a thick medium.
We wrap up the introduction with
a very simple argument for the parametric size of
the rate.

After the introduction, section \ref{sec:calculation} is given
over to the calculation itself, where the most important issue
will be the technical implementation of our method for isolating the
corrections due to overlapping divergences.  (Details of the
calculation which do not involve new methods but instead closely
follow those established in ref.\ \cite{2brem} are relegated to
appendices.)  Final formulas for the case of a thick medium are
summarized in section \ref{sec:summary} in terms of a single integral, which
may be computed numerically.  Section \ref{sec:conclusion} offers
our conclusion, including comments on the sign of the result.


\subsection {Simplified Monte Carlo versus overlapping formation times}
\label{sec:MC}

\subsubsection{Overview}

In this paper, we will ultimately present results by giving the
{\it correction}
$\Delta \, d\Gamma/dx\,dy$ to double splitting, due to overlapping
formation times, instead of giving the
double-splitting rate $d\Gamma/dx\,dy$ by itself.
We begin by explaining why this is the physically
sensible choice for the calculation that we do.

\begin {figure}[t]
\begin {center}
  \includegraphics[scale=0.7]{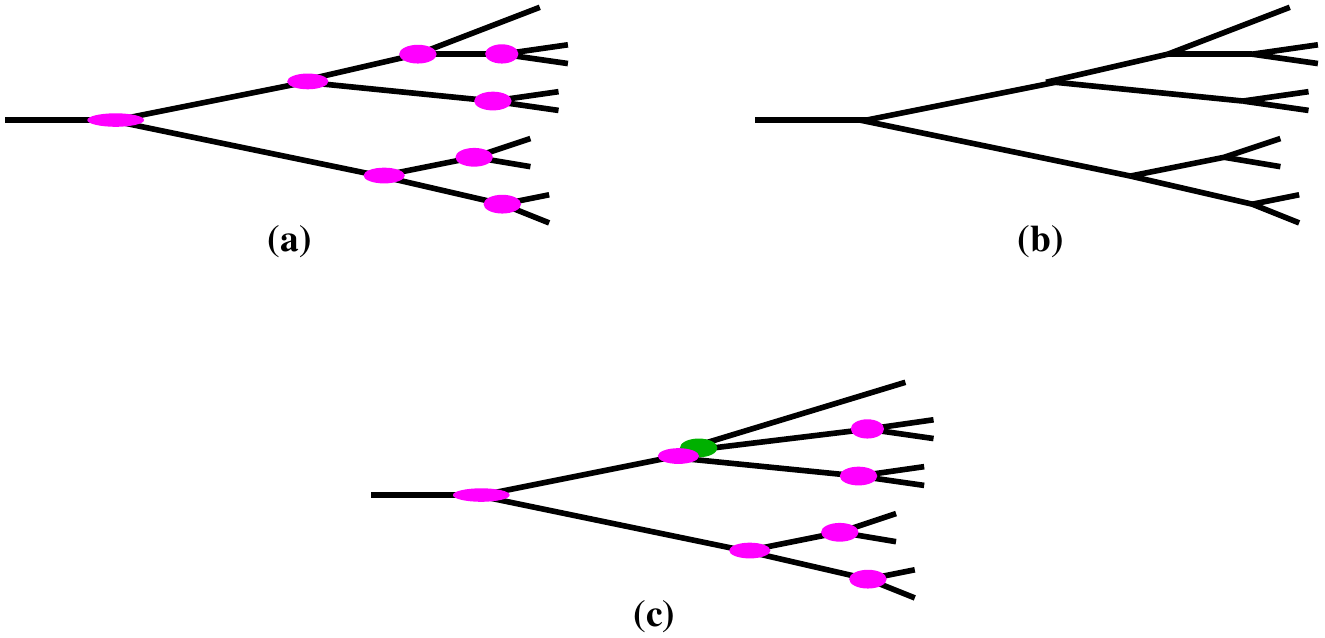}
  \caption{
     \label{fig:MC}
     (a) A depiction of a high-energy particle showering in a medium
     as it moves from left to right.
     The magenta ovals crudely
     depict the formation lengths (transversely as well as
     longitudinally) associated with each splitting.
     We show here a case where consecutive splittings are well separated
     compared to formation times [see text].
     (b) A corresponding depiction of the approximation made in a simple
     idealized Monte Carlo (IMC), where the formation times are treated
     as effectively zero.
     (c) shows the case where two consecutive splittings (colored magenta and
     green) happen to overlap.
     In all of these figures, we have exaggerated the transverse direction:
     for high-energy particles (and high-energy daughters),
     splittings will be very nearly collinear.
  }
\end {center}
\end {figure}

In order to simplify calculations in this paper, we are going to assume
that the medium is thick---much wider than any of the formation times
for splitting.  Now consider an (approximately on-shell)
high-energy particle that showers in the
medium.  That is, imagine that the medium is thick enough that there
are several splittings in the medium, as shown in fig.\ \ref{fig:MC}a.
In the situation
pertaining to jet energy loss in quark-gluon plasmas formed
in relativistic heavy ion collisions, this could apply to
an energetic particle (not necessarily the primary energetic particle)
that showers and stops in the medium.
Imagine approximating the development of this shower by
an idealized in-medium ``Monte Carlo'':
Start with a calculation or model of the rate
for each single splitting, assume that consecutive splittings
are independent, and evolve through time, rolling dice after each
small time increment to decide whether each high-energy
particle splits then.  Even for purely in-medium development,
this description is simplistic and is not intended to describe
the many effects that are included in actual Monte Carlos used
for phenomenology.%
\footnote{
  For a description of phenomenological Monte Carlos in the context
  of quark-gluon plasmas, see, for example,
  refs. \cite{JET,JET2,MajumderSummary}.
  Such Monte Carlos typically deal with initial-state (vacuum) radiation,
  handle particles that are high enough energy to escape the medium
  (unlike our case described above, chosen to simplify our initial
  investigations), treat hadronization of high-energy particles that escape
  the medium, deal with finite medium-size effects in cases of
  extremely large formation lengths for extremely high-energy partons,
  account for collisional energy loss and collisional
  $\p_\perp$ broadening, and much more.
  Some (see later discussion) also attempt to heuristically
  model corrections to independent treatment of splittings
  \cite{JEWEL,Martini}. 
  In contrast to Monte Carlos used for detailed phenomenology,
  for some examples of theoretical insight gained
  from studying various characteristics of the idealized in-medium showers we
  focus on here, see refs.\ \cite{JeonMoore,stop,BIM,BMreview}.
}
We will refer to this idealized calculation, based just on formulas for
single-splitting probabilities, as the ``idealized Monte Carlo (IMC)''
result.  The assumption that the splittings
are quantum-mechanically independent is equivalent to saying that
this idealized Monte Carlo treats the formation times as effectively zero.
That is, the picture of fig.\ \ref{fig:MC}a is treated as
fig.\ \ref{fig:MC}b.

We want to compute how to account for what happens when two of the
splittings are close enough that formation times overlap, such as in
fig.\ \ref{fig:MC}c, in which case the idealized Monte Carlo
assumption that the
splittings are independent breaks down.
Let's ignore very-soft bremsstrahlung for now, since it is more-democratic
splittings that naively dominate energy loss; the
(nonetheless important) effects
of very-soft bremsstrahlung have been treated elsewhere
\cite{Blaizot,Iancu,Wu}.
So imagine that in each splitting the daughters both carry a
non-negligible fraction of the parent's energy.%
\footnote{
  As will be discussed later in section \ref{sec:picture},
  the specific assumption here is
  (parametrically) that $E_{\rm daughter} \gg \alphas^2 E_{\rm parent}$
  for each daughter.
}
In this case, the separation between splittings is on average large
compared to the formation times, provided that $\alphas$
(evaluated at the scale that characterizes high-energy splittings%
\footnote{
  For hard splitting in a thick medium, this
  is $\alphas$ at scale $Q_\perp \sim (\hat q E)^{1/4}$.  See,
  for example, the brief discussion in section I.E of
  ref.\ \cite{2brem}.
}%
) is small.  The chances that three or more consecutive (democratic)
splittings
happen to occur with overlapping formation times is then even smaller
than the chance that two consecutive splittings overlap.  So,
to compute the effect of overlapping formation times, it is enough to
focus on two consecutive collisions.

\begin {figure}[t]
\begin {center}
  \includegraphics[scale=0.38]{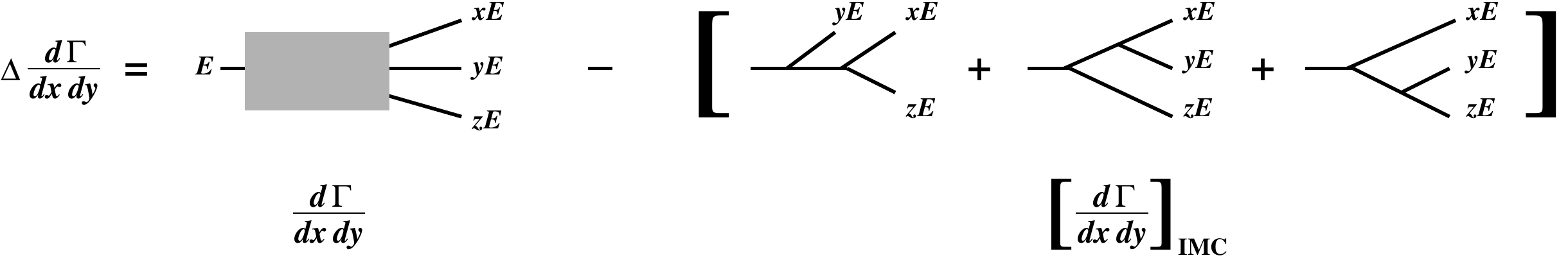}
  \caption{
     \label{fig:correct}
     A pictorial version of the definition of $\Delta\,d\Gamma/dx\,dy$
     as the difference between (i) the double-splitting rate (represented
     by the gray box) and (ii) the
     comparable rate given by idealized Monte Carlo (IMC)
     restricted to two splittings.
     Above, $z \equiv 1{-}x{-}y$.
  }
\end {center}
\end {figure}

\begin {figure}[t]
\begin {center}
  \includegraphics[scale=0.38]{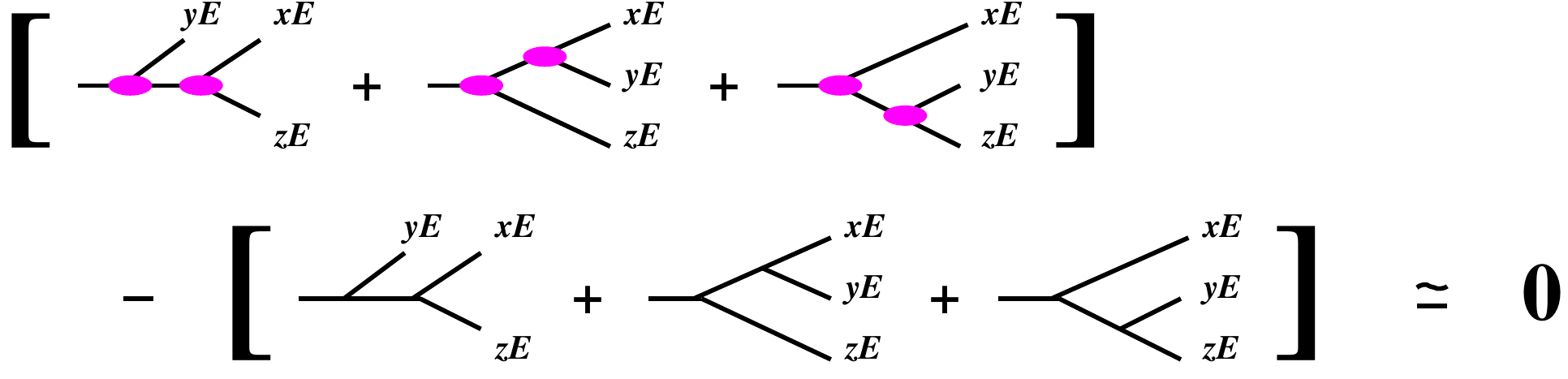}
  \caption{
     \label{fig:cancel}
     A pictorial summary of the cancellation of the contributions
     to fig.\ \ref{fig:correct} for splittings that
     are well-separated in time.
  }
\end {center}
\end {figure}

A cartoon of the corresponding
correction is
given in fig.\ \ref{fig:correct}.  Let $\delta H$ be the part of the
Hamiltonian of the theory that includes the splitting vertices for
high-energy particles.  The first term on the right of
fig.\ \ref{fig:correct} represents a calculation of the double-bremsstrahlung
rate $d\Gamma/dx\,dy$ in medium, where we have formally expanded to
second order in $\delta H$, even though the real-world situation
may be that there are eventually many more splittings (such as in
fig.\ \ref{fig:MC}c).  The other terms, subtracted on the right-hand
side of fig.\ \ref{fig:correct}, represent the result
$(d\Gamma/dx\,dy)_{\rm IMC}$ that one would obtain from an idealized
Monte Carlo if one formally expanded that result to second order in
the splitting probability.  (More on what we mean by that in a moment.)
Individually, both $d\Gamma/dx\,dy$ and $(d\Gamma/dx\,dy)_{\rm IMC}$
receive contributions from consecutive splittings that are separated in time
by much more
than the formation times, but those contributions cancel in the
difference, as depicted pictorially in fig.\ \ref{fig:cancel}.
Indeed, individually, both
$d\Gamma/dx\,dy$ and $(d\Gamma/dx\,dy)_{\rm IMC}$ as defined above
would be formally infinite in the case of an infinite medium.
(More on that in a moment as well.)
In contrast, the
result for
\begin {equation}
   \Delta\, \frac{d\Gamma}{dx\,dy} \equiv
   \frac{d\Gamma}{dx\,dy}
   -
   \left[ \frac{d\Gamma}{dx\,dy} \right]_{\rm IMC}
\label{eq:DeltaDef}
\end {equation}
is finite and depends only on separations $\lesssim$ the relevant formation
time.

\subsubsection{Some simple analogies}

Since the above points are important, we will try to illuminate them
with some analogies.  First, as a warm-up, consider the decay of a
particle in quantum mechanics.  The generic way to compute the decay
rate is to formally compute the {\it probability}\/ $P$ for decay, to
first order in the process that causes the decay.  One finds a result
that formally grows proportional to the total time $\Time$ as
$P = \Gamma\,\Time$, from which we extract the rate $\Gamma$.
But of course the probability of decay can't really be $P = \Gamma\,\Time$
because that probability would exceed 1 for large enough $\Time$.  Instead,
the probability is $P = 1 - e^{-\Gamma\Time}$.
The formula $P = \Gamma\,\Time$ is analogous to what
we mean above when we
say to formally compute or expand a result to a given order in perturbation
theory.
This example is a rough analogy to constructing an idealized Monte Carlo based
on the single bremsstrahlung rate: $\Gamma$ is analogous to the single
bremsstrahlung rate, whereas the result $P = 1 - e^{-\Gamma\Time}$ is
analogous to what you would get if you actually used a result for $\Gamma$
in an idealized Monte Carlo (as opposed to discussing
an idealized Monte Carlo result that had been ``formally
expanded'' to some fixed order in perturbation theory).
Let's now turn to an example that is more analogous to our double
bremsstrahlung problem.

Consider the classical analogy of a very tiny device (``particle'')
that has a certain probability $\Gamma_1$ per unit time of emitting
a flash of light.  If we formally expand to first order in $\Gamma_1$,
then the probability of emitting exactly one flash of light in time
$\Time$ is (formally) $P_1 = \Gamma_1\Time$.
If we formally expand to {\it second} order in $\Gamma_1$,
then the probability of emitting exactly two flashes of light%
\footnote{
  Whether or not one treats the two flashes as distinguishable or
  indistinguishable is inessential to our analogy; and so the factors
  of $\frac12$ here and in the rest of the argument are inessential
  to the point we want to make.
}
in time
$\Time$ is $P_2 = \frac12 (\Gamma_1\Time)^2$.  If we naively divided $P_2$
by $\Time$ to get a rate, then we would awkwardly
say that the rate
for two flashes is $\Gamma_2 = \tfrac12 \Gamma_1^2 \Time$, which
(unlike $\Gamma_1 = P_1/\Time$) diverges as
$\Time \to \infty$.  So the ``rate for two flashes'' (analogous to
the $d\Gamma/dx\,dy$ for double splitting) is not, by itself,
a very meaningful quantity.

But now suppose that the device had the additional
property that, for some interval
$\delta t$ after emitting one flash, the rate for emitting another
flash was temporarily changed to $\Gamma_1+\delta\Gamma$.
We might then ask for the correction
to the previous result.  We could again formally expand to second
order in flash rates (which are now correlated as just
described) to find the probability $P_2$ in this
new situation, which would roughly%
\footnote{
  We say ``roughly'' because, if one wants an exact answer, then
  there are boundary issues having to do
  with the end of time at $t=\Time$.  These sorts of boundary issues
  will be
  important to address later on in our discussion but are not
  important for the purpose of this analogy.
}
 have the form
$P_2 = \tfrac12 (\Gamma_1\Time)^2
 + \Gamma_1 \, \delta\Gamma \, \delta t \, \Time$.
If we divided by $\Time$ to define a
$\Gamma_2 = \tfrac12 \Gamma_1^2\Time + \Gamma_1 \, \delta\Gamma \, \delta t $,
we would again have something ill-defined as $\Time\to\infty$.
However, the {\it correction} to $\Gamma_2$ due to the
change is perfectly well defined as
$\Delta \Gamma_2 \equiv \Gamma_2 - \tfrac12 \Gamma_1^2\Time
 = \Gamma_1 \, \delta\Gamma \, \delta t$.
In this analogy, $\Gamma_2$ is like our double-bremsstrahlung rate
$d\Gamma/dx\,dy$; $\frac12 \Gamma_1^2 \Time$ is like $(d\Gamma/dx\,dy)_{\rm IMC}$;
$\delta t$ is like the formation time; and
the correction $\Delta\Gamma_2$ is like the
$\Delta\,d\Gamma/dx\,dy$ of (\ref{eq:DeltaDef}).

To further illuminate the importance and relevance of the subtraction
(\ref{eq:DeltaDef}), we will present in a moment an analogy
with the importance of similar subtractions in kinetic theory.
But it will be useful to first discuss what one should do with
(\ref{eq:DeltaDef}) once one has calculated it.


\subsubsection{Uses}

What can one do with a calculation of the correction
$\Delta\,d\Gamma/dx\,dy$?
First note that its definition (\ref{eq:DeltaDef})
is as a difference of two positive
quantities.  {\it A priori}, that difference might
have either sign: negative if
the effect of overlapping formation times suppresses the
double bremsstrahlung rate and positive if it enhances it.

\begin {figure}[t]
\begin {center}
  \includegraphics[scale=0.7]{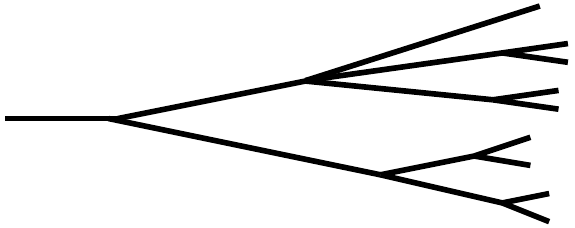}
  \caption{
     \label{fig:MC4}
     An example of a process in a corrected Monte Carlo that could
     easily be implemented if
     $\Delta\,d\Gamma/dx\,dy$
     is positive.  Here, the $1\to3$ splitting, representing
     the inclusion of an additional possible process in the Monte
     Carlo with probability distribution $\Delta\,d\Gamma/dx\,dy$,
     would account for
     the correction due to the possibility of fig.\ \ref{fig:MC}c.
  }
\end {center}
\end {figure}

If the correction is positive, there is
a relatively easy way to implement the correction to an idealized Monte Carlo
simulation: Simply interpret $\Delta\,d\Gamma/dx\,dy$ as
the probability distribution of an additional type of local splitting
process that ``instantly'' produces three daughters (instead of just
two daughters) from one parent.
This would allow for Monte Carlo showers such as
fig.\ \ref{fig:MC4}.
In contrast, if $\Delta\,d\Gamma/dx\,dy$ is negative,
one has to work harder.
(Examples in the context of relativistic heavy ion collisions:
the Monte Carlo generators JEWEL \cite{JEWEL} and MARTINI \cite{Martini}
implement heuristic models for
a reduction in multiple-splitting rates due to overlapping formation
times.%
\footnote{
  On a related note:
  See ref.\ \cite{POWHEG} for
  a discussion of implementing negative-weight corrections in the
  context of {\it vacuum} Monte Carlo.
}%
)

In this paper, we focus on the calculation of $\Delta\,d\Gamma/dx\,dy$
and will not pursue how to incorporate it into Monte Carlo.
The earlier discussion of (uncorrected) idealized
Monte Carlo was necessary, however, for the definition
(\ref{eq:DeltaDef}) of $\Delta\,d\Gamma/dx\,dy$.
We will discuss the sign of the correction shortly, when presenting
numerical results in section \ref{sec:results}.


\subsubsection{A kinetic theory analogy}

The same issues that lead to focusing on $\Delta\Gamma$
rather than $\Gamma$ (for higher-order processes)
arise in kinetic theory problems as well.
We start by briefly reviewing
a kinetic theory example from the literature.
Then we'll give an example more closely analogous to our double
bremsstrahlung problem (and to our earlier flashing device analogy).

\begin {figure}[t]
\begin {center}
 \includegraphics[scale=0.4]{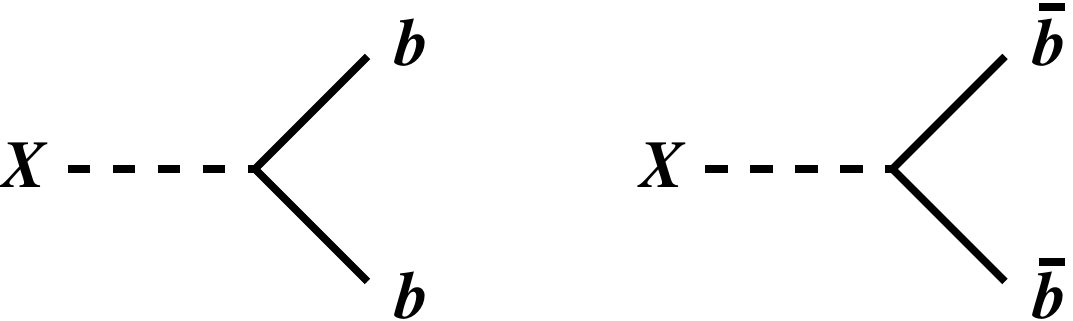}
  \caption{
     \label{fig:KW}
     A depiction of decay processes $X \to bb$ and $X \to \bar b\bar b$
     for our first kinetic theory analogy, based on ref.\ \cite{KolbWolfram}.
  }
\end {center}
\end {figure}

\begin {figure}[t]
\begin {center}
 \includegraphics[scale=0.4]{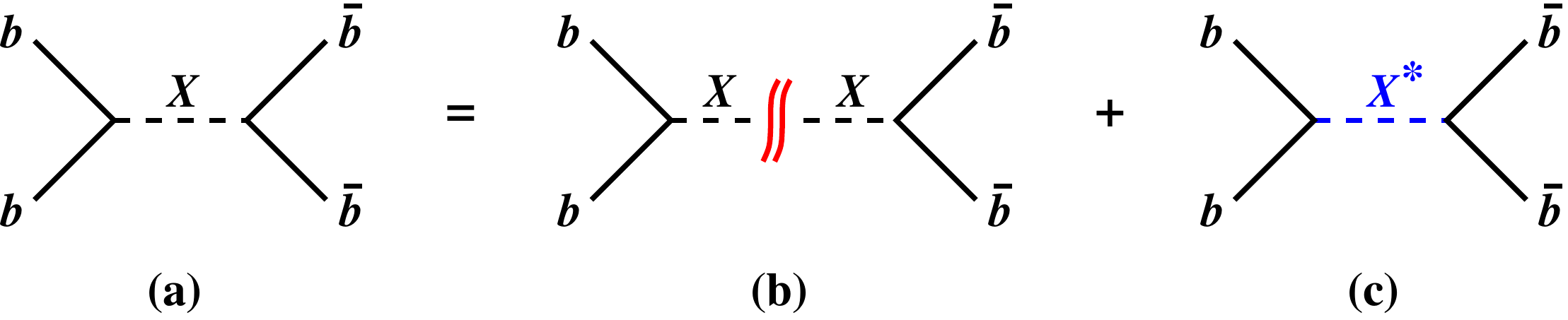}
  \caption{
     \label{fig:KW2}
     A depiction of (a) the process $bb \to X \to \bar b\bar b$
     separated into contributions (b) where the $X$ is on-shell plus
     (c) the rest.  The wavy double-line cut in the middle of
     (b) is used to indicate that the intermediate particle
     $X$ is on-shell.
  }
\end {center}
\end {figure}

\begin {figure}[t]
\begin {center}
 \includegraphics[scale=0.4]{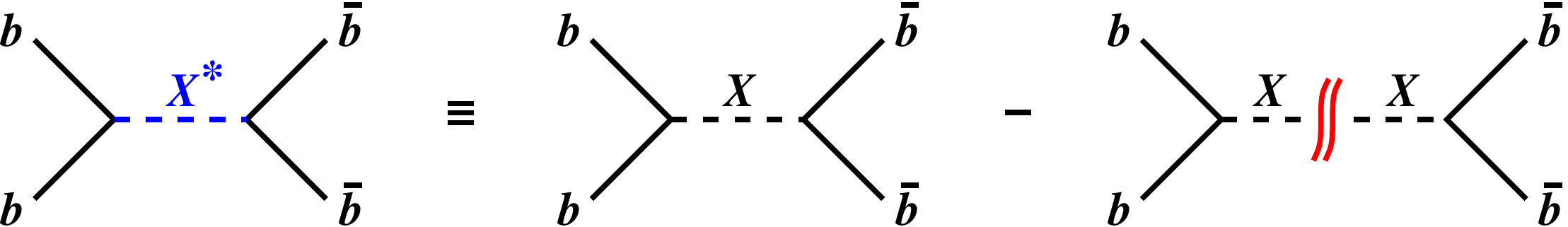}
  \caption{
     \label{fig:KW3}
     Reorganization of fig.\ \ref{fig:KW2}
     to show the subtracted quantity
     that should be used in the collision term of a kinetic theory
     analysis.
  }
\end {center}
\end {figure}

Consider a kinetic theory discussion of the following simple
model, analyzed by Kolb and Wolfram \cite{KolbWolfram}%
\footnote{
  See in particular the discussion surrounding eq.\ (2.3.12) of
  ref.\ \cite{KolbWolfram}.
}
in the context of baryogenesis in grand unified theories.
The toy model has
a stable, nearly-massless particle $b$ and a massive, unstable boson
$X$ which can decay by both $X \to bb$ and $X \to \bar b\bar b$,
as depicted in fig.\ \ref{fig:KW}.
If one writes a kinetic theory for these particles, one would
include the processes of fig.\ \ref{fig:KW} (and their inverses)
in the collision term.
Now consider additionally including
the process $bb \to X \to \bar b\bar b$, depicted in
fig.\ \ref{fig:KW2}a.  There is a problem of double counting
because the Feynman diagram for $bb \to X \to \bar b\bar b$ includes
two different types of physical processes.  One of these
is the case where the intermediate $X$ boson is approximately on
shell, which we depict by fig.\ \ref{fig:KW2}b.
The other is what's left: the case where the $X$ boson is off-shell,
which we depict in fig.\ \ref{fig:KW2}c with an asterisk
on the label $X$.  The first case (fig.\ \ref{fig:KW2}b) is {\it already}
accounted for by solving kinetic theory using a collision term based
of fig.\ \ref{fig:KW}, and so supplementing the collision term by
the full result of fig.\ \ref{fig:KW2}a would double count
the case of on-shell $X$.  Instead, the collision term should
contain just (i) fig.\ \ref{fig:KW} and (ii) the off-shell contribution
of fig.\ \ref{fig:KW2}c.
How precisely does one define the ``off-shell'' part
of the contribution?  By rearranging the terms of fig.\ \ref{fig:KW2}
to make it a definition, as in fig.\ \ref{fig:KW3}.
This subtraction is (crudely) analogous to our problem's
subtraction (\ref{eq:DeltaDef}).  The fact that fig.\ \ref{fig:KW3}
and not fig.\ \ref{fig:KW2}a is the correct thing to use in the collision term
in Kolb and Wolfram's problem is analogous to our discussion of
adding $\Delta\,d\Gamma/dx\,dy$ rather than $d\Gamma/dx\,dy$
to a Monte Carlo description originally based on
single-splitting rates,
as in fig.\ \ref{fig:MC4}.

\begin {figure}[t]
\begin {center}
  \includegraphics[scale=0.4]{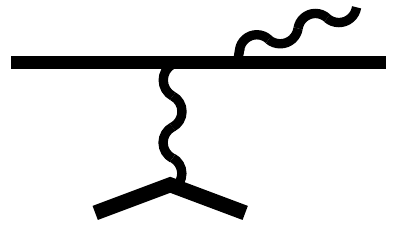}
  \caption{
     \label{fig:toy}
     A depiction of the leading-order bremsstrahlung process
     that could be used in the collision term for
     a kinetic theory description of a QED plasma, in cases where the
     LPM effect can be ignored.  (There are other Feynman diagrams
     that contribute to this process at this order,
     but we just show the one for simplicity.)
  }
\end {center}
\end {figure}

\begin {figure}[t]
\begin {center}
  \includegraphics[scale=0.35]{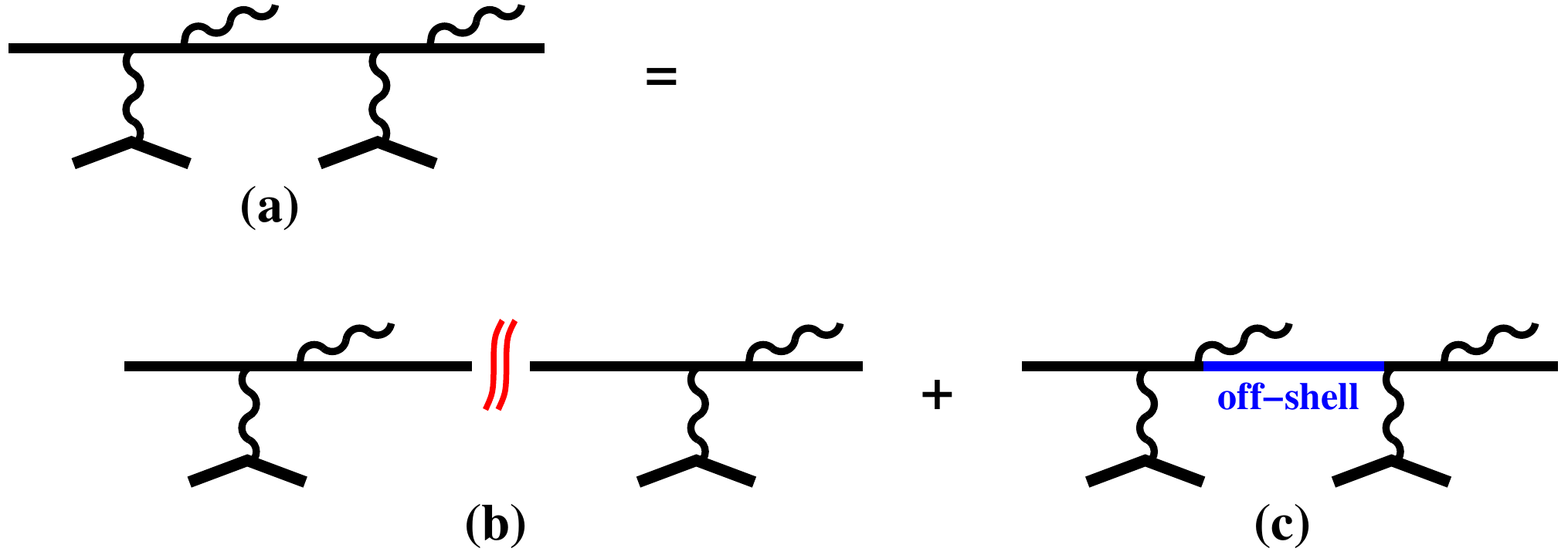}
  \caption{
     \label{fig:toy2}
     A depiction of (a) double bremsstrahlung in the kinetic theory
     example of fig.\ \ref{fig:toy},
     separated into contributions (b) where the intermediate particle
     is on-shell plus
     (c) the rest.
  }
\end {center}
\end {figure}

We now give a kinetic theory example that is somewhat more analogous to the
current problem.  Ignore the LPM effect, but consider a kinetic
theory description of a QED plasma that includes the leading-order,
$2{\to}3$ process for bremsstrahlung, depicted by fig.\ \ref{fig:toy}.
What if we now want to systematically include higher-order processes
in the collision term?  Consider in particular the $3{\to}5$ process
depicted in fig.\ \ref{fig:toy2}a.  Just like fig.\ \ref{fig:KW2},
this process contains two types of contributions.  Including the
contribution with an on-shell intermediate line, depicted by
fig.\ \ref{fig:toy2}b, would be double counting.  Instead, one
should only add to the collision term the remaining piece, fig.\
\ref{fig:toy2}c, which is defined as the difference between
figs.\ \ref{fig:toy2}a and \ref{fig:toy2}b.
Here, figs.\ \ref{fig:toy2}a, b, and c are analogous to
our problem's
$d\Gamma/dx\kern1pt dy$,
$[d\Gamma/dx\kern1pt dy]_{\rm IMC}$, and
$\Delta\,d\Gamma/dx\kern1pt dy$ respectively.

These are not perfect analogies (they do not involve the LPM
effect), but we hope they help illuminate the importance of
the subtraction (\ref{eq:DeltaDef}).


\subsection {What we compute (and what we do not)}
\label {eq:what}

\begin {figure}[t]
\begin {center}
  \includegraphics[scale=0.5]{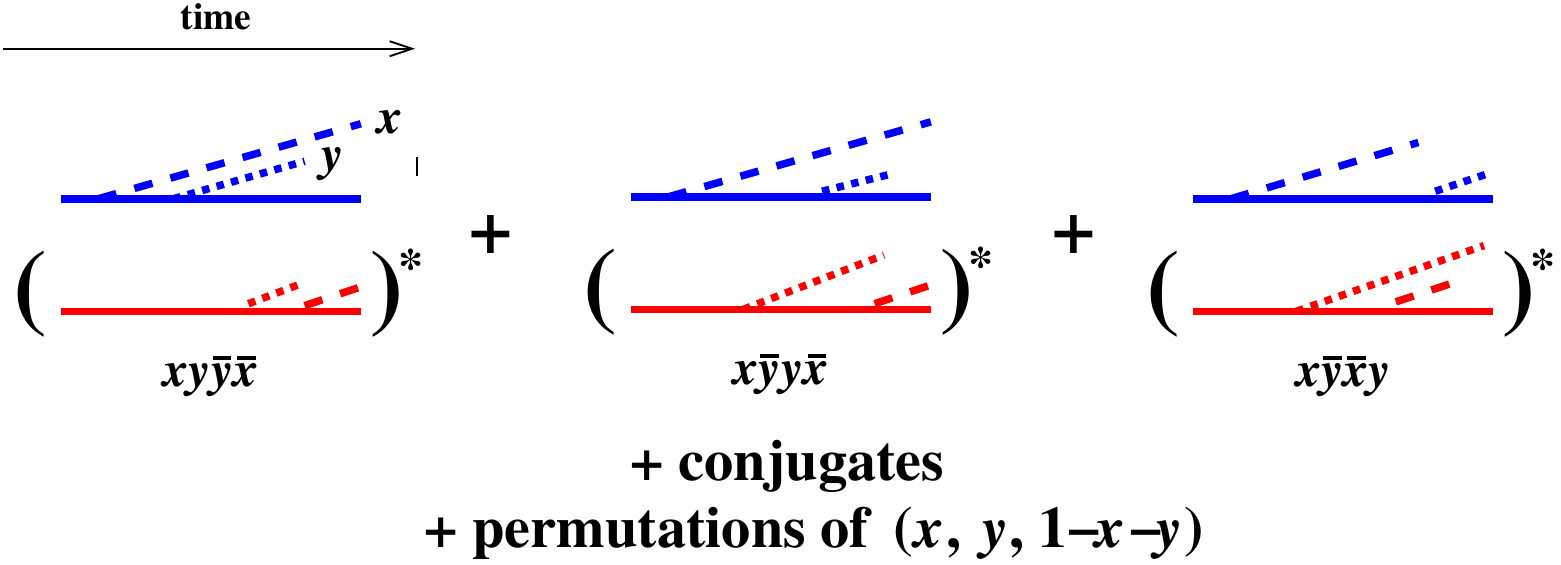}
  \caption{
     \label{fig:subset}
     The subset of interference contributions to double splitting
     previously evaluated in ref.\ \cite{2brem}:
     the ``crossed'' diagrams.
     To simplify the drawing, all particles, including
     bremsstrahlung gluons, are indicated by straight lines.
     The long-dashed and short-dashed lines
     are the daughters with momentum fractions $x$ and $y$
     respectively. 
     The naming of the diagrams indicates the time order
     in which emissions occur in the amplitude and conjugate amplitude.
     For instance, $x\bar y y \bar x$ means first
     (i) $x$ emission in the amplitude, then (ii) $y$ emission in the
     conjugate amplitude, then (iii) $y$ emission in the amplitude,
     and then (iv) $x$ emission in the conjugate amplitude.
  }
\end {center}
\end {figure}

\begin {figure}[t]
\begin {center}
  \includegraphics[scale=0.5]{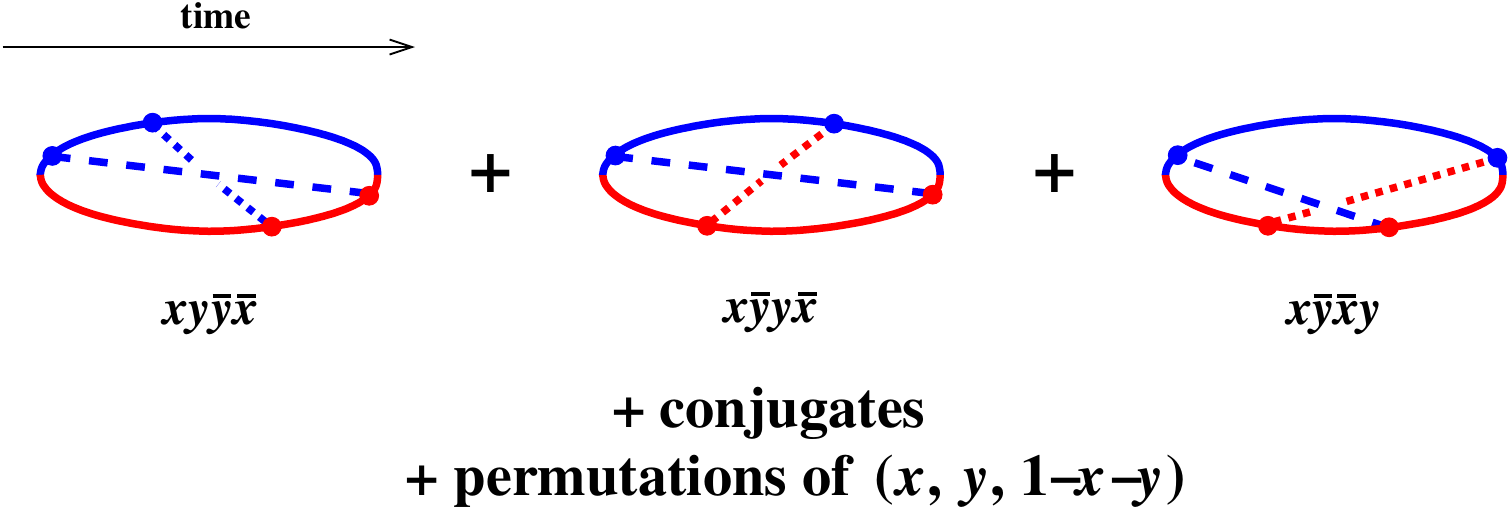}
  \caption{
     \label{fig:subset2}
     An alternative depiction of fig.\ \ref{fig:subset}, with
     amplitudes (blue) and conjugate amplitudes (red) sewn together.
     The dashed lines are colored according to whether they
     were first emitted in the amplitude or conjugate amplitude.
  }
\end {center}
\end {figure}

The preceding work \cite{2brem} developed most of the formalism we will
need for carrying out calculations and then (in approximations
reviewed below) computed the subset of
contributions to the double-bremsstrahlung rate
depicted in fig.\ \ref{fig:subset}.  It is very convenient to alternatively
represent these contributions as in fig.\ \ref{fig:subset2},
where the upper (blue) part of the diagrams depict a contribution to
the amplitude and the lower (red) part depict a contribution to the
conjugate amplitude.  Ref.\ \cite{2brem} referred to these as the
``crossed'' contributions to the rate because the interior lines
are crossed in the representations of fig.\ \ref{fig:subset2}.
In both figs.\ \ref{fig:subset} and \ref{fig:subset2}, we explicitly
show only the high-energy particles; the (many) interactions of those
high-energy particles with the medium are implicit.
(See the introduction of ref.\ \cite{2brem} for more
description.)

\begin {figure}[t]
\begin {center}
  \includegraphics[scale=0.5]{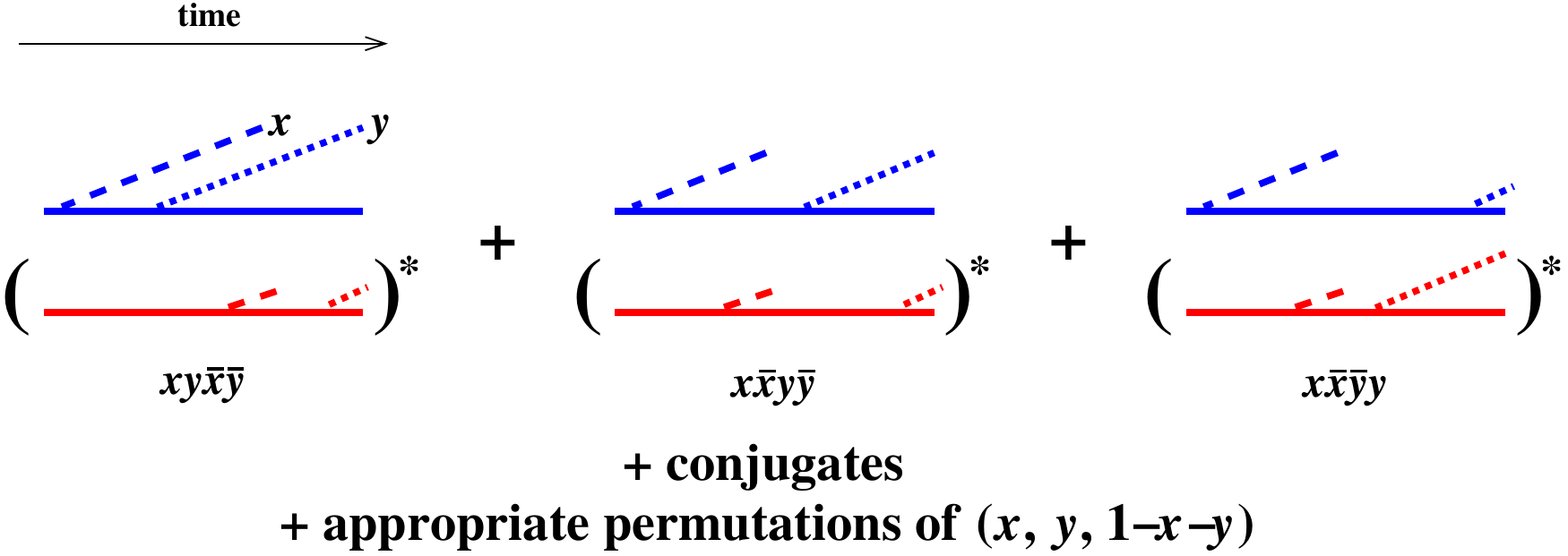}
  \caption{
     \label{fig:seqs}
     The interference contributions evaluated in this paper:
     the ``sequential'' diagrams.
  }
\end {center}
\end {figure}

\begin {figure}[t]
\begin {center}
  \includegraphics[scale=0.5]{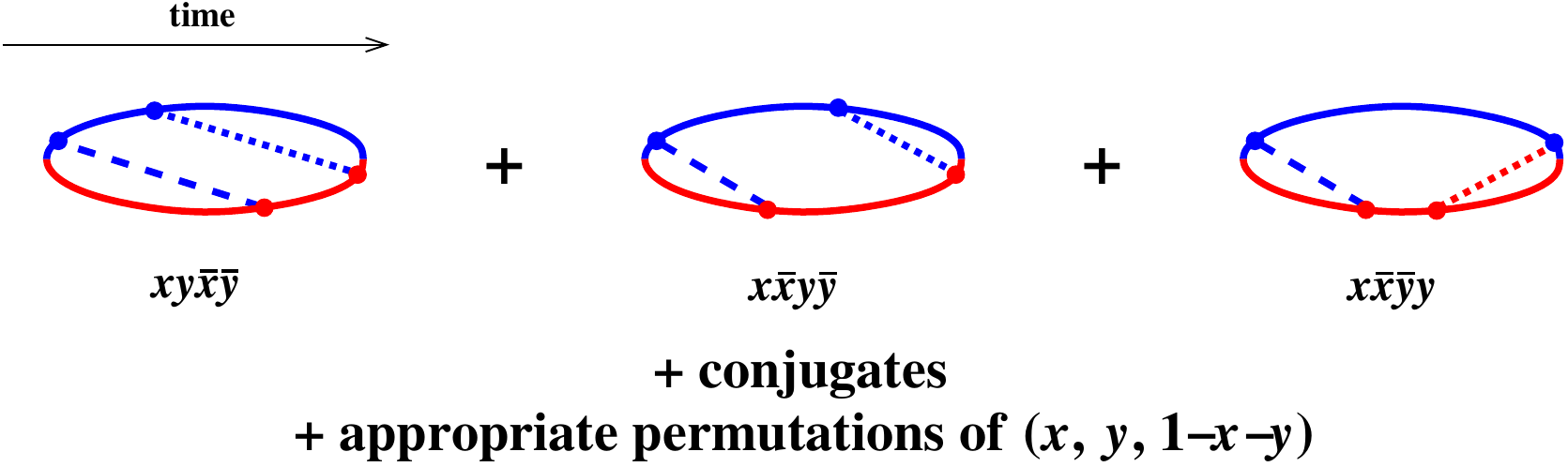}
  \caption{
     \label{fig:seqs2}
     An equivalent representation of fig.\ \ref{fig:seqs}.
  }
\end {center}
\end {figure}

In this paper, we will now evaluate the diagrams of figs.\ \ref{fig:seqs}
and \ref{fig:seqs2},
which we refer to as ``sequential'' contributions because the two
bremsstrahlung emissions happen in the same order in both the amplitude
and conjugate amplitude.
To compute the desired correction $\Delta\,d\Gamma/dx\,dy$ to double
bremsstrahlung due to overlapping formation times, we will need
to subtract from our results the naive calculation of double bremsstrahlung
as two consecutive, quantum-mechanically independent splitting processes,
as in (\ref{eq:DeltaDef}).
In the last two diagrams
($x\bar x y \bar y$ and $x\bar x\bar y y$) of figs.\ \ref{fig:seqs} and
\ref{fig:seqs2},
the $x$ and $y$ bremsstrahlung processes do not overlap
in time.  We will see later that these diagrams roughly, {\it but not
quite}, match up with the idealized Monte Carlo calculation.
Figuring out how to correctly compute the difference will be the
main new technical development required for this paper.

As discussed in the preceding work \cite{2brem}, it is possible to set
up the formalism in a quite general way that would require both
highly non-trivial numerics and a non-trivial treatment of color dynamics
to implement, but
one can proceed much further analytically by making a few
additional approximations.  Though the methods we will discuss in this
paper can be applied more generally, it behooves us to keep things simple
in this first treatment.  So we will follow ref.\ \cite{2brem} when
it comes to explicit calculations, by making the following approximations.
\begin {itemize}
\item
  We will assume that the medium is static, uniform and infinite (which
  in physical terms means approximately uniform over the formation time
  and corresponding formation length).
\item
  We take the large-$\Nc$ limit of QCD to simplify the color dynamics.
  [The specialization of our general result for $\Delta\,d\Gamma/dx\,dy$
  to the soft limiting case $y \ll x \ll 1$ will not depend on
  this assumption.]
\item
  We make the multiple-scattering approximation to interactions with
  the medium, appropriate for very high energies and also known as the
  harmonic oscillator or $\hat q$
  approximation.%
\footnote{
  For a discussion (in different formalism) of double bremsstrahlung
  in the opposite limit---media thin enough
  that the physics is dominated by
  a single interaction with
  the medium---see ref.\ \cite{FOV}.
  See also the related discussion in ref.\ \cite{CPT}.
}
\end {itemize}

\begin {figure}[t]
\begin {center}
  \includegraphics[scale=0.5]{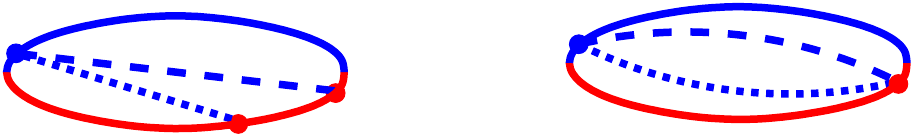}
  \caption{
     \label{fig:4ptExamples}
     Examples of interference contributions involving 4-gluon
     vertices. Such contributions are not evaluated in this paper
     but will be needed for
     a complete calculation of double bremsstrahlung from an initial
     gluon for arbitrary $x$ and $y$.
  }
\end {center}
\end {figure}

Also as in ref.\ \cite{2brem}, we will focus on the case
where the initial high-energy particle
is a gluon (and so, in large-$\Nc$, the final high-energy
particles are also all gluons),
as the resulting final-state permutation symmetries make for
fewer diagrams to consider so far.
However, there is a downside.  In the case of gluons, one must also
consider the 4-gluon interaction, which gives rise to additional
interference contributions.
Examples are given in fig.\ \ref{fig:4ptExamples}.
Because
the calculation of these
additional diagrams would distract from the main point of this paper,
which is how to treat the sequential diagrams in the calculation of
$\Delta d\Gamma/dx\,dy$, we will leave
fig.\ \ref{fig:4ptExamples} for future work \cite{4point}.
It turns out that these contributions are small whenever at least
one of the three final gluons is soft, so the still-incomplete results
that we derive here will have some range of applicability.
But we will need fig.\ \ref{fig:4ptExamples} for
a complete calculation for the case of arbitrary $x$ and $y$.

\begin {figure}[t]
\begin {center}
  \includegraphics[scale=0.5]{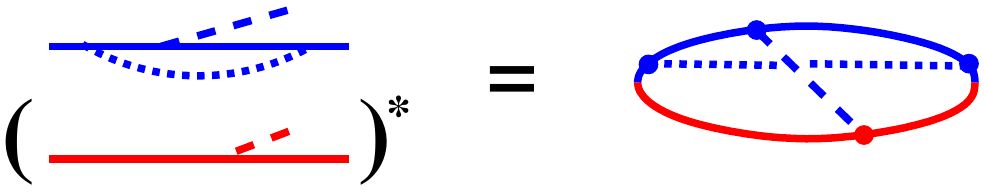}
  \caption{
     \label{fig:virtual}
     An example of a virtual loop correction to single splitting.
  }
\end {center}
\end {figure}

Another problem that we defer for another time is
the change in the {\it single}-bremsstrahlung rate
due to virtual corrections, such
as the one shown in fig.\ \ref{fig:virtual}.  This has been worked
out in the limiting case $y \ll x \ll 1$ in the context of leading parton
average energy loss in refs.\ \cite{Blaizot,Iancu,Wu} and is related
to anomalous scaling of the effective medium parameter $\hat q$
with energy.

Finally, we should mention that the relative transverse momentum of daughters
immediately following a double-splitting has been integrated
in our results.
Though we will make some qualitative comments regarding transverse momenta
later on, we have not calculated transverse momentum distributions.
(If we did not integrate the double-splitting rate
over transverse momenta, the calculation would
be much harder, including the necessity of accounting for decaying color
coherence effects \cite{Blaizot0} occurring {\it after}
the last splitting time in, for example, each of the
diagrams of fig.\ \ref{fig:subset}.
See appendix \ref{app:Blaizot0} for a brief discussion.)


\subsection {Preview of Results}
\label {sec:results}

\begin {figure}[tp]
\begin {center}
  \includegraphics[scale=0.7]{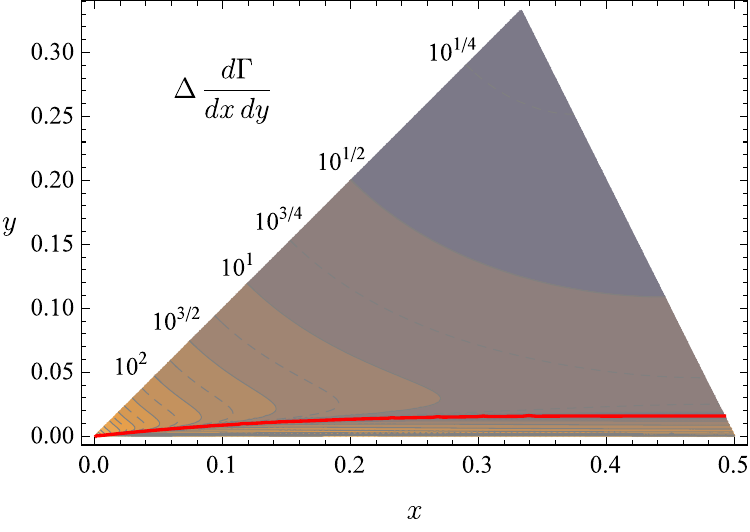} \\[20pt]
  \includegraphics[scale=0.7]{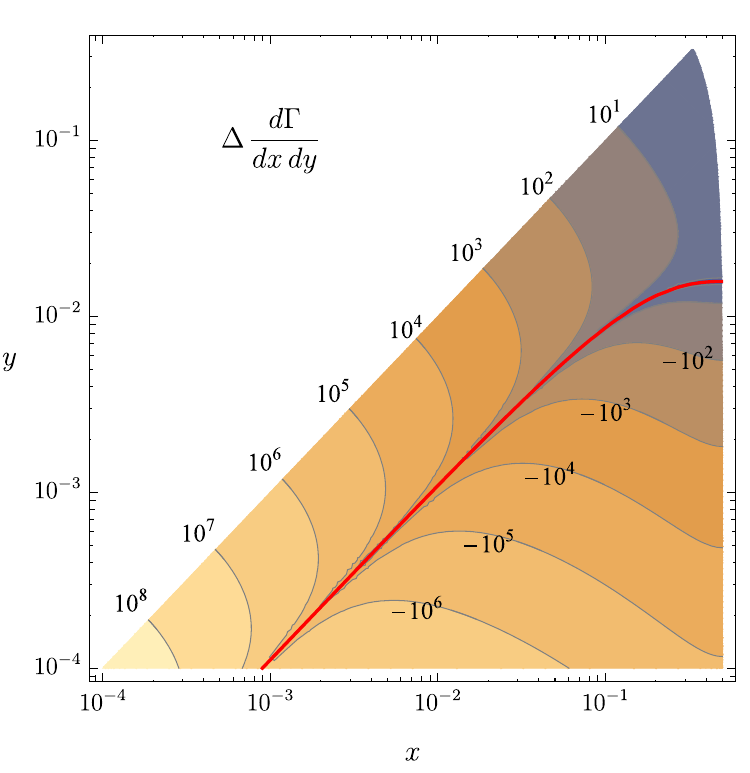}
  \caption{
     \label{fig:result}
     Results for $\Delta\, d\Gamma/dx\,dy$ in units of
     $\CA^2\alphas^2 (\hat q_{\rm A}/E)^{1/2}$, including both crossed and
     sequential contributions (but not yet the 4-gluon vertex contributions
     such as fig.\ \ref{fig:4ptExamples}).
     Since all three final state particles are gluons and so are identical
     particles, we only show results
     for the region $y < x < z \equiv 1{-}x{-}y$.
     (All other orderings are equivalent by permutation.)
     The red line shows where the
     result vanishes, dividing the sub-region of positive corrections from
     the sub-region of negative corrections.
  }
\end {center}
\end {figure}

Numerical results are given in fig.\ \ref{fig:result} for the sum of
the crossed and sequential contributions to the correction
$\Delta\,\Gamma/dx\,dy$ for real double gluon bremsstrahlung from
an initial gluon.%
\footnote{
  As noted in the figure caption, the three final state gluons are identical
  particles.
  Here and throughout the paper, our $\Delta\,\Gamma/dx\,dy$ is normalized
  so that rates are (formally) related to differential rates by
  \[
    \Gamma
    = \int_{y<x<z} dx\>dy\>dz\>\delta(1{-}x{-}y{-}z)\> \frac{d\Gamma}{dx\,dy}
    = \frac{1}{3!}
      \int_0^1 dx\>dy\>dz\>\delta(1{-}x{-}y{-}z)\> \frac{d\Gamma}{dx\,dy}
    \,.
  \]
  We say ``(formally)'' because the correction $\Delta\Gamma$
  to the total rate
  based on $\Delta\,d\Gamma/dx\,dy$ is infrared divergent.
}
As
mentioned above, the effects of 4-gluon vertices (such as in
fig.\ \ref{fig:4ptExamples}) have not yet been included.  We do not
expect these to be important when one of the final gluons is soft, and
so we can already draw the conclusion from fig.\ \ref{fig:result} that
sometimes the correction $\Delta\,\Gamma/dx\,dy$ is positive and
sometimes it is negative, with the corresponding implications for the
relative ease or difficulty of Monte Carlo implementation discussed
earlier at the end of section \ref{sec:MC}.  However, as a pragmatic
matter, note from the figure that one final gluon has to have about 10
times smaller energy than the other two in order to get a negative
correction.  So, if one did implement a correction to real
double bremsstrahlung in Monte Carlo, the most important effects
for shower development (that
are not simply absorbable into the running of $\hat q$ from soft
emissions as discussed in \cite{Blaizot,Iancu,Wu}) may correspond
to the cases of positive correction, which is the more straightforward
case to implement.  However, the most interesting region is where
all three final gluons carry substantial momentum fractions, and
for that case we will need those 4-gluon vertex contributions,
which we have left for later work.

\begin {figure}[tp]
\begin {center}
  \includegraphics[scale=0.7]{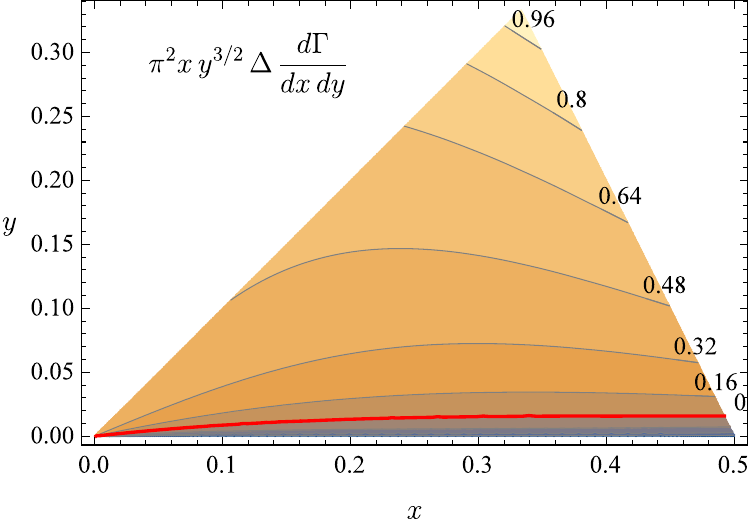}
  \includegraphics[scale=0.7]{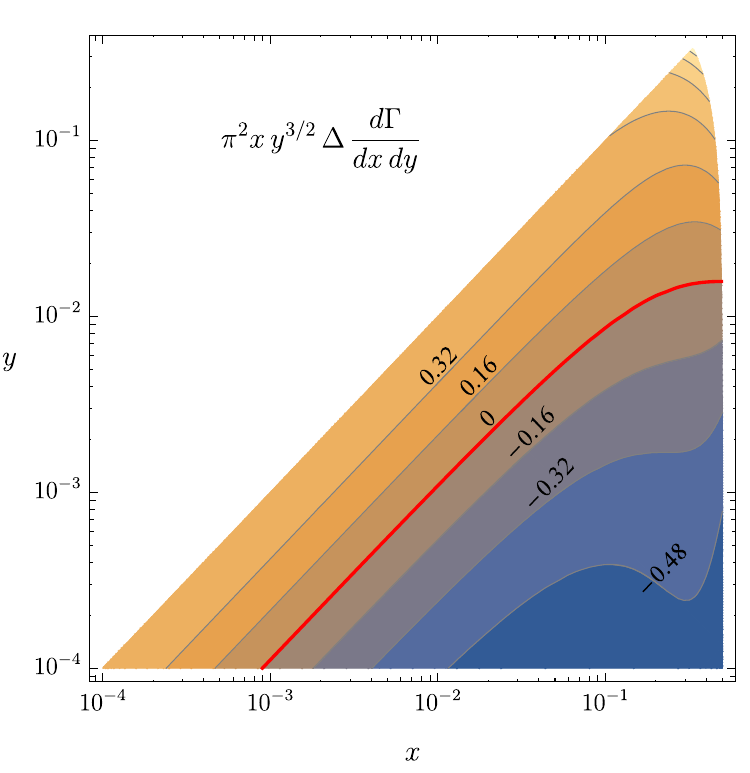}
  \caption{
     \label{fig:resultScaled}
     As fig.\ \ref{fig:result} but removing a factor of
     $1/(\pi^2 x y^{3/2})$ from the answer.
     The apex $(x,y)=(\frac13,\frac13)$ of the triangular region
     above
     corresponds to
     $\pi^2 x y^{3/2} \Delta\,d\Gamma/dx\,dy = 1.05$
     [in units of $\CA^2\alphas^2 (\hat q_{\rm A}/E)^{1/2}$].
  }
\end {center}
\end {figure}

\begin {figure}[tp]
\begin {center}
  \includegraphics[scale=0.6]{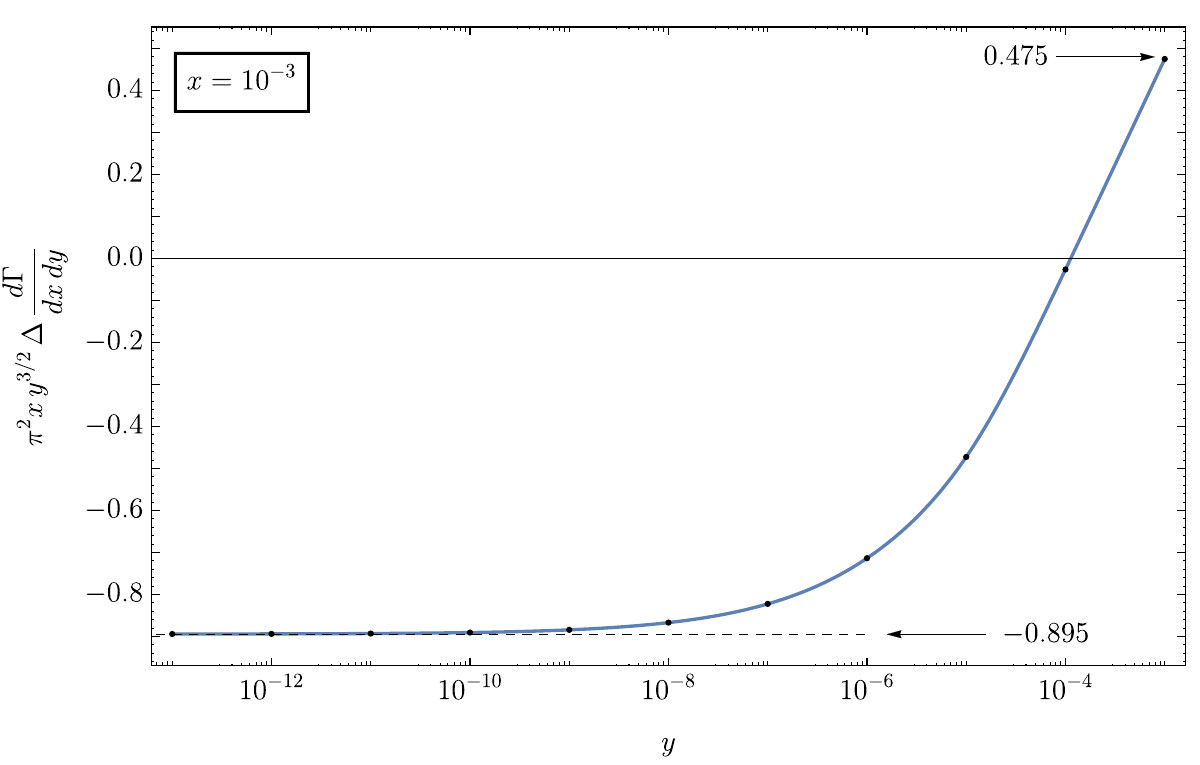}
  \caption{
     \label{fig:result1}
     The same result as fig.\ \ref{fig:resultScaled} but showing the $y$
     dependence for fixed $x = 10^{-3}$.
  }
\end {center}
\end {figure}

In the special limiting case of $y \lll x \ll 1$, the numerical
results approach
\begin {equation}
  \Delta \, \frac{d\Gamma}{dx\,dy}
  \simeq
  - 0.895 \, \frac{\CA^2\alphas^2}{\pi^2 x y^{3/2}}
  \sqrt{\frac{\hat q_{\rm A}}{E}} \,.
\label {eq:smallyx}
\end {equation}
This result is negative, as in the corresponding region of
fig.\ \ref{fig:result}.  In a moment, we will give a simple
general argument about why $\Delta\,d\Gamma/dx\,dy$ could
be expected to scale
parametrically like $1/x y^{3/2}$, as above.  This scaling suggests that
a smoother way to plot our final results, even outside of the
$y \ll x \ll 1$ limit, is to pull out a factor of $1/\pi^2 x y^{3/2}$ from
the answer.
Fig.\ \ref{fig:resultScaled} presents our results that way.
One needs to see a larger range of $y$ to verify the (somewhat slow)
approach to the limit (\ref{eq:smallyx}), which is shown in
fig.\ \ref{fig:result1} for a particular, small fixed value of
$x$.

We should note that the soft limit (\ref{eq:smallyx}) for
the correction $\Delta\,d\Gamma/dx\,dy$ to the double real gluon
bremsstrahlung rate cannot be extracted from early work
\cite{Blaizot,Iancu,Wu} on soft multiple bremsstrahlung because
that work used approximations that are only valid for the
sum of (i) real double gluon emission with (ii) virtual corrections
to single gluon emission and are not valid for (i) and (ii) individually.%
\footnote{
  Though refs.\ \cite{Blaizot,Iancu,Wu}
  give final results that are integrated over
  $y$, it is possible to extract the integrand and so extract
  something one might be tempted to call $d\Gamma/dx\,dy$.
  In particular, it's possible to identify the parts of the calculation that
  correspond to diagrams representing real double emission from those
  representing virtual corrections to single emission.  See
  Appendix F of ref.\ \cite{2brem} for a particular example showing
  how one of our diagrammatic contributions to $d\Gamma/dx\,dy$
  exactly matches, in the $y \ll x \ll 1$ limit,
  a corresponding piece of the calculation of Wu \cite{Wu}.
  However, this cannot be done for all contributions to
  $d\Gamma/dx\,dy$ because the methods of earlier work such
  as \cite{Iancu,Wu} assumed that the colors of the
  highest-momentum particle in the
  amplitude and the highest-momentum particle in the conjugate amplitude
  jointly combine into the adjoint representation ($A \otimes A \to A$).
  This is a valid assumption (at leading-log order in the $y \ll x \ll 1$
  limit) only if
  one adds together real and virtual emission diagrams.  See
  Appendix F of ref.\ \cite{2brem}.
}
Also, our result (\ref{eq:smallyx}) depends on diagrams
that were not evaluated in refs.\ \cite{Blaizot,Iancu,Wu}, such as the
$x\bar y\bar x y$ diagram of figs.\ \ref{fig:subset} and \ref{fig:subset2}.

The integral formula we will derive for $\Delta\,d\Gamma/dx\,dy$ in
the general case is
a complicated expression that is painstaking to implement.  In
Appendix \ref{app:approx}, we provide, as an alternative,
a relatively simple analytic formula that has been fitted to approximate
fig.\ \ref{fig:resultScaled} very well.


\subsection {Why \boldmath$1/x y^{3/2}$?}
\label {sec:picture}

Before moving on to the details of calculations, we give one
crude but simple explanation of why the rate for overlapping emissions
has parametric dependence
$(d\Gamma/dx\,dy)_{\rm overlap} \sim 1/x y^{3/2}$ when $y < x < z \equiv 1{-}x{-}y$.

First, let's review some features of the LPM effect for
the usual case of {\it single} bremsstrahlung, with $x$ representing
the less energetic of the two daughters.
In the QCD case, the formation time for this bremsstrahlung process
is (for thick media) parametrically%
\footnote{
  See, for example, Baier \cite{BaierNote} for the QCD formation
  time expressed in terms of $\hat q$, or the discussion leading
  up to eq.\ (4.15) in the review \cite{PeigneSmilga} (where
  $\mu^2/\lambda \sim \hat q$ and $\omega = x E$).
  Here's one quick argument: The least energetic daughter is the most
  easily deflected.  If it picks up transverse momentum of order
  $Q_\perp$ from the medium during the splitting process, the angular
  separation between it and the other daughter will be of order the
  ratio of its transverse and longitudinal momenta:
  $\theta \sim Q_\perp/xE$.  Over time $t$, the transverse separation
  of the $x$ daughter from the trajectory the parent would have
  followed then grows to be of order $b \sim \theta t \sim Q_\perp t/xE$.
  The quantum coherence of the emission process ceases when this
  separation can first be resolved by the transverse momentum
  $Q_\perp$ of the splitting process---that is, when $b \sim 1/Q_\perp$.
  This condition defines the formation time $\tform$.  By
  definition of $\hat q$, the average $Q_\perp^2 \sim \hat q t$.
  Combining the above relations using $t = \tform$ yields
  (\ref{eq:tform}).
}
\begin {equation}
   \tformx \sim \sqrt{\frac{x E}{\hat q}} \,.
\label {eq:tform}
\end {equation}
Each formation time offers one opportunity for splitting, with
probability of order $\alphas$.
So the rate for emission of a daughter with momentum fraction of order
$x$ is
\begin {equation}
   \Gamma_x \sim \frac{\alphas}{\tformx}
   \sim \alphas \sqrt{\frac{\hat q}{x E}} \,,
\label {eq:Gammax}
\end {equation}
corresponding to%
\footnote{
   \label {foot:IR}
   A quick note on infrared cut-offs:
   The power-law divergence of (\ref{eq:Gammax}) as $x \to 0$ may
   at first sight look like the LPM effect is causing an enhancement
   of the splitting rate in this limit, but the LPM effect is always
   a suppression.  A useful way to understand this is to rewrite
   (\ref{eq:Gammax}) as
   \[
      \Gamma_x \sim \frac{\alphas}{\tau_{\rm el}} \times \frac{1}{N}
   \]
   where $\tau_{\rm el}$ is the mean free time between elastic collisions
   with the medium and $N \sim \tform / \tau_{\rm el}$ is the number
   of elastic collisions during the formation time.
   The first factor $\alphas/\tau_{\rm el}$ is (parametrically)
   the rate in the absence
   of the LPM effect, in which case each collision offers an independent
   opportunity for bremsstrahlung.  The $1/N$ factor is the LPM suppression,
   and the above analysis {\it assumed} $N \gg 1$
   (i.e.\ $\tform \gg \tau_{\rm el}$) so that we could, for instance, take
   $Q_\perp \sim \sqrt{\hat q \tform}$ in the preceding footnote.
   When this assumption fails completely ($t_{\rm form} \ll \tau_{\rm el}$),
   the rate is simply the unsuppressed rate
   $\Gamma_x \sim \alphas/\tau_{\rm el}$.
}
\begin {equation}
   \frac{d\Gamma_x}{dx} \sim \frac{\alphas}{x \tformx}
   \sim \frac{\alphas}{x} \, \sqrt{\frac{\hat q}{x E}} \,.
\label {eq:dGammax}
\end {equation}

Now consider double bremsstrahlung with $y < x$.
From (\ref{eq:tform}), the $y$ formation time will be shorter than
the $x$ formation time.
The probability that
an emission with momentum fraction of order $y$ happens
{\it during}\/ the $x$ formation
time $\tformx$ is then given by
\begin {equation}
   \Gamma_y \, \tformx
   \sim \frac{\alphas}{\tformy} \, \tformx ,
\label {eq:Proby}
\end {equation}
corresponding to
\begin {equation}
   \frac{d\Gamma_y}{dy} \, \tformx
   \sim \frac{\alphas}{y \tformy} \, \tformx .
\label {eq:dProby}
\end {equation}
Multiplying (\ref{eq:dGammax}) and (\ref{eq:dProby}) gives
the rate for overlapping $x$ and $y$ emissions:
\begin {equation}
   \left[ \frac{d\Gamma}{dx\,dy} \right]_{\rm overlap}
   \sim \frac{d\Gamma_x}{dx} \times \frac{d\Gamma_y}{dy} \, \tformx
   \sim \frac{\alphas^2}{x y \tformy}
   \sim \frac{\alphas^2}{x y^{3/2}} \sqrt{\frac{\hat q}{E}}
   \,.
\end {equation}
If the presence of the $y$ emission has a significant effect on
the $x$ emission (or vice versa),
then the correction $\Delta\,d\Gamma/dx\,dy$ defined
by (\ref{eq:DeltaDef}) would then also be
\begin {equation}
   \Delta \, \frac{d\Gamma}{dx\,dy}
   \sim \frac{\alphas^2}{x y^{3/2}} \sqrt{\frac{\hat q}{E}}
   \,.
\label {eq:parametric}
\end {equation}

This is indeed the parametric behavior (\ref{eq:smallyx}) of
our result in this paper.  It turns out that, separately, the
crossed and sequential diagrams are each
logarithmically enhanced,
\begin {equation}
   \left[ \frac{d\Gamma}{dx\,dy} \right]_{\rm crossed}
   \sim \left[ \Delta \frac{d\Gamma}{dx\,dy} \right]_{\rm sequential}
   \sim \frac{\alphas^2}{x y^{3/2}} \sqrt{\frac{\hat q}{E}} \,
        \ln\left( \frac{x}{y} \right)
\label {eq:LogEnhanced}
\end {equation}
for $y \ll x \ll 1$, but the logarithmically-enhanced contributions
cancel each other in the total (\ref{eq:parametric}).
Appendix \ref{app:log} discusses how the logarithmic enhancement of
various individual contributions can
be related to collinear logarithms from DGLAP evolution and fragmentation,
and how there is a Gunion-Bertsch-like \cite{GB} cancellation of those
logarithms in the total result.
This is not to say that collinear logarithms are never relevant,
just that they do not appear at the order $\alphas^2/x y^{3/2}$
of (\ref{eq:parametric}).  They do appear, notably, in calculations
\cite{Blaizot,Iancu,Wu} of energy loss.  In that case, the effect of the
leading contribution (\ref{eq:parametric}) to energy loss from
double bremsstrahlung is canceled by the leading
contribution from virtual corrections to the single bremsstrahlung rate,
and so it is the sub-leading behavior of each which becomes important.

Finally, note using (\ref{eq:tform})
that the total probability (\ref{eq:Proby})
of emitting a $y$ during the
formation time of $x$ is of order
\begin {equation}
   \Gamma_y \, \tformx
   \sim \alphas \sqrt{\frac{x}{y}} .
\end {equation}
So one would need to deal with resumming multiple $O(y)$ emissions during
the $x$ formation time if interested in $y \lesssim \alphas^2 x$.
We will not treat that case in this paper.%
\footnote{
   Resummation of small $y$ corrections has been discussed in
   refs.\ \cite{Wu0,Blaizot1,Blaizot,Iancu,Wu,Iancu2} in the
   context of the effective running of $\hat q$ and the
   average energy loss of the leading parton.
   The kinematics is different for
   that than for the isolated double bremsstrahlung rate analyzed
   in this paper,
   due to canceling contributions to energy loss (in the soft limit)
   from virtual corrections to single bremsstrahlung.
   For the running of $\hat q$ and leading-parton energy loss,
   resummation is only necessary
   to address the contribution
   from $y$'s that are so small that $\alphas \ln^2(x/y) \gg 1$.
}
(When discussing the mathematical behavior of our results, we will
nonetheless discuss some absurdly small values of $y$, such as in
fig.\ \ref{fig:result1}.  Our results for such small $y$
directly apply only to the formal limit that the $\alphas$ associated with
splitting is arbitrarily small.%
\footnote{
  Another reason that our results for such very small $y$ are only of
  formal interest is that the $y$ formation time
  $t_{{\rm form},y} \sim \sqrt{yE/\hat q}$ must remain large compared
  to the elastic mean free path $\tau_{\rm el}$ in order for the
  $\hat q$ approximation to be valid.  (See footnote \ref{foot:IR}.)
}%
)


\section {The Calculation}
\label {sec:calculation}

In this section, we turn to the specifics of calculating the
sequential diagrams of fig.\ \ref{fig:seqs2}.  The first
interference diagram in the figure, $xy\bar x\bar y$, can
be evaluated by mostly-straightforward application of the
methods of the preceding work \cite{2brem}.  We will leave
the details for later, in section \ref{sec:xyxy} and appendix
\ref{app:xyxy}.
The only new subtlety of this diagram, as opposed to
the crossed diagrams evaluated in ref.\ \cite{2brem},
is that there are two inequivalent ways that color can be routed
in the large-$\Nc$ approximation, which must be accounted for
appropriately.

To begin, we will instead focus on the other explicit
diagrams of fig.\ \ref{fig:seqs2},
$x\bar x y \bar y + x\bar x \bar y y$, as these are the
diagrams that involve the most significant new issue: careful attention
to the subtraction of $(d\Gamma/dx\,dy)_{\rm IMC}$ from
$d\Gamma/dx\,dy$.


\subsection {\boldmath$2\Re(x \bar xy\bar y+x\bar x \bar y y)$ vs.\
  idealized Monte Carlo}
\label {sec:xxyy}

\begin {figure}[t]
\begin {center}
  \begin{picture}(250,150)(0,0)
  \put(0,0){\includegraphics[scale=0.5]{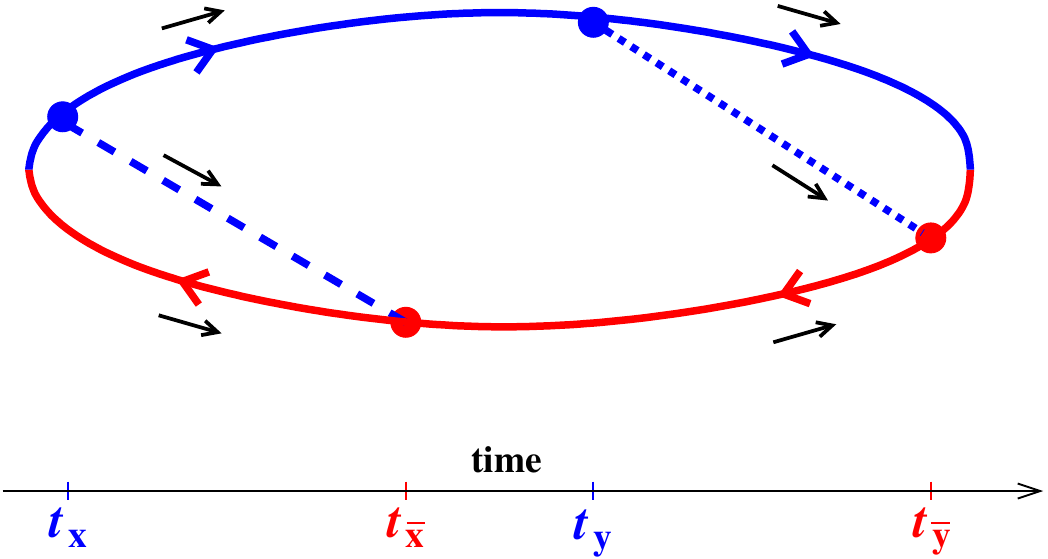}}
  \put(32,44){${-}1$}
  \put(45,100){$x$}
  \put(30,138){$1{-}x$}
  \put(180,80){$y$}
  \put(170,138){$z\equiv 1{-}x{-}y$}
  \put(165,44){${-}(1{-}x)$}
  \end{picture}
  \caption{
     \label{fig:xxyy1}
     The $x\bar x y\bar y$ interference, showing longitudinal momentum
     fractions $x_i$ and our notation for the vertex times.
  }
\end {center}
\end {figure}
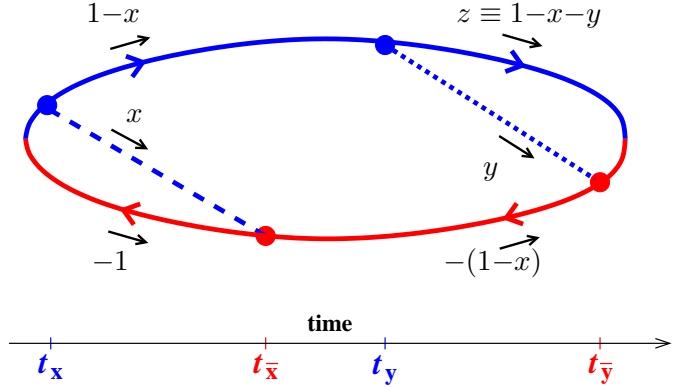

Start by considering
$x\bar x y\bar y$.  Our convention for labeling time
in this diagram is shown in fig.\ \ref{fig:xxyy1}.
Following the philosophy of the preceding paper \cite{2brem}, we
may interpret the evolution between vertex times to be described
by 2-dimensional non-relativistic non-Hermitian quantum mechanics,
with the appropriate number of particles.  In the case of
fig.\ \ref{fig:xxyy1}, it is a
(i) 3-particle problem for $t_\xx < t < t_\xbx$, (ii) 2-particle problem
for $t_\xbx < t < t_\yx$, and (iii) 3-particle problem for
$t_\yx < t < t_\ybx$.  Also following ref.\ \cite{2brem}, we may use
symmetry to reduce the $N$ particle problems to $(N{-}2)$ particle
problems, leaving us with (i) 1-particle, (ii) 0-particle, and
(iii) 1-particle problems respectively.

At this point, we could implement the methods of the preceding paper
\cite{2brem} to turn fig.\ \ref{fig:xxyy1} into explicit equations and then
start turning the crank.  We do this in appendix \ref{app:xxyy} for
the sake of concreteness.
However, many of these details can be sidestepped if we take
a slightly looser approach, which is how we will proceed
here in the main text.

The important point is that,
for the
$x\bar x y\bar y$ diagram of fig.\ \ref{fig:xxyy1}
(and similarly for the $x\bar x \bar y y$ diagram in fig.\ \ref{fig:seqs2}),
the time interval $(t_\xx, t_\xbx)$
of the $x$ emission does not overlap with the time interval
$(t_\yx,t_\ybx)$ of the $y$ emission.
So the times of these events are unrelated---there is no reason the
$y$ emission cannot occur a very long time after the $x$ emission.
In the formalism, this is because evolution during
the intermediate time interval $t_\xbx < t < t_\yx$
corresponds to a problem with effectively {\it zero}\/ particles
and so does not have any time dependence: the $x\bar x y \bar y$
interference contribution does not care how far apart
$t_\xbx$ and $t_\yx$ are.
This is consistent with interpreting this diagram as
representing two consecutive splittings that are completely independent
from each other, as in an idealized Monte Carlo calculation.


\subsubsection {A crude correspondence}

\begin {figure}[t]
\begin {center}
  \includegraphics[scale=0.7]{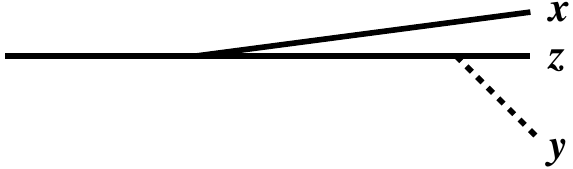}
  \caption{
     \label{fig:xxyyMC}
     Idealized Monte Carlo contribution related to
     $x\bar x y\bar y + x\bar x \bar y y + \bar xx \bar yy + \bar x xy\bar y
      = 2\Re(x\bar x y\bar y + x\bar x \bar y y)$.
  }
\end {center}
\end {figure}

\begin {figure}[t]
\begin {center}
  \includegraphics[scale=0.5]{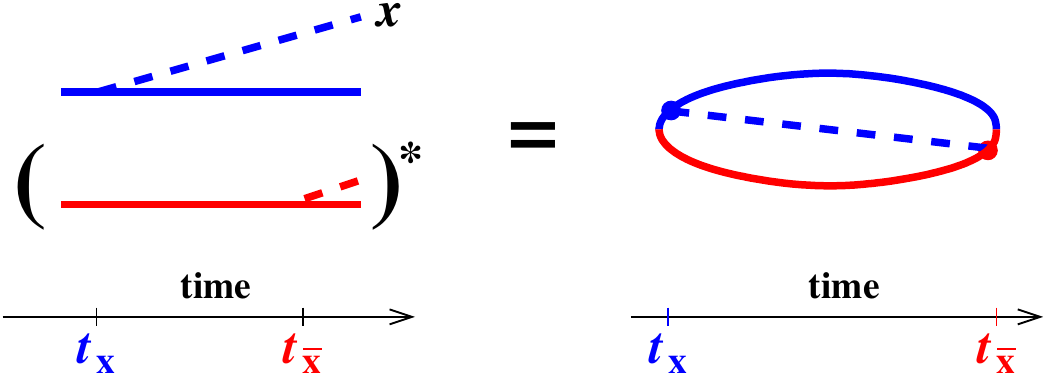}
  \caption{
     \label{fig:xxrate}
     The $x\bar x$ interference contribution to single splitting.
  }
\end {center}
\end {figure}

In particular,
the $x\bar x y \bar y + x\bar x\bar y y$ cases (plus conjugates)
in figs.\ \ref{fig:seqs} and \ref{fig:seqs2}
are related to the idealized Monte Carlo contribution
of fig.\ \ref{fig:xxyyMC},
where a first splitting,
$E \to (1{-}x) E$ and $xE$, is
followed later by an independent second splitting,
$(1{-}x)E \to zE$ and $yE$.
In what follows, it will be convenient to introduce the longitudinal
momentum fraction $\yfrak$
of the $y$ daughter with respect to its immediate parent
in the second splitting,
\begin {equation}
   \yfrak \equiv \frac{y}{1-x} \,.
\label {eq:yfrak}
\end {equation}
That is, the $y$ daughter has energy $y E = \yfrak(1-x)E$.
In the language that we use for labeling diagrams,
single-splitting rates are given by
$2 \Re(x\bar x) = x\bar x + \bar x x$,
with $x\bar x$ depicted in fig.\ \ref{fig:xxrate}.
The correspondence here between double-splitting diagrams and Monte
Carlo is, crudely speaking,
\begin {equation}
   \left[\frac{d\Gamma}{dx\,d\yfrak}\right]_{
        \begin{subarray}{} x\bar x y\bar y + x\bar x \bar y y \\
                        + \bar xx \bar yy + \bar x xy\bar y \end{subarray}
   }
   \approx
   \left[\frac{d\Gamma}{dx\,d\yfrak}\right]_{\rm corresponding~IMC}
   =
           \left[\frac{d\Gamma}{dx}\right]_E \times
           \Time \,
           \left[\frac{d\Gamma}{d\yfrak}\right]_{(1-x)E} ,
\label {eq:xxyyvMC}
\end {equation}
where $\Time$ formally represents the amount of time between
the first splitting and the end of eternity (a regularization that
we will soon have to treat more carefully).
The right-hand side of (\ref{eq:xxyyvMC}) is the single-splitting rate
$[d\Gamma/dx]_E$ for the first splitting (the $x$ emission)
times the single-splitting
probability $\Time \, [d\Gamma/d{\mathfrak y}]_{(1-x)E}$
for the second splitting ($y$ emission).
The subscripts $E$ and $(1-x)E$ on the
single-splitting rates indicate the energy of the parent.
Using (\ref{eq:yfrak}), the rate
(\ref{eq:xxyyvMC}) can be recast in the form
\begin {equation}
   \left[\frac{d\Gamma}{dx\,dy}\right]_{
        \begin{subarray}{} x\bar x y\bar y + x\bar x \bar y y \\
                        + \bar xx \bar yy + \bar x xy\bar y \end{subarray}
   }
   \approx
   \left[\frac{d\Gamma}{dx\,dy}\right]_{\rm corresponding~IMC}
   = \frac{\Time}{1-x} \,
           \left[\frac{d\Gamma}{dx}\right]_E \,
           \left[\frac{d\Gamma}{d\yfrak}\right]_{(1-x)E} .
\label {eq:xxyyvMC2}
\end {equation}
As in the earlier discussion in section \ref{sec:MC}, the infinite
quantity $\Time$ will only make a temporary appearance along the
way towards finding the correction $\Delta\, d\Gamma/dx\,dy$ to
idealized Monte Carlo.

Let's now use some more explicit formulas for the single-splitting rates
appearing on the right-hand side of (\ref{eq:xxyyvMC2}).
In notation similar to that of the preceding paper \cite{2brem},%
\footnote{
  See in particular
  eq.\ (2.37) of ref.\ \cite{2brem}.
}
\begin {equation}
   \left[\frac{d\Gamma}{dx}\right]_E
   =
   \frac{\alpha P(x)}{[x(1-x)E]^2}
   \Re \int_0^\infty d(\Delta t) \>
   \grad_{\B^\xbx} \cdot \grad_{\B^\xx}
   \langle \B^\xbx,\Delta t | \B^\xx,0 \rangle_{E,x}
   \Bigr|_{\B^\xbx = \B^\xx = 0} ,
\label {eq:1brem}
\end {equation}
corresponding to twice the real part of the $x\bar x$ diagram for
single splitting, shown in fig.\ \ref{fig:xxrate}.  The $x\bar x$
contribution alone is%
\footnote{
   It will not be important here in the main text, but, for some
   normalization issues associated with applying (\ref{eq:1brem})
   and (\ref{eq:1bremxx}) to $[d\Gamma/d\yfrak]_{(1-x)E}$,
   see appendix \ref{app:xxyy}.
}
\begin {equation}
   \left[\frac{d\Gamma}{dx}\right]_{x\bar x,E}
   =
   \frac{\alpha P(x)}{2[x(1-x)E]^2}
   \int_0^\infty d(\Delta t) \>
   \grad_{\B^\xbx} \cdot \grad_{\B^\xx}
   \langle \B^\xbx,\Delta t | \B^\xx,0 \rangle_{E,x}
   \Bigr|_{\B^\xbx = \B^\xx = 0} .
\label {eq:1bremxx}
\end {equation}
Here, $P(x)$ is the relevant (helicity-averaged) DGLAP splitting function.
$\B$ is the single transverse position associated with the
``effectively 1-particle'' description of this process.
The subscripts at the end of $\langle \B^\xbx,\Delta t | \B^\xx,0 \rangle_{E,x}$
are a reminder (which will be useful in a moment)
of what energy and branching fraction should be used in
the calculation of the propagator for $\B$.
The integration variable
$\Delta t$ in (\ref{eq:1brem}) represents the time difference
$t_\xbx - t_\xx$ in fig.\ \ref{fig:xxrate}.
We will later give more explicit formulas for (\ref{eq:1brem})
in the multiple scattering
($\hat q$) approximation, but for the moment there is no reason not to
be general.  To simplify notation in what follows,
we will rewrite (\ref{eq:1brem}) generically as
\begin {equation}
   \left[\frac{d\Gamma}{dx}\right]_E
   =
   \Re \int_0^\infty d(\Delta t) \>
   \left[\frac{d\Gamma}{dx\,d(\Delta t)}\right]_E \,,
\label {eq:1bremB}
\end {equation}
where the complex-valued $d\Gamma/dx\,d(\Delta t)$ is defined by
\begin {equation}
   \left[\frac{d\Gamma}{dx\,d(\Delta t)}\right]_E \equiv
   \frac{\alpha P(x)}{[x(1-x)E]^2}
   \grad_{\B^\xbx} \cdot \grad_{\B^\xx}
   \langle \B^\xbx,\Delta t | \B^\xx,0 \rangle_{E,x}
   \Bigr|_{\B^\xbx = \B^\xx = 0} .
\label {eq:dGdDt}
\end {equation}

Putting the integral form (\ref{eq:1bremB}) of the single bremsstrahlung rate
into the right-hand
side of (\ref{eq:xxyyvMC2}), we can write the idealized Monte Carlo
result for double bremsstrahlung as
\begin {align}
   \left[\frac{d\Gamma}{dx\,dy}\right]_{\rm corresponding~IMC}
   = \frac{\Time}{(1-x)} &
     \int_0^\infty d(\Delta t_\xx)
     \int_0^\infty d(\Delta t_\yx)
\nonumber\\& \times
     \Re \left[\frac{d\Gamma}{dx\,d(\Delta t_\xx)}\right]_E \,
     \Re \left[\frac{d\Gamma}{d\yfrak\,d(\Delta t_\yx)}\right]_{(1-x)E}
\label {eq:MC1}
\end {align}
with the interpretations
\begin {equation}
   \Delta t_\xx \equiv |t_\xbx-t_\xx|
   \quad \mbox{and} \quad
   \Delta t_\yx \equiv |t_\ybx - t_\yx| .
\label {eq:Deltats}
\end {equation}
As we verify explicitly in appendix \ref{app:xxyy}, the above
idealized Monte Carlo contribution corresponds to the result for the
diagrams
$x\bar x y\bar y + x\bar x \bar y y + \bar xx \bar yy + \bar x xy\bar y
 = 2\Re(x\bar x y\bar y + x\bar x \bar y y)$,
with one critically important difference.
To explain that difference, look at the piece of
(\ref{eq:MC1}) corresponding just to $x\bar x$ followed by $y\bar y$
[as opposed to the full $2\Re(x\bar x) = x\bar x+\bar x x$ followed by
$2\Re(y\bar y) = y\bar y+\bar y y$]:
\begin {align}
   \left[\frac{d\Gamma}{dx\,dy}\right]_{
          \begin{subarray}{}
             \rm IMC~corresponding\\ to~x\bar xy\bar y\
          \end{subarray}}
   = \frac{\Time}{1-x} &
     \int_0^\infty d(\Delta t_\xx)
     \int_0^\infty d(\Delta t_\yx)
\nonumber\\& \times
     \frac12 \left[\frac{d\Gamma}{dx\,d(\Delta t_\xx)}\right]_E \,
     \frac12 \left[\frac{d\Gamma}{d\yfrak\,d(\Delta t_\yx)}\right]_{(1-x)E}
   .
\label {eq:MC2}
\end {align}
The actual result from the $x\bar x y\bar y$ diagram is the same
{\it except}\/ for the constraints on the time integration
in fig.\ \ref{fig:xxyy1}:
\begin {align}
   \Time \int_0^\infty \! d(\Delta t_\xx) \int_0^\infty \! d(\Delta t_\yx)
   \>\cdots
   \qquad &\mbox{in (\ref{eq:MC2})}
\nonumber\\
   \longrightarrow \quad
   \int_{t_\xx < t_\xbx < t_\yx < t_\ybx} \! dt_\xbx \, dt_\yx \, dt_\ybx
   \>\cdots
   \qquad &\mbox{in $x\bar xy\bar y$ (fig.\ \ref{fig:xxyy1}).}
\label {eq:timeints}
\end {align}
(There are only three time integrals on the right-hand side because of
our focus on computing rates rather than probabilities in this paper,
and our associated assumption of time translation invariance over
relevant time scales.)

If we are sloppy, we might be tempted to conclude that these two types
of time integrations are the same by (i) rewriting the right-hand side
of (\ref{eq:timeints}) as
\begin {equation}
  \int_0^\infty \! d(t_\xbx-t_\xx)
  \int_0^\infty \! d(t_\yx-t_\xbx)
  \int_0^\infty \! d(t_\ybx-t_\yx) \>\cdots ;
\label {eq:timeint2}
\end {equation}
(ii) noticing that the integrand in (\ref{eq:MC2})
does not depend on $t_\yx - t_\xbx$,
just as previously discussed concerning the
propagation of the effectively 0-particle intermediate state in
the diagram for $x\bar x y\bar y$;
and so (iii) replacing the $\int d(t_\yx-t_\xbx)$ in (\ref{eq:timeint2})
by $\Time$ to get the left-hand side of (\ref{eq:timeints}).
The last step is not quite correct, however, because the maximum allowed
duration of the intermediate
time interval $(t_\xbx,t_\yx)$ in fig.\ \ref{fig:xxyy1}
will depend, for example, on how much of the available time
$\Time$ is being used up by the following time interval
$(t_\yx,t_\ybx)$, since $t_\ybx$ must also occur before the end of eternity.
The integrand in (\ref{eq:MC2}) will fall off exponentially when the
duration $t_\ybx-t_\yx$ of the last interval becomes large compared
to the $y$ emission formation time, and so that interval's effect on the
sloppy result
\begin {equation}
  \int_0^\infty \! d(t_\yx-t_\xbx) \approx \Time
\end {equation}
is only an edge effect, changing the right-hand side by subtracting
an amount of order $\Time^0$.
Such an edge effect would be negligible (relatively speaking)
as $\Time \to \infty$
were we interested only
individually in
the formal result for $x\bar xy\bar y$ or the corresponding
idealized Monte Carlo
contribution (results which are individually
physically nonsensical as $\Time\to\infty$).
However, we are instead interested in
the difference $\Delta\, d\Gamma/dx\,dy$, for which the
leading $O(\Time)$ pieces {\it cancel}, leaving behind the
$O(\Time^0)$ corrections, which then have a sensible
$\Time \to \infty$ limit.

We could attempt to proceed by figuring out how to somehow consistently
carry through the calculation with a sharp cut-off on time, as
we were sloppily considering above.  This introduces many headaches.%
\footnote{
  One headache is that there is usually difficulty with radiation
  fields whenever
  you allow a charged particle to suddenly appear or disappear.
  Another is that we've just seen that edge effects will contribute to
  our answer, but if a single splitting takes place right at
  the edge of time, the presence of the edge will affect its rate.
  For instance, if $t_\yx$ in fig.\ \ref{fig:xxyy1}
  occurs a third of a formation time before
  the end of time, then there is no room for $t_\ybx-t_\yx$ to stretch
  out as far as one formation time.  So you would not reproduce the
  single splitting rate at the edge, making comparison with
  idealized Monte Carlo
  problematic.
}
The path
of least confusion is to introduce a smooth and physically-realizable
cut-off.  One regularization method is to imagine a situation where the strength
$\hat q$ of medium interactions very slowly falls off at large times to reach
the vacuum.  For example, take
\begin {equation}
   \hat q(t) = \hat q_0 \, e^{-t^2/\Time^2} .
\label {eq:qhatt0}
\end {equation}
The prescription that then arises in the $\Time\to\infty$
limit turns out to be reasonably intuitive and straightforward to implement.


\subsubsection {The prescription}
\label {sec:prescription}

Here, we will state the prescription and use it to find
$\Delta\,d\Gamma/dx\,dy$.
We leave to appendix \ref{app:justify}
a more thorough justification,
based on slowly-varying choices of $\hat q(t)$ like (\ref{eq:qhatt0}).

Let $\tau_x$ and $\tau_y$ be the (instantaneous) times of the
$x$ and $y$ emission
in the idealized Monte Carlo treatment of fig.\ \ref{fig:xxyyMC}.
When comparing to the the calculation of
$2\Re(x\bar xy\bar y + x\bar x\bar y y)$, the prescription is to
identify these Monte Carlo times
with the {\it midpoints}\/ of the corresponding time intervals
$(t_\xx,t_\xbx)$ and $(t_\yx,t_\ybx)$ of the emissions.  That is,
\begin {equation}
   \tau_x = \frac{t_\xx + t_\xbx}{2} \,,
   \qquad
   \tau_y = \frac{t_\yx + t_\ybx}{2} \,.
\label {eq:tauxy}
\end {equation}
Now remember that the $\Time$ in the idealized Monte Carlo formula
(\ref{eq:xxyyvMC}) was just an attempt to regularize the integral
\begin {equation}
   \int_0^\infty d(\tau_\yx-\tau_\xx) \approx \Time .
\end {equation}
Instead of giving that integral a name $\Time$, let's
just step back and write this integral in place of $\Time$.
We can then combine it with the $\Delta t$ integrals in
the more explicit idealized Monte Carlo result (\ref{eq:MC2}) to
recast (\ref{eq:MC2}) as
\begin {align}
   \left[\frac{d\Gamma}{dx\,dy}\right]_{
          \begin{subarray}{}
             \rm IMC~corresponding\\ to~x\bar xy\bar y\
          \end{subarray}}
   = \frac{1}{1-x} &
     \int_{\tau_\xx < \tau_\yx} \! dt_\xbx \, dt_\yx \, dt_\ybx
\nonumber\\& \times
     \frac12 \left[\frac{d\Gamma}{dx\,d(\Delta t_\xx)}\right]_E \,
     \frac12 \left[\frac{d\Gamma}{d\yfrak\,d(\Delta t_\yx)}\right]_{(1-x)E}
\label {eq:MC3}
\end {align}
with the $\Delta t$'s and $\tau$'s defined here by
(\ref{eq:Deltats}) and (\ref{eq:tauxy}).
In contrast, the result of the $x\bar x y\bar y$ diagram is given
using the integral on the right-hand side of (\ref{eq:timeints}),
\begin {align}
   \left[\frac{d\Gamma}{dx\,dy}\right]_{x\bar x y\bar y}
   = \frac{1}{1-x} &
     \int_{t_\xx < t_\xbx < t_\yx < t_\ybx} \! dt_\xbx \, dt_\yx \, dt_\ybx
\nonumber\\& \times
     \frac12 \left[\frac{d\Gamma}{dx\,d(\Delta t_\xx)}\right]_E \,
     \frac12 \left[\frac{d\Gamma}{d\yfrak\,d(\Delta t_\yx)}\right]_{(1-x)E}
   .
\label {eq:xxyyA}
\end {align}
The {\it difference}, which contributes to $\Delta\,d\Gamma/dx\,dy$,
corresponds to
\begin {equation}
   \int_{t_\xx < t_\xbx < t_\yx < t_\ybx} \! dt_\xbx \, dt_\yx \, dt_\ybx
   \>\cdots
   -
   \int_{\tau_\xx < \tau_\yx} \! dt_\xbx \, dt_\yx \, dt_\ybx
   \>\cdots
\end {equation}
which may be reorganized as
\begin {equation}
   -
   \int_0^\infty \! d(\Delta t_\xx)
   \int_0^\infty \! d(\Delta t_\yx)
   \int_{\tau_\xx}^{\tau_\xx+\frac12(\Delta t_\xx+\Delta t_y)} \! d\tau_\yx
   \>\cdots .
\label {eq:ints}
\end {equation}
The integrand in (\ref{eq:MC3}), however, depends only on
the $\Delta t$'s and not separately on $\tau_\yx$,
so the $d\tau_\yx$ integration
in (\ref{eq:ints}) is trivial and leaves
\begin {equation}
   -
   \int_0^\infty \! d(\Delta t_\xx)
   \int_0^\infty \! d(\Delta t_\yx) \>
   \tfrac12(\Delta t_\xx+\Delta t_y)
   \>\cdots .
\end {equation}
Letting the notation $[\Delta\,d\Gamma/dx\,dy]_{x\bar x y\bar y}$
represent the difference between $x\bar x y\bar y$ and the corresponding
piece of the idealized Monte Carlo calculation, (\ref{eq:MC3}) becomes
\begin {align}
   \left[\Delta \frac{d\Gamma}{dx\,dy}\right]_{x\bar xy\bar y}
   = - \frac{1}{1-x} &
   \int_0^\infty \! d(\Delta t_\xx)
   \int_0^\infty \! d(\Delta t_\yx) \>
   \tfrac12(\Delta t_\xx+\Delta t_y)
\nonumber\\& \times
     \frac12 \left[\frac{d\Gamma}{dx\,d(\Delta t_\xx)}\right]_E \,
     \frac12 \left[\frac{d\Gamma}{d\yfrak\,d(\Delta t_\yx)}\right]_{(1-x)E}
   .
\end {align}
Treating $x\bar x \bar y y$ similarly
(as well as the conjugates $\bar xx \bar yy + \bar x xy\bar y$),
\begin {align}
   \left[\Delta \frac{d\Gamma}{dx\,dy}\right]_{
        \begin{subarray}{} x\bar x y\bar y + x\bar x \bar y y \\
                        + \bar xx \bar yy + \bar x xy\bar y \end{subarray}
   }
   = - \frac{1}{1-x} &
   \int_0^\infty \! d(\Delta t_\xx)
   \int_0^\infty \! d(\Delta t_\yx) \>
   \tfrac12(\Delta t_\xx+\Delta t_y)
\nonumber\\& \times
     \Re \left[\frac{d\Gamma}{dx\,d(\Delta t_\xx)}\right]_E \,
     \Re \left[\frac{d\Gamma}{d\yfrak\,d(\Delta t_\yx)}\right]_{(1-x)E}
   .
\label {eq:Dxxyyetc}
\end {align}
The important thing about these integrals is that there are no
large-time issues because $\Delta\,d\Gamma/dx\,dy$ cares only about the
cases where the formation times overlap.
The integrals above are infrared ($\Delta t\to\infty$) convergent.
The ultraviolet ($\Delta t \to 0$) is another issue, which we will
address later, similar to the small-$\Delta t$ divergences
encountered for the crossed diagrams in the preceding paper \cite{2brem}.


\subsubsection {Multiple scattering ($\hat q$) approximation}

Now specialize to the multiple scattering approximation, where
the quantum mechanics problems are harmonic oscillator problems.
As reviewed in our notation in the preceding paper \cite{2brem},%
\footnote{
  See in particular
  eqs.\ (7.1--5) of ref.\ \cite{2brem}.
}
the single-splitting result is then specifically
\begin {equation}
   \left[\frac{d\Gamma}{dx}\right]_E
   =
   - \frac{\alpha P(x)}{\pi} \,
   \Re \int_0^\infty d(\Delta t) \>
   \Omega_{E,x}^2 \csc^2(\Omega_{E,x} \, \Delta t)
   =
   \frac{\alpha P(x)}{\pi} \,
   \Re(i \Omega_{E,x}) .
\label {eq:Gamma1}
\end {equation}
For simplicity of notation, we specialize now to the specific
case treated in this paper, where all of the high-energy particles
are gluons.  In that case, the complex harmonic oscillator
frequency appearing above is
\begin {equation}
   \Omega_{E,x} =
   \sqrt{
     - \frac{i \hat q_{\rm A}}{2 E}
     \left( -1 + \frac{1}{1{-}x} + \frac{1}{x} \right)
   } .
\label {eq:Omegai}
\end {equation}
The earlier result (\ref{eq:Dxxyyetc}) is then
\begin {align}
   \left[\Delta \frac{d\Gamma}{dx\,dy}\right]_{
        \begin{subarray}{} x\bar x y\bar y + x\bar x \bar y y \\
                        + \bar xx \bar yy + \bar x xy\bar y \end{subarray}
   }
   &= \frac{\alphas^2 \, P(x) \, P(\yfrak)}{2\pi^2(1-x)}
\nonumber\\ & \quad \times
   \Bigl[
     \Re(i \Omega_{E,x})
     \Re \int_0^\infty \! d(\Delta t_\yx) \>
         \Omega_{(1-x)E,\yfrak}^2 \, \Delta t_\yx
         \csc^2(\Omega_{(1-x)E,\yfrak} \, \Delta t_\yx)
\nonumber\\ & \quad\qquad
     +
     \Re(i \Omega_{(1-x)E,\yfrak})
     \Re \int_0^\infty \! d(\Delta t_\xx) \>
         \Omega_{E,x}^2 \, \Delta t_\xx
         \csc^2(\Omega_{E,x} \, \Delta t_\xx)
   \Bigr] ,
\label {eq:DxxyyetcHO}
\end {align}
where
\begin {equation}
   \Omega_{(1-x)E,\yfrak} =
   \sqrt{
     - \frac{i \hat q_{\rm A}}{2 (1{-}x)E}
     \left( -1 + \frac{1}{1{-}\yfrak} + \frac{1}{\yfrak} \right)
   } =
   \sqrt{
     - \frac{i \hat q_{\rm A}}{2 E}
     \left( -\frac{1}{1{-}x} + \frac{1}{1{-}x{-}y} + \frac{1}{y} \right)
   } .
\label {eq:Omegafseq}
\end {equation}


\subsubsection {Small-$\Delta t$ divergence}

The integrals in the result (\ref{eq:DxxyyetcHO}) have logarithmic
divergences associated with $\Delta t \to 0$:
\begin {align}
   \left[\Delta \frac{d\Gamma}{dx\,dy}\right]_{
        \begin{subarray}{} x\bar x y\bar y + x\bar x \bar y y \\
                        + \bar xx \bar yy + \bar x xy\bar y \end{subarray}
   }
   &= \frac{\alphas^2 \, P(x) \, P(\yfrak)}{2\pi^2(1-x)} \,
     \Re(i \Omega_{E,x} + i \Omega_{(1-x)E,\yfrak})
     \Re \int_0 \frac{d(\Delta t)}{\Delta t} 
\nonumber\\ & \quad\qquad
         + {\rm UV~convergent} .
\label {eq:xxyysmalldt}
\end {align}
We will see in section \ref{sec:xyxy} that this divergence
cancels a similar small-time divergence of the first diagram
in fig.\ \ref{fig:seqs2},
$xy\bar x\bar y$ (plus its conjugate).

As discussed in the preceding paper \cite{2brem}, one must be careful about pole
contributions from $\Delta t = 0$ even when $1/\Delta t$ divergences cancel.
We will defer these pole contributions to section \ref{sec:pole}.


\subsubsection {Permutations}

\begin {figure}[t]
\begin {center}
  \includegraphics[scale=0.5]{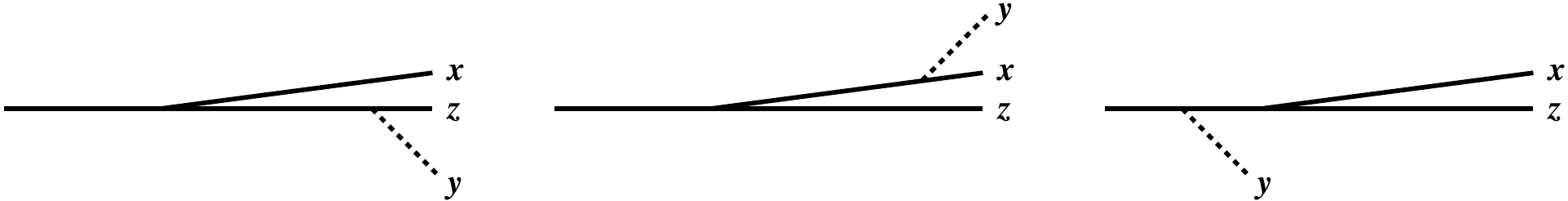}
  \caption{
     \label{fig:allMC}
     All idealized
     Monte Carlo contributions producing three daughters $(x,y,z)$.
  }
\end {center}
\end {figure}

So far, we have only talked about subtracting away one contribution
to the idealized Monte Carlo result---the one shown in fig.\ \ref{fig:xxyyMC}.
What about the others?  A complete summary of the ways that Monte
Carlo can produce three daughters $(x,y,z)$ is shown in
fig.\ \ref{fig:allMC}.
Let's focus on the case where all high-energy particles are gluons.
Then all the idealized Monte Carlo contributions in fig.\ \ref{fig:allMC} are
related by permutations of
$(x,y,z)$, and they will be subtracted in the corresponding permutations
of (\ref{eq:DxxyyetcHO}).  The first Monte Carlo contribution in
fig.\ \ref{fig:allMC} is subtracted from
$2\Re(x\bar x y\bar y+x\bar x\bar y y)$, as we have been discussing.
The second is subtracted from the $x\leftrightarrow z$ permutation
$2\Re(z\bar z y\bar y+z\bar z\bar y y)$.  The third is subtracted
from the $x \leftrightarrow y$ permutation
$2\Re(y\bar y x\bar x + y\bar y \bar x x)$.

Note that the $x\bar xy\bar y$ diagram of fig.\ \ref{fig:xxyy1} maps
into itself under $y \leftrightarrow z$, given the identity of our
final state particles.  So we should not include
this permutation when we sum up all the diagrams.  Later, though,
it will be slightly convenient if we arrange our notation to
pretend that $y \leftrightarrow z$ is a distinct permutation.
So, looking ahead, define $\Fint(x,y,z)$ by
\begin {equation}
   2 \Fint(x,y,z) \equiv
   \left[\Delta \frac{d\Gamma}{dx\,dy}\right]_{
        \begin{subarray}{} x\bar x y\bar y + x\bar x \bar y y \\
                        + \bar xx \bar yy + \bar x xy\bar y \end{subarray}
   } ,
\end {equation}
which is given by (\ref{eq:DxxyyetcHO}).  Then we can write the
desired sum of permutations not only as
\begin {equation}
   2 \Fint(x,y,z) + 2 \Fint(z,y,x) + 2 \Fint(y,x,z)
\end {equation}
but also as
\begin {equation}
   \Fint(x,y,z) + \Fint(y,z,x) + \Fint(z,x,y)
 + \Fint(y,x,z) + \Fint(x,z,y) + \Fint(z,y,x) .
\label {eq:Fperms}
\end {equation}


\subsection {Discussion of \boldmath$x y\bar x\bar y$}
\label {sec:xyxy}

We now turn to the remaining sequential
diagram: the first diagram ($xy\bar x\bar y$) of figs.\ \ref{fig:seqs}
and \ref{fig:seqs2}.  Evaluating this diagram is mostly an
exercise in applying the methods of the previous paper \cite{2brem},
and we leave most of the details to appendix \ref{app:xyxy}.
However, there is one new issue that we touch on here in the main
text: the different ways one may route color in the
$xy\bar x\bar y$ diagram in the large-$\Nc$ limit.


\subsubsection {Color routings}

\begin {figure}[t]
\begin {center}
  \includegraphics[scale=0.8]{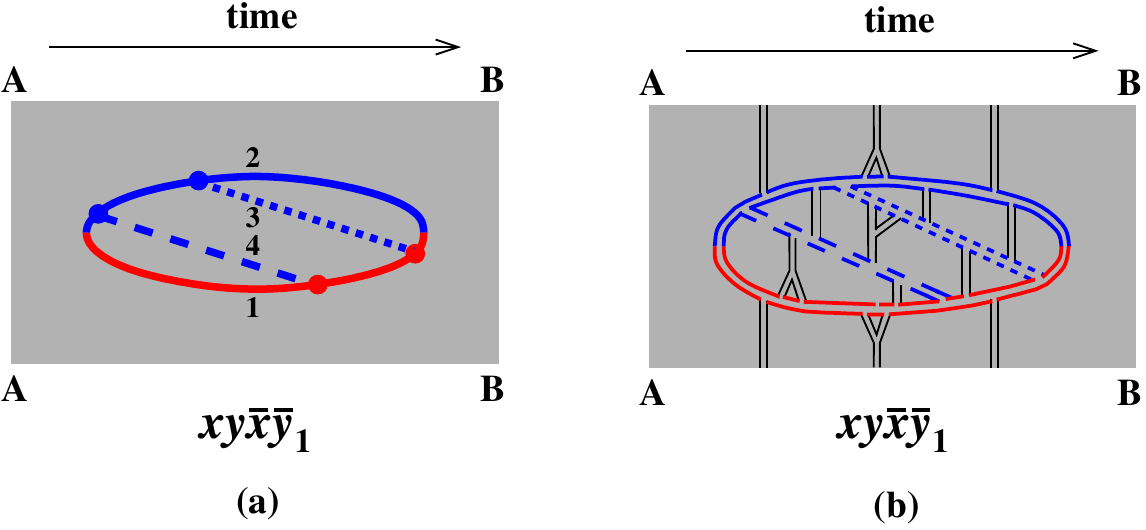}
  \caption{
     \label{fig:cylinder2A}
     One of the two distinct large-$\Nc$
     color routings of the $xy\bar x\bar y$ interference
     diagram drawn on a cylinder.
     The top edge AB of the shaded region is to be identified with the
     bottom edge AB.
     (b) explicitly shows the corresponding color flow for an
     example of
     medium background field correlations (black) that gives a planar
     diagram (and so leading-order in $1/\Nc$).
     In our notation, this interference contribution could be referred
     to as either $xy\bar x\bar y_1$ or $xz\bar x\bar z_2$.
  }
\end {center}
\end {figure}

\begin {figure}[t]
\begin {center}
 \includegraphics[scale=0.8]{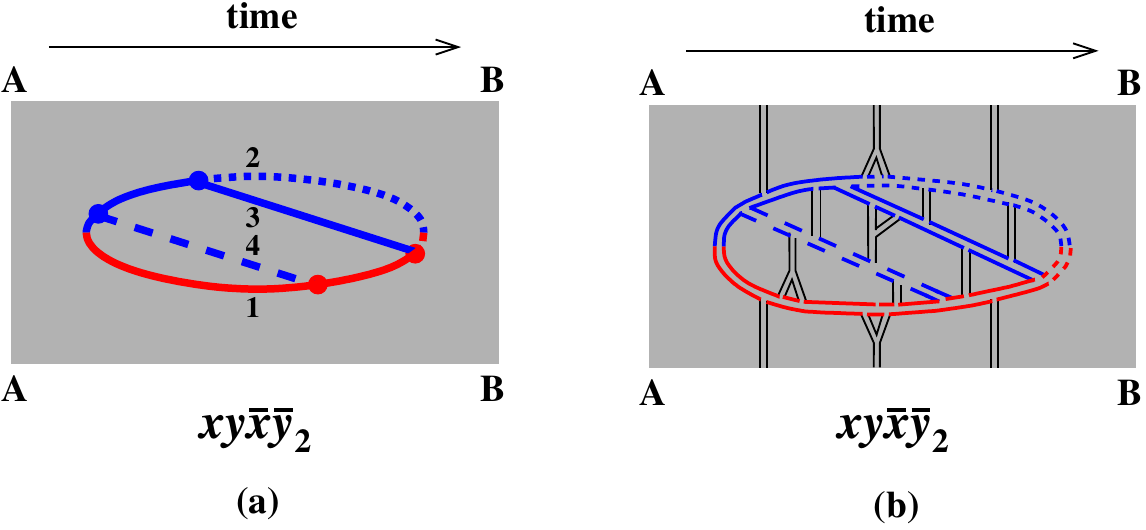}
  \caption{
     \label{fig:cylinder2B}
     As fig.\ \ref{fig:cylinder2A} but showing the other distinct
     color routing of $xy\bar x\bar y$.
     In our notation, this interference contribution could be referred
     to as either $xy\bar x\bar y_2$ or $xz\bar x\bar z_1$.
  }
\end {center}
\end {figure}

Like our discussion of $x\bar xy\bar y$ above,
the $xy\bar x\bar y$ diagram of fig.\ \ref{fig:seqs2} technically
remains the same
if one permutes $y\leftrightarrow z$.
However, in this paper we are working in the large-$\Nc$ limit, and,
in that limit, there are two distinct color routings of the
$xy\bar x\bar y$ diagram which are not individually
$y\leftrightarrow z$ symmetric.
We show these two large-$\Nc$ color routings in figs.\ \ref{fig:cylinder2A}
and \ref{fig:cylinder2B}, which we will refer to as $xy\bar x\bar y_1$ and
$xy\bar x\bar y_2$ respectively.
In the figures, we follow the convention of the preceding paper \cite{2brem}
of drawing our large-$\Nc$, time-ordered diagrams on a cylinder.
In large $\Nc$, correlations of high-energy particles' interactions
with the plasma only exist between high-energy particles that are
neighbors as one goes around the cylinder, which is why in
{\it this}\/ context
the diagrams of figs.\ \ref{fig:cylinder2A}a and \ref{fig:cylinder2B}a
represent different diagrams.
Note that $xy\bar x\bar y_1$ and $xy\bar x\bar y_2$ are related by
$y\leftrightarrow z$, and so we could also call them
$xz\bar x\bar z_2$ and $xz\bar x\bar z_1$ respectively.

The distinguishing
difference between the calculations of the two color routings will,
in the language of the preceding paper \cite{2brem}, be the assignments
of the longitudinal momentum fractions $x_i$ for the 4-particle part of the
evolution $t_\yx < t < t_\xbx$.  Going around the cylinder,
the first routing $xy\bar x\bar y_1$
has (as labeled in fig.\ \ref{fig:cylinder2A})
\begin {equation}
   (x_1,x_2,x_3,x_4) = (-1,1{-}x{-}y,y,x) ,
\label {eq:order1}
\end {equation}
whereas the second routing $xy\bar x\bar y_2$ has (as labeled in
fig.\ \ref{fig:cylinder2B}),
\begin {equation}
   (x_1,x_2,x_3,x_4) = (-1,y,1{-}x{-}y,x)
   \equiv (\hat x_1,\hat x_2,\hat x_3,\hat x_4) .
\label {eq:order2}
\end {equation}
Because this last assignment is identical to the one used for
the canonical diagram analyzed in ref.\ \cite{2brem}, we will focus
on the evaluation of $xy\bar x\bar y_2$ to simplify comparison
with previous work.
Then, if we define
\begin {equation}
   \BnBint(x,y,z) \equiv 
   2\Re
   \left[\Delta \frac{d\Gamma}{dx\,dy}\right]_{xy\bar x\bar y_2} ,
\end {equation}
the full result that we want from $xy\bar x\bar y$ plus its distinct
permutations,
including all distinct color routings, is
\begin {equation}
   \BnBint(x,y,z) + \BnBint(y,z,x) + \BnBint(z,x,y)
 + \BnBint(y,x,z) + \BnBint(x,z,y) + \BnBint(z,y,x) .
\label {eq:BnBperms}
\end {equation}

We mention in passing that the reason we could sidestep discussion of
color routings for the $x\bar xy\bar y$ diagram analyzed in section
\ref{sec:xxyy} is because in that case there was no interval of
4-particle evolution and so no distinction like (\ref{eq:order1})
vs.\ (\ref{eq:order2}).  Unlike 4-particle evolution,
the choice of ordering of the $x_i$ for 3-particle
evolution makes no difference to the calculation since
3 high-energy particles are all neighbors when drawn on our
time-ordered cylinder.


\subsubsection {Result and small-$\Delta t$ behavior}

In notation similar to that used
in the preceding paper \cite{2brem},
the final result for $xy\bar x\bar y_2$ is
\begin {align}
  \left[\frac{d\Gamma}{dx\,dy}\right]_{xy\bar x\bar y_2} =
   \int_0^\infty & d(\Delta t) \>
   \frac{\CA^2 \alphas^2 M_\ix M_\fx^\seq}{32\pi^4 E^2} \, 
   ({-}\hat x_1 \hat x_2 \hat x_3 \hat x_4)
   \Omega_+\Omega_- \csc(\Omega_+\Delta t) \csc(\Omega_-\Delta t)
\nonumber\\ &\times
   \Bigl\{
     (\bar\beta Y_\yx^\seq Y_\xbx^\seq
        + \bar\alpha \Ybar_{\yx\xbx}^{\,\seq} Y_{\yx\xbx}^\seq) I_0^\seq
     + (\bar\alpha+\bar\beta+2\bar\gamma) Z_{\yx\xbx}^\seq I_1^\seq
\nonumber\\ &\quad
     + \bigl[
         (\bar\alpha+\bar\gamma) Y_\yx^\seq Y_\xbx^\seq
         + (\bar\beta+\bar\gamma) \Ybar_{\yx\xbx}^{\,\seq} Y_{\yx\xbx}^\seq
        \bigr] I_2^\seq
\nonumber\\ &\quad
     - (\bar\alpha+\bar\beta+\bar\gamma)
       (\Ybar_{\yx\xbx}^{\,\seq} Y_\xbx^\seq I_3^\seq + Y_\yx^\seq Y_{\yx\xbx}^\seq I_4^\seq)
   \Bigl\}
\label{eq:xyxyresult}
\end {align}
where formulas for the various symbols are given in appendix
\ref{app:xyxy} and $\Delta t$ represents the intermediate time
interval $t_\xbx-t_\yx$.
This formula is identical to that for
$xy\bar y\bar x$ in ref.\ \cite{2brem}
except for the addition of a superscript ``\seq''
on some symbols (standing for ``sequential'' interference diagram
as opposed to crossed),
the bars on $(\bar\alpha,\bar\beta,\bar\gamma)$,
and subscript labels $\bar y$ there replaced by $\bar x$ here.
These modifications are explained in the appendix.

The small $\Delta t$ limit of the integrand is also discussed in the
appendix, with result
\begin {align}
  \left[\frac{d\Gamma}{dx\,dy}\right]_{xy\bar x\bar y_2}
   &=  \frac{\alphas^2 \, P(x) \, P(\yfrak)}{8\pi^2(1-x)} \,
     \int_0 d(\Delta t) \left(
          \frac{1}{(\Delta t)^2}
          - \frac{i( \Omega_{E,x} + \Omega_{(1-x)E,\yfrak} )}{\Delta t}
     \right)
\nonumber\\ & \quad\qquad
         + {\rm UV~convergent} .
\label {eq:xyxy2smalldtA}
\end {align}
As in ref.\ \cite{2brem}, the $1/(\Delta t)^2$ divergence may be
eliminated by subtracting out the vacuum calculation (which must
total to zero when summed over all interference processes).
If we now add in the conjugate diagram $\bar x\bar y x y_2$, that
leaves
\begin {align}
  2\Re\left[\frac{d\Gamma}{dx\,dy}\right]_{xy\bar x\bar y_2}
   &\to  -\frac{\alphas^2 \, P(x) \, P(\yfrak)}{4\pi^2(1-x)} \,
     \Re\Bigl[
     (i \Omega_{E,x} + i \Omega_{(1-x)E,\yfrak})
     \int_0 \frac{d(\Delta t)}{\Delta t}
     \Bigr]
\nonumber\\ & \quad\qquad
         + {\rm UV~convergent} .
\label {eq:xyxy2smalldt}
\end {align}
This result is invariant under $y\leftrightarrow z$ and so will be the same
for the other color routing $xy\bar x\bar y_1$.  In total, then
\begin {align}
  2\Re\left[\frac{d\Gamma}{dx\,dy}\right]_{xy\bar x\bar y}
   &\to -\frac{\alphas^2 \, P(x) \, P(\yfrak)}{2\pi^2(1-x)} \,
     \Re\Bigl[
     (i \Omega_{E,x} + i \Omega_{(1-x)E,\yfrak})
     \int_0 \frac{d(\Delta t)}{\Delta t}
     \Bigr]
\nonumber\\ & \quad\qquad
         + {\rm UV~convergent} .
\label {eq:xyxysmalldt}
\end {align}
For real $\Delta t$, this indeed cancels the small-$\Delta t$ behavior
of (\ref{eq:xxyysmalldt}), as promised.  So our final $\Delta t$ integrals
for the sum of all sequential diagrams (fig.\ \ref{fig:seqs2}) will be
convergent.


\subsection {Pole Contributions}
\label {sec:pole}

As mentioned earlier and as discussed in the previous paper \cite{2brem},
one needs to be careful about the cancellation of $1/\Delta t$ divergences
between different diagrams, such as for the sum of
(\ref{eq:xxyysmalldt}) and (\ref{eq:xyxysmalldt}) in the present case.
The problem is that there can be contributions coming from the
pole at $\Delta t{=}0$ that need to be accounted for.
The previous paper \cite{2brem} attempted to isolate these
pole contributions by appropriate replacements of the form
\begin {equation}
   \int_0^\infty \frac{d(\Delta t)}{\Delta t} \, \cdots
   ~\to~
   \int_0^\infty \frac{d(\Delta t)}{\Delta t \pm i\epsilon} \, \cdots
\end {equation}
for each $\Delta t$ integral.  The same method applied to the
sequential diagrams would give a pole contribution to
$\Phi+\Psi \equiv
\frac12 [2\Re(x\bar xy\bar y+x\bar x\bar y y) - {\rm IMC}]
+ 2\Re(xy\bar x\bar y_2)$
of
\begin {equation}
     \frac{\alphas^2 \, P(x) \, P(\yfrak)}{8\pi(1-x)} \,
     \Re(\Omega_{E,x} + \Omega_{(1-x)E,\yfrak}) .
\label {eq:xyxy2poleNaive}
\end {equation}
We will refer to this as a ``$1/\pi$'' contribution because of the
single factor of $\pi$ in the denominator.
It turns out that both this result for the pole contribution to
the sequential diagrams
and the previous result \cite{2brem} for the
pole contribution to the crossed diagrams
are incomplete: They miss some additional terms that involve
$1/\pi^2$.  The proper calculation of the pole terms is a
lengthy enough
issue that it would distract from our focus in this paper, and
so here we will just quote results.  The full discussion of why
the previous analysis was incomplete, and of how to compute the
full pole pieces (using dimensional regularization), is left to
ref.\ \cite{dimreg}.

With regard to sequential diagrams,
the result is that (\ref{eq:xyxy2poleNaive}) should be
supplemented by the additional contribution
\begin {equation}
   - \frac{\alphas^2 \, P(x) \, P(\yfrak)}{4\pi^2(1-x)}
           \Re(i \Omega_{E,x} + i \Omega_{(1-x)E,\yfrak}) .
\end {equation}
For crossed diagrams,
which are also needed for our total results of figs.\ \ref{fig:result}
and \ref{fig:resultScaled}, the corrected pole contribution will be
given in section \ref{sec:crossedpole} below.
This correction is, in fact, critical to
the Gunion-Bertsch-like cancellation of logarithms
(\ref{eq:LogEnhanced}) discussed in Appendix \ref{app:log}.%
\footnote{
   Without the correct treatment of the poles,
   the coefficient in the third line of
   Table \ref{tab:logs} would have come out $-\frac32$ instead of
   $-\frac12$.
}


\section {Summary of Formula}
\label {sec:summary}

\subsection{Sequential Diagrams}

We now give a summary of our final formulas for the sequential diagrams
of fig.\ \ref{fig:seqs2},
in the same style as the summary of the crossed-diagram contribution
in section VIII of the preceding paper
\cite{2brem}.  The two should be added together (along with the
contributions involving 4-gluon vertices,
like fig.\ \ref{fig:4ptExamples}, which we have left for the future).

The main result of this paper is%
\begin {align}
   \left[ \Delta \frac{d\Gamma}{dx\>dy} \right]_{\rm sequential}
   = \quad
   & {\cal A}_\seq(x,y) + {\cal A}_\seq(1{-}x{-}y,y) + {\cal A}_\seq(x,1{-}x{-}y)
\nonumber\\
   + ~ &
   {\cal A}_\seq(y,x) + {\cal A}_\seq(y,1{-}x{-}y) + {\cal A}_\seq(1{-}x{-}y,x)
\label {eq:dGammaseq}
\end {align}
where ${\cal A}_{\rm seq}(x,y)$ is
the result for
(i) $2\Re(xy\bar x\bar y_2)$, for the large-$\Nc$ color routing
of fig.\ \ref{fig:cylinder2B}, plus {\it half}\/ of
(ii) $2\Re(x\bar x y\bar y + x\bar x \bar y y)$
minus the corresponding idealized Monte Carlo result.
That is, ${\cal A}_{\rm seq}$ corresponds to $\BnBint + \Fint$ in the
notation of (\ref{eq:Fperms}) and (\ref{eq:BnBperms}).
We will write this as
\begin {equation}
   {\cal A}_\seq(x,y)
   =
   {\cal A}^{\rm pole}_\seq(x,y)
   + \int_0^{\infty} d(\Delta t) \>
     \Bigl[
        2 \Re \bigl( B_{\rm seq}(x,y,\Delta t) \bigr)
        + F_{\rm seq}(x,y,\Delta t)
     \bigr] ,
\label {eq:Aseq}
\end {equation}
where
\begin {align}
   B_\seq(x,y,\Delta t) &=
       C_\seq(\{\hat x_i\},\bar\alpha,\bar\beta,\bar\gamma,\Delta t)
\nonumber\\
   &=
       C_\seq({-}1,y,1{-}x{-}y,x,\bar\alpha,\bar\beta,\bar\gamma,\Delta t)
\end {align}
corresponds to $x y\bar x\bar y_2$.
The other term, $F_\seq$,
corresponds to half of $2\Re(x\bar x y\bar y + x\bar x \bar y y)$
minus the corresponding idealized Monte Carlo result.
$C$ is defined to have the vacuum result subtracted, so write
\begin {equation}
   C_\seq = D_\seq - \lim_{\hat q\to 0} D_\seq .
\end {equation}
$D_\seq$, defined as the integrand for $xy\bar x\bar y_2$, is given
by
\begin {align}
   D_\seq(x_1,&x_2,x_3,x_4,\bar\alpha,\bar\beta,\bar\gamma,\Delta t) =
\nonumber\\ &
   \frac{\CA^2 \alphas^2 M_\ix M_\fx^\seq}{32\pi^4 E^2} \, 
   ({-}x_1 x_2 x_3 x_4)
   \Omega_+\Omega_- \csc(\Omega_+\Delta t) \csc(\Omega_-\Delta t)
\nonumber\\ &\times
   \Bigl\{
     (\bar\beta Y_\yx^\seq Y_\xbx^\seq
        + \bar\alpha \Ybar_{\yx\xbx}^{\,\seq} Y_{\yx\xbx}^\seq) I_0^\seq
     + (\bar\alpha+\bar\beta+2\bar\gamma) Z_{\yx\xbx}^\seq I_1^\seq
\nonumber\\ &\quad
     + \bigl[
         (\bar\alpha+\bar\gamma) Y_\yx^\seq Y_\xbx^\seq
         + (\bar\beta+\bar\gamma) \Ybar_{\yx\xbx}^{\,\seq} Y_{\yx\xbx}^\seq
        \bigr] I_2^\seq
\nonumber\\ &\quad
     - (\bar\alpha+\bar\beta+\bar\gamma)
       (\Ybar_{\yx\xbx}^{\,\seq} Y_\xbx^\seq I_3^\seq + Y_\yx^\seq Y_{\yx\xbx}^\seq I_4^\seq)
   \Bigl\} .
\label {eq:Dseq}
\end {align}
$(\bar\alpha,\bar\beta,\bar\gamma)$, the $(X^\seq,Y^\seq,Z^\seq)$ and the
$I^\seq$ are given in appendix \ref{app:xyxy}
by (\ref{eq:abcbar}), (\ref{eq:XYZseqdef}), and
(\ref{eq:Iseq}) respectively.
The 4-particle evolution frequencies $\Omega_\pm$
are given by eq.\ (5.21) of ref.\ \cite{2brem}.
The $M$ and other $\Omega$ here are
\begin {equation}
   M_\ix = x_1 x_4 (x_1{+}x_4) E ,
   \qquad
   M_\fx^\seq = x_2 x_3 (x_2{+}x_3) E
\label {eq:Mifseq}
\end {equation}
and
\begin {equation}
   \Omega_\ix
   = \sqrt{ 
     -\frac{i \hat q_{\rm A}}{2E}
     \left( \frac{1}{x_1} 
            + \frac{1}{x_4} - \frac{1}{x_1{+}x_4} \right)
   } ,
   \qquad
   \Omega_\fx^\seq
   = \sqrt{ 
     -\frac{i \hat q_{\rm A}}{2E}
     \left( \frac{1}{x_2} + \frac{1}{x_3}
            - \frac{1}{x_2 + x_3}
     \right)
   }
\end {equation}
[which equal $\Omega_{E,x}$ (\ref{eq:Omegai}) and
$\Omega_{(1-x)E,\yfrak}$ (\ref{eq:Omegafseq}) when $\{x_i\} = \{\hat x_i\}$].%
\footnote{
  In contrast, beware that our $M_\fx^\seq$ in (\ref{eq:Mifseq}) is
  not the same as $M_{(1-x)E,\yfrak}$ when $\{x_i\} = \{\hat x_i\}$.
  Instead, $M_\fx^\seq = (1{-}x)^2 M_{(1-x)E,\yfrak}$.
}
Next,
\begin {align}
   F_\seq(x,y,\Delta t) =
   \frac{\alphas^2 P(x) P(\frac{y}{1-x})}{4\pi^2(1-x)}
   \Bigl[ &
      \Re(i\Omega_\ix) \,
      \Re\bigl( \Delta t \, (\Omega_\fx^\seq)^2
                \csc^2(\Omega_\fx^\seq \, \Delta t) \bigr)
\nonumber\\
      + &
      \Re(i\Omega_\fx^\seq) \,
      \Re\bigl( \Delta t \, \Omega_\ix^2 \csc^2(\Omega_\ix \, \Delta t) \bigr)
   \Bigr] .
\label {eq:Fseq}
\end {align}
Finally, the pole piece is
\begin {equation}
   {\cal A}_\seq^{\rm pole}(x,y)
   = \frac{\alphas^2 \, P(x) \, P(\yfrak)}{2\pi^2(1-x)}
   \biggl(
      - \tfrac12
           \Re(i \Omega_{E,x} + i \Omega_{(1-x)E,\yfrak})
      + \tfrac{\pi}{4}
           \Re(\Omega_{E,x} + \Omega_{(1-x)E,\yfrak})
   \biggr) .
\end {equation}
In the last formulas, $P(x)$ refers to the spin-averaged
$g{\to}gg$ DGLAP splitting function%
\footnote{
   We do not bother to show the usual singularity prescriptions
   and contributions
   in (\ref{eq:Pggg}) because we do not integrate over longitudinal
   momentum fractions in this paper.
}
\begin {equation}
   P_{g\to gg}(x) =
   \CA \, \frac{[1 + x^4 + (1{-}x)^4]}{x (1{-}x)} \,.
\label {eq:Pggg}
\end {equation}


\subsection{Crossed Diagrams}
\label {sec:crossedpole}

The formulas for crossed diagrams are summarized in section VIII of
ref.\ \cite{2brem},
except that the pole contribution there needs
to be corrected \cite{dimreg} by replacing the term
\begin {equation}
   2 \Re\left[ \frac{d\Gamma}{dx\,dy} \right]_{\rm fig.~23}^{\rm pole}
\end {equation}
in the formula (8.2) for $A(x,y)$ in ref.\ \cite{2brem} by
\begin {align}
   A^{\rm pole}(x,y) &\equiv
   2\Re \Biggl[
    \frac{\CA^2 \alphas^2}{16\pi^2} \,
    x y (1{-}x)^2 (1{-}y)^2(1{-}x{-}y)^2
\nonumber\\ & \qquad~ \times
    \biggl\{
       -i
       [\Omega_{-1,1-x,x} + \Omega_{-(1-y),1-x-y,x}
        - \Omega_{-1,1-y,y}^* - \Omega_{-(1-x),1-x-y,y}^*]
\nonumber\\ & \qquad~ \qquad \times
       \biggl[
       \left(
         (\alpha + \beta)
         + \frac{(\alpha + \gamma) xy}{(1{-}x)(1{-}y)}
       \right) \ln \left( \frac{1{-}x{-}y}{(1{-}x)(1{-}y)} \right)
       + \frac{2(\alpha+\beta+\gamma) x y}{(1{-}x)(1{-}y)}
       \biggr]
\nonumber\\ &\qquad~ \quad
     - \pi
       [\Omega_{-1,1-x,x} + \Omega_{-(1-y),1-x-y,x}
        + \Omega_{-1,1-y,y}^* + \Omega_{-(1-x),1-x-y,y}^*]
\nonumber\\ & \qquad~ \qquad \times
       \left(
         (\alpha + \beta)
         + \frac{(\alpha + \gamma) xy}{(1{-}x)(1{-}y)}
       \right)
     \biggr\}
   \Biggr] .
\end {align}
The $\pi/\pi^2$ terms above are the same as in ref.\ \cite{2brem}
but are corrected by the addition of the $1/\pi^2$ terms also included above.


\section {Conclusion}
\label {sec:conclusion}

We have still not computed contributions from 4-gluon vertices such
as fig.\ \ref{fig:4ptExamples}, which we expect will be important when
no final gluon is soft, i.e.\ when $y \sim x \sim z$.  (Conversely,
we believe
that these contributions will not be important when at least one
final gluon {\it is}\/ soft.)  To use our results for energy loss
calculations, we will also need to consider virtual corrections to
single gluon bremsstrahlung.  Both of these calculations have been
left for future work.

One of the interesting aspects of our results in fig.\ \ref{fig:resultScaled}
is the sign of the correction $\Delta\, d\Gamma/dx\,dy$
due to overlapping formation times.
We finish below with a qualitative picture of why the correction should be
negative for $y \ll x,z$.  Unfortunately, we do not have any
simple, compelling, qualitative argument for the other, more
important case: How
can we simply understand why the correction should be positive
for $y \sim x \ll z$ (and possibly also for $y \sim x \sim z$,
depending on the size and sign of 4-gluon vertex contributions)?


\subsection*{A picture of why \boldmath$\Delta\,d\Gamma/dx\,dy$
             is negative for \boldmath$y \ll x,z$}

\begin {figure}[t]
\begin {center}
  \includegraphics[scale=0.7]{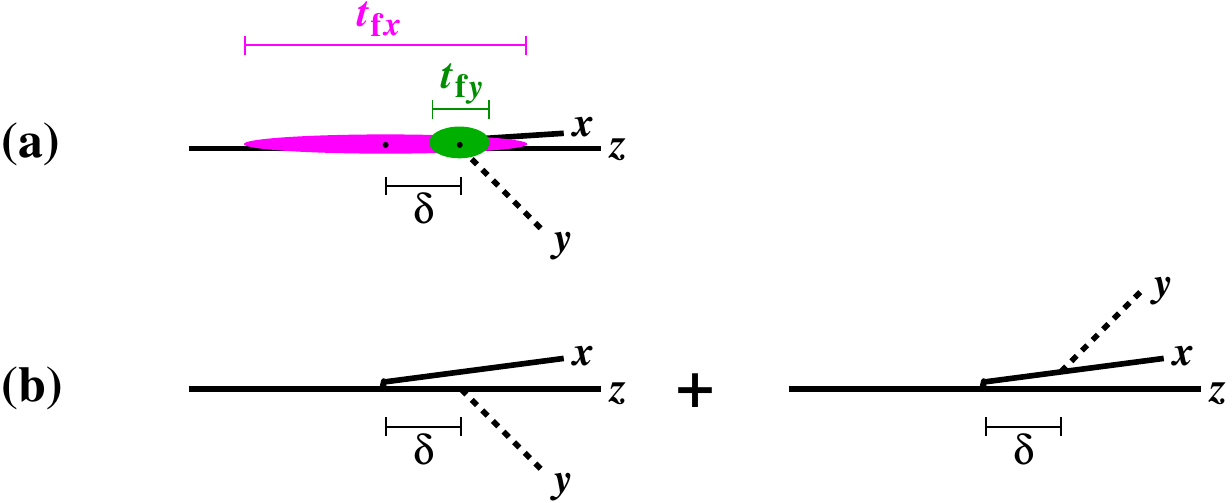}
  \caption{
     \label{fig:picture}
     (a) A double-bremsstrahlung process with $y \ll x$ and
     overlapping formation times.
     $t_{\rm f x}$ and $t_{\rm f y}$ indicate the scale of the formation times
     for $x$ and $y$ emission respectively.
     (b) The corresponding idealized Monte Carlo contributions.
     $\delta$ is the time separation between the $x$ emission and
     the $y$ emission, defined in (a) as the separation between
     the midpoint times $\tau_x$ and $\tau_y$ of (\ref{eq:tauxy}).
     As usual, transverse separations are highly exaggerated in the
     drawing.
     [There is no significance to (i) $y$ being drawn angled up rather
     than down in the last term or (ii) the
     slightly exaggerated transverse separation of $x$ in the drawing
     of (b) compared to (a).  Both were done just to make the diagram
     clearer and less crowded.]
  }
\end {center}
\end {figure}

Consider $y \ll x < z$.
As reviewed earlier, QCD
formation times for softer emissions are smaller than formation
times for harder emissions;
so the formation time for emitting $y$ is small compared
to the formation time for emitting $x$.  Now consider the case
where these formation times overlap, as shown in
fig.\ \ref{fig:picture}a.  We've chosen to look at a case
where the $y$ emission happens some time $\delta$
after the midpoint of the $x$ emission.  The analogous contribution
in the idealized Monte Carlo calculation is shown in fig.\ \ref{fig:picture}b.
In the idealized Monte Carlo calculation, which ignores formation times,
we have treated the ``time'' of the $x$ and $y$ emissions to
be the midpoints of the corresponding formation times in
fig.\ \ref{fig:picture}a.
(This is the natural choice.  Remember that it is also the
choice we used in section \ref{sec:prescription} and technically
justify in appendix \ref{app:justify}.)

In the idealized Monte Carlo calculation, the chance of $y$ emission from the
$x$ daughter is treated as completely independent from the chance of
$y$ emission from the $z$ daughter, and these two probabilities are
added together.  But
the separation of the $x$ and $z$ daughters is actually
so small on these time scales that the $y$ emission cannot resolve
them as distinct particles when $y \ll x$.

Specifically, $x$ emission is associated with transverse momenta of
order $Q_{\perp,x} \sim \sqrt{\hat q t_{\rm form,x}}$.  Using the formation
time (\ref{eq:tform}), the corresponding scale of transverse separation
of $x$ from $z$ during that formation time is
$b_x \sim 1/Q_{\perp,x} \sim (x E \hat q)^{-1/4}$.  When $y \ll x$,
this is indeed very small compared to the similar transverse scale
characterizing the $y$ emission during its formation time:
$b_y \sim (y E \hat q)^{-1/4}$.

So, instead of seeing two daughter
gluons ($x$ and $z$), the $y$ emission effectively sees a single adjoint-charge
particle.
(This resolution issue is similar to work by Mehtar-Tani,
Salgado, and Konrad \cite{resolution}.%
\footnote{
  The $q \bar q$ antenna in ref.\ \cite{resolution} is the
  analog of our $x$ and $z$ daughters, and the radiation from
  that antenna is the analog of our $y$ emission.
}%
)
On these time scales, there is therefore effectively only one
particle providing a chance for initiating $y$ emission rather
than the two particles in the idealized Monte Carlo calculation associated
with fig.\ \ref{fig:picture}b.  That is, idealized Monte Carlo overcounts
the probability for the $y$ emission, and so the correction to
idealized Monte Carlo should be negative for $y \ll x$.


\begin{acknowledgments}

This work was supported, in part, by the U.S. Department
of Energy under Grant No.~DE-SC0007984.
We thank Diana Vaman and Sangyong Jeon for useful discussions.

\end{acknowledgments}

\appendix

\section {Approximate analytic formula fitted to result}
\label {app:approx}

The following approximation reproduces the results of fig.\
\ref{fig:resultScaled} with a maximum absolute error%
\footnote{
  We quote absolute error rather than relative error because the result
  is zero along the red curve in fig.\ \ref{fig:resultScaled}.
  Any numerical approximation will have infinite relative error exactly
  on this curve, which is irrelevant to the question of how useful the
  approximation is.
}
of 0.017 for
all $y > 10^{-4}$ (assuming one permutes the final state gluons to choose
$y < x < z$, just as in fig.\ \ref{fig:resultScaled}):

\begin{equation}
  \pi^2 \,x \,y^{\frac{3}{2}} \,\Delta\,\frac{d \Gamma}{dx dy} =
  \sum^3_{m=0}\sum^4_{n=0}
  \left(a_{m n} + b_{m n} \Big(\frac{y}{x}\Big)^{\frac{1}{3}} \right) s^m t^n ,
\label {eq:approx}
\end {equation}
where the parameters
\begin {equation}
  s \equiv \frac{2 (x - y)}{t},
\qquad
  t \equiv 2 x + y
\end{equation}
each vary independently from 0 to 1.
The numerical coefficients $a_{mn}$ and $b_{mn}$ are given
in tables \ref{tab:amn} and \ref{tab:bmn}.  We have made no
effort to make the approximation work well for $y < 10^{-4}$.

\begin{table}[t]
\begin{tabular}{|l||*{5}{c|}}\hline
\backslashbox{m}{n}
&\makebox[2em]{0}&\makebox[2em]{1}&\makebox[2em]{2}
&\makebox[2em]{3}&\makebox[2em]{4}\\\hline\hline
0& -4.98232& 40.5485& -198.390& 351.600& -202.365\\ \hline 
 1& 6.33241& -81.4351& 409.898& -730.480& 419.607\\ \hline 
 2& -2.31929& 49.1060& -251.070& 448.653& -257.366\\ \hline 
 3& 0.0211282& -7.27248& 38.4361& -68.9343& 39.6011\\ \hline 
\end{tabular}
\caption{
  \label{tab:amn}
  The coefficients $a_{mn}$ in eq.\ (\ref{eq:approx}).
}
\end{table}

\begin{table}[t]
\begin{tabular}{|l||*{5}{c|}}\hline
\backslashbox{m}{n}
&\makebox[2em]{0}&\makebox[2em]{1}&\makebox[2em]{2}
&\makebox[2em]{3}&\makebox[2em]{4}\\\hline\hline
0& 5.46263& -40.7646& 199.438& -353.153& 203.648\\ \hline
 1& -3.79756& 61.5337& -312.829& 558.682& -321.248\\ \hline
 2& 0.227523& -19.0923& 100.471& -180.559& 103.988\\ \hline
 3& 0.399152& -3.44293& 16.6646& -29.2812& 16.5223\\ \hline
\end{tabular}
\caption{
  \label{tab:bmn}
  The coefficients $b_{mn}$ in eq.\ (\ref{eq:approx}).
}
\end{table}


\section {Logarithms and their cancellation}
\label {app:log}

\begin {figure}[t]
\begin {center}
  \includegraphics[scale=0.5]{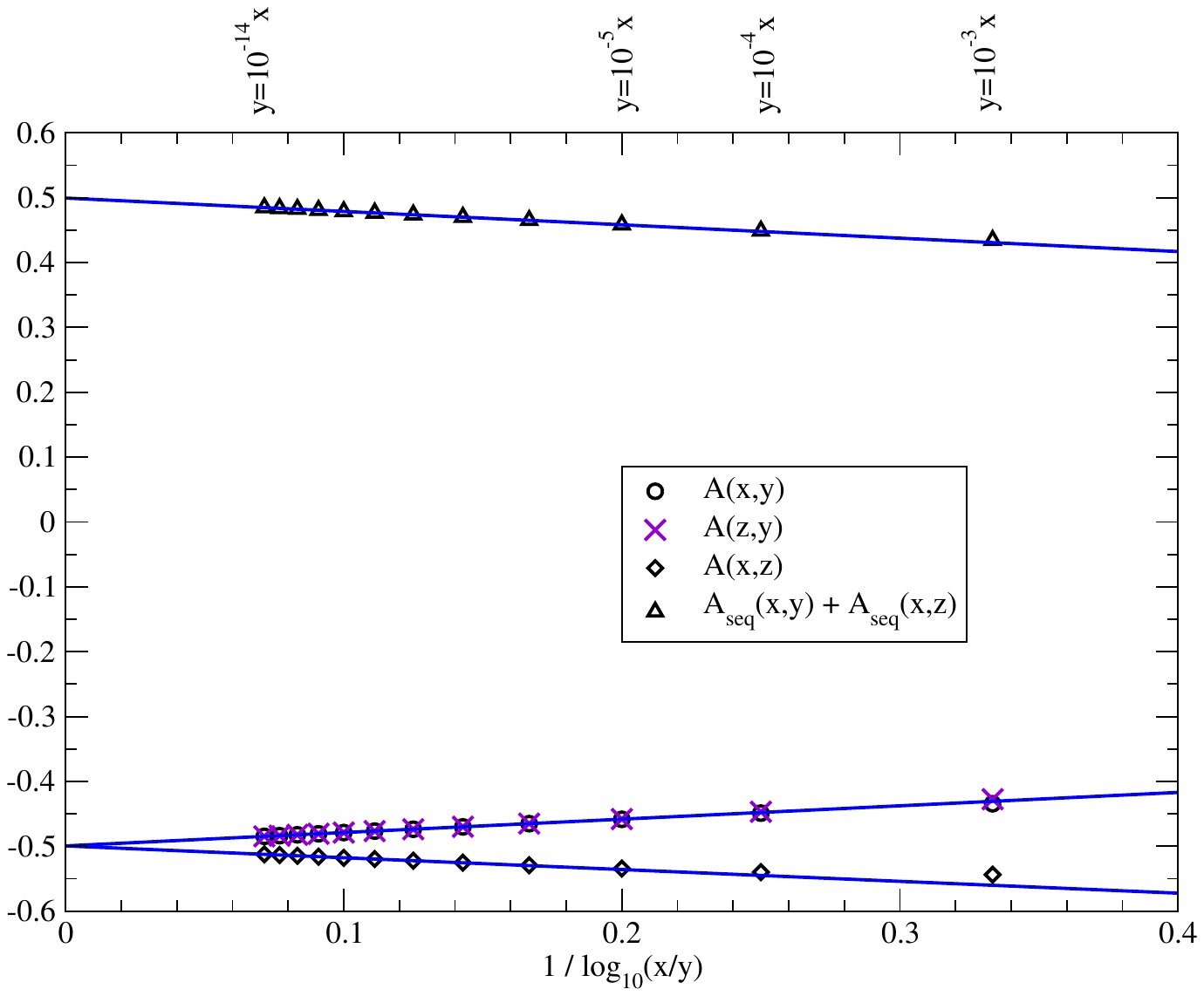}
  \caption{
     \label{fig:Nlogs}
     Various pieces of $\Delta\,d\Gamma/dx\,dy$ (corresponding to
     Table\ \ref{tab:logs}), in units of
     $\frac{\CA^2 \alphas^2}{\pi^2 x y^{3/2}} \bigl(\frac{\hat q}{E}\bigr)^{1/2} \,
        \ln\bigl( \frac{x}{y} \bigr)$.
     The results are plotted against $1/\log_{10}(x/y)$,
     which allows for a simple
     linear extrapolation of the $y\to 0$ limit for fixed $x$.
     Data points were calculated for $x=10^{-3}$, and we checked that
     results for $x=10^{-4}$ would be indistinguishable to the eye.
     Note that the $\times$ and $\circ$ data points are hard to distinguish
     because they are almost on top of each other.
     We have not shown points for
     ${\cal A}_{\rm seq}(y,x)+{\cal A}_{\rm seq}(y,z)$ and
     ${\cal A}_{\rm seq}(z,y)+{\cal A}_{\rm seq}(z,x)$: these would both lie
     right on top of the 
     ${\cal A}_{\rm seq}(x,y)+{\cal A}_{\rm seq}(x,z)$ points.
  }
\end {center}
\end {figure}

In this appendix, we discuss the behavior of {\it individual}
contributions to $\Delta\,d\Gamma/dx\,dy$ in the $y \ll x \ll 1$
limit.  The individual contributions are logarithmically enhanced
compared to the total, as in (\ref{eq:LogEnhanced}).
In more detail,
\begin {subequations}
\label {eq:Blogs}
\begin {align}
   \left[ \frac{d\Gamma}{dx\,dy} \right]_{\rm crossed}
   &\approx -\frac32 \,
        \frac{\CA^2 \alphas^2}{\pi^2 x y^{3/2}} \sqrt{\frac{\hat q}{E}} \,
        \ln\left( \frac{x}{y} \right) ,
\\
   \left[ \Delta \frac{d\Gamma}{dx\,dy} \right]_{\rm seq\phantom{ssed}}
   &\approx +\frac32 \,
        \frac{\CA^2 \alphas^2}{\pi^2 x y^{3/2}} \sqrt{\frac{\hat q}{E}} \,
        \ln\left( \frac{x}{y} \right) ,
\end {align}
\end {subequations}
where $\approx$ indicates that we are only showing the leading logarithm.
We can break the contributions down somewhat further, as shown in
Table\ \ref{tab:logs}.  (One way to extract the coefficients
in this table is from numerical extrapolation,
as shown in fig.\ \ref{fig:Nlogs}.)
\begin {table}[t]

\tabcolsep 10pt
\begin {tabular}{|c|c|l|}
\hline
  $2\Re(xy\bar y\bar x + x\bar yy\bar x)$
      & $A(x,y)$
      & $-\tfrac12$
\\
  $2\Re(zy\bar y\bar z + z\bar yy\bar z)$
      & $A(z,y)$
      & $-\tfrac12$
\\
  $2\Re(x\bar zz\bar x + x\bar z\bar x z
      + z\bar xx\bar z + z\bar x\bar z x)$
      & $A(x,z)$
      & $-\tfrac12$
\\
\hline
  $2\Re(x\bar x y\bar y + x\bar x \bar y y) - {\rm IMC}$
      & ${\cal A}_\seq(x,y)+{\cal A}_\seq(x,z)$
      & $+\tfrac12$
\\
  $2\Re(z\bar z y\bar y + z\bar z \bar y y) - {\rm IMC}$
      & ${\cal A}_\seq(z,y)+{\cal A}_\seq(z,x)$
      & $+\tfrac12$
\\
  $2\Re(y\bar y x\bar x + y\bar y \bar x x) - {\rm IMC}$
      & ${\cal A}_\seq(y,x)+{\cal A}_\seq(y,z)$
      & $+\tfrac12$
\\
\hline
\end {tabular}
\caption
    {
    \label {tab:logs}
    Coefficients of logarithms.
    The first column lists the subset of diagrams that are responsible
    for generating logarithms, using the notation of
    figs.\ \ref{fig:subset}--\ref{fig:seqs2} (to include permutations
    of $x$, $y$, and $z$).  The second column identifies the UV-safe
    combinations that contain those diagrams, with ${\cal A}_{\rm seq}$
    given by (\ref{eq:Aseq}) of this paper.  $A$ is given by
    eq.\ (8.2) of ref.\ \cite{2brem}, as corrected by section
    \ref{sec:crossedpole}
    of this paper.
    The last column is the corresponding leading-log contribution to
    $\Delta\, d\Gamma/dx\,dy$ in units of
    $\frac{\CA^2 \alphas^2}{\pi^2 x y^{3/2}} \bigl(\frac{\hat q}{E}\bigr)^{1/2} \,
        \ln\bigl( \frac{x}{y} \bigr)$.
    }
\end {table}

\subsection {DGLAP}

\begin {figure}[t]
\begin {center}
  \includegraphics[scale=0.4]{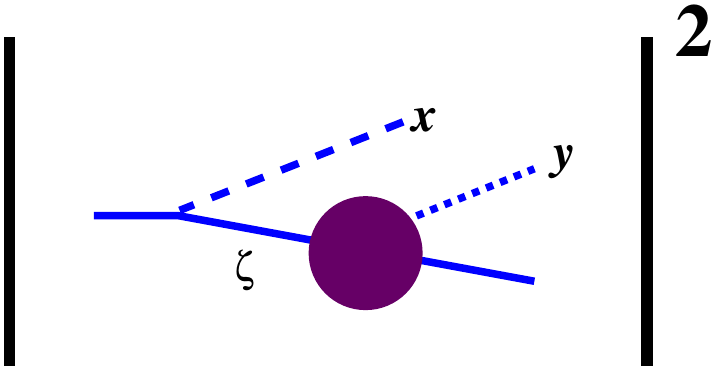}
  \caption{
     \label{fig:DGLAP}
     Magnitude squared of a Feynman diagram, representing
     $2 \Re(x\bar xy\bar y + x\bar x \bar y y)$ if
     interactions with the medium are only considered for the
     $y$ emission.  The blob represents those medium interactions,
     and everything preceding the blob is approximated
     as effectively in vacuum.
  }
\end {center}
\end {figure}

There is a relatively simple way to understand some of these logarithms.
Let's consider the $2 \Re(x\bar xy\bar y + x\bar x \bar y y)$ entry
of Table \ref{tab:logs}, which corresponds to the second and third
diagrams of figs.\ \ref{fig:seqs} and \ref{fig:seqs2}.
Since $y$ is the softest high-energy particle in the process, it is
the one most affected by the medium (which is, recall, why its formation
time is the shortest).  So focus attention on the contributions to this process
where the only interaction with the medium occurs during the
$y$ emission.  That is, consider fig.\ \ref{fig:DGLAP}.
In this picture, the $x$ emission looks like an initial-state radiation
correction to the underlying process of the $y$ emission.
Initial-state radiation generates a factorizable logarithm associated
with DGLAP evolution.  Let's calculate it.  For the sake of
familiarity, we'll start from the usual leading-order DGLAP evolution equation
(though that is a somewhat backwards way to start given how
DGLAP evolution is derived in the first place).

Consider the (vacuum) DGLAP evolution of an initial parton distribution
function $f(\zeta,Q^2)$ for a parton of energy $\zeta E$.
(We use the symbol $\zeta$ for momentum fraction here instead of the
more traditional $x$ or $z$ just to avoid confusion with the
specific use of $x$, $y$, and $z \equiv 1{-}x{-}y$ in this paper as
the momentum fractions of the three final-state gluons.)
The basic DGLAP evolution equation
takes the form
\begin {equation}
   \frac{df(\zeta,Q^2)}{d \ln(Q^2)}
   = \frac{\alphas}{2\pi} \int_\zeta^1 \frac{d\xi}{\xi} \> P(\xi) \,
       f\bigl( \frac{\zeta}{\xi}, Q^2 \bigr) ,
\end {equation}
where $P$ is the relevant DGLAP splitting function.
Now formally solve perturbatively by iteration, writing
$f = f_0 + f_1 + \cdots$ with $f_0(\zeta) = \delta(1-\zeta)$ representing
the case of no initial radiation.
Then $f_1$ (corresponding to a single DGLAP emission from the initial state)
is given by
\begin {equation}
   \frac{df_1(\zeta,Q^2)}{d \ln(Q^2)}
   = \frac{\alphas}{2\pi} \int_\zeta^1 \frac{d\xi}{\xi} \> P(\xi) \,
       f_0\bigl( \frac{\zeta}{\xi}, Q^2 \bigr)
   = \frac{\alphas}{2\pi} \, P(\zeta) .
\end {equation}
The solution is
\begin {equation}
  f_1(\zeta,Q^2) = \frac{\alphas}{2\pi} \, P(\zeta) \, \ln \frac{Q^2}{Q_0^2} \,,
\end {equation}
where we'll discuss what the reference momentum $Q_0^2$ should be in
a moment.  The result we want combines this probabilistic
description of the initial $x$ emission with the rate for
the sub-process associated with the blob in fig.\ \ref{fig:DGLAP}.
That sub-process represents the rate of $y$ emission in the medium,
and so the leading-log estimate of the contribution to $d\Gamma/dx\,dy$
associated with fig.\ \ref{fig:DGLAP} should be
\begin {equation}
   \left[ \frac{d\Gamma}{dx\,dy} \right]_{{\rm initial}~x}
     \approx f_1(1{-}x,Q^2) \, \frac{d\Gamma}{dy}
   = \frac{\alphas}{2\pi} \, P(1{-}x) \, \ln\Bigl( \frac{Q^2}{Q_0^2} \Bigr)
     \,\frac{d\Gamma}{dy} .
\label {eq:DGLAPlog}
\end {equation}
In the main text and in the preceding paper \cite{2brem}, our convention
has usually been to identify the argument of the DGLAP splitting function
with the momentum fraction $x$ of the emitted gluon instead of with
$\zeta = 1{-}x$ in fig.\ \ref{fig:DGLAP}.  Since the relevant splitting
$g \to gg$ for this paper is symmetric,
it doesn't matter in any case.  So we'll henceforth
rewrite $P(1{-}x)$ above as $P(x)$.

The remaining job is to figure out the scales of
the relevant upper and lower
kinematic bounds on DGLAP
virtuality $Q^2$ as applies to our problem, in order to know what
ratio appears in the logarithm in (\ref{eq:DGLAPlog}).
In the high-energy, nearly-collinear limit,
virtuality is related to off-shellness
$\Delta E$ of energy by $Q^2 \sim E_Q \, \Delta E$.  In the case
of fig.\ \ref{fig:DGLAP}, $E_Q \simeq E$ for small $x$, where
$E$ is the initial parton energy.  $\Delta E$
is related to time by the uncertainty principle, and so
$Q^2 \sim E/\Delta t_x$, where $\Delta t_x$ is the time scale associated
with the $x$ emission.  So, (\ref{eq:DGLAPlog}) can be translated to
\begin {equation}
   \left[ \frac{d\Gamma}{dx\,dy} \right]_{{\rm initial}~x}
   \approx \frac{\alphas}{2\pi} \, P(x) \,
     \ln\Bigl( \frac{\Delta t_x^{\rm max}}{\Delta t_x^{\rm min}} \Bigr)
     \, \frac{d\Gamma}{dy} .
\label {eq:DGLAPlog2}
\end {equation}
The time scale $\Delta t_y$
of the underlying $y$ emission is the formation time
$t_{{\rm form},y}$, which provides the lower limit on the
time scale $\Delta t_x$ for generating a DGLAP logarithm.

In the above DGLAP analysis based on fig.\ \ref{fig:DGLAP},
we assumed that medium effects on the
$x$ emission could be ignored.  This assumption is only valid for
time scale small compared to the formation time $t_{{\rm form},x}$
for $x$ emission.  Another way to explain this infrared cut-off is
to note that the logarithm in (\ref{eq:DGLAPlog}) is a {\it collinear}
logarithm.  In vacuum, the $x$ and $\zeta$ particle trajectories
can be arbitrarily collinear, and there is no small-angle cut-off
for collinear logarithms if we ignore the masses of particles.
In medium, however, particles constantly change their direction over
time, and so the $x$ and $\zeta$ trajectories cannot remain collinear over
arbitrarily large times.  Formation times tell you how long you can wait
before such changes destroy quantum coherence.  The upshot is
that the relevant range of $\Delta t$ for which a vacuum DGLAP analysis
applies is
\begin {equation}
   t_{{\rm form},y} \lesssim \Delta t_x \lesssim t_{{\rm form},x} .
\label {eq:dtxrange}
\end {equation}
Correspondingly the logarithm in (\ref{eq:DGLAPlog}) is
$\ln(t_{{\rm form},x}/t_{{\rm form},y})$.
Combining $t_{{\rm form},x} \propto \sqrt{x}$ from (\ref{eq:tform})
with the soft limits
$P(x) \simeq 2\CA/x$ from (\ref{eq:Pggg}) and
\begin {equation}
   \frac{d\Gamma}{dy} \simeq
   \frac{\CA \alphas}{\pi y^{3/2}} \sqrt{\frac{\hat q_{\rm A}}{E}}
\end {equation}
from (\ref{eq:Gamma1}) and (\ref{eq:Omegai}) then gives
\begin {equation}
   \left[ \frac{d\Gamma}{dx\,dy} \right]_{{\rm initial}~x}
   \approx \frac{\CA^2 \alphas^2}{2\pi^2 x y^{3/2}} \,
     \ln\Bigl( \frac{x}{y} \Bigr) .
\end {equation}
As promised, this agrees with the
$2 \Re(x\bar xy\bar y + x\bar x \bar y y)$ entry
of Table \ref{tab:logs}.

Readers may wonder why, in our discussion above (and especially in our
discussion of Gunion-Bertsch cancellation below), we have focused on
(i) vacuum $x$ radiation from an underlying medium-induced soft
$y$-emission process, rather than (ii) soft, vacuum $y$ radiation from an
underlying medium-induced $x$-emission process.  The reason is that
the latter contribution to real double bremsstrahlung $d\Gamma/dx\,dy$
is sub-leading compared to the former
by a factor of $\sqrt{y/x}$.  To understand
this, consider the $x{\leftrightarrow}y$
analog of (\ref{eq:DGLAPlog2}),
which would contribute to $d\Gamma/dx\,dy$ something of order
$\alphas \, P(y) \, \frac{d\Gamma}{dx} \times \log$.  Since
$d\Gamma/dx \sim \alphas/x^{3/2}$ in units of $\sqrt{\hat q/E}$,
this gives a contribution to $d\Gamma/dx\,dy$ of order
$(\alphas^2/x^{3/2} y)\times\log$, which is sub-leading compared to
(\ref{eq:Blogs}) and
(\ref{eq:parametric}).


\subsection {Gunion-Bertsch}

\begin {figure}[t]
\begin {center}
  \includegraphics[scale=0.4]{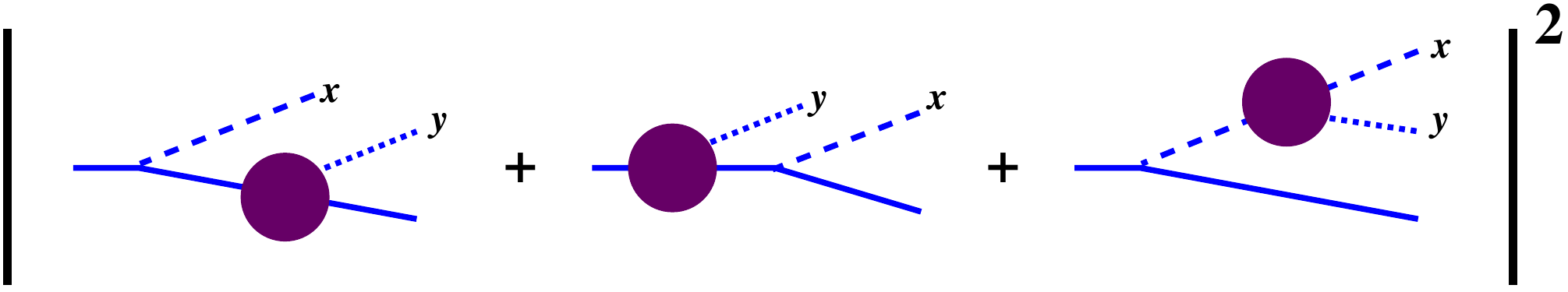}
  \caption{
     \label{fig:GB1}
     Generalization of Fig.\ \ref{fig:DGLAP} to the total rate.
  }
\end {center}
\end {figure}

\begin {figure}[t]
\begin {center}
  \includegraphics[scale=0.4]{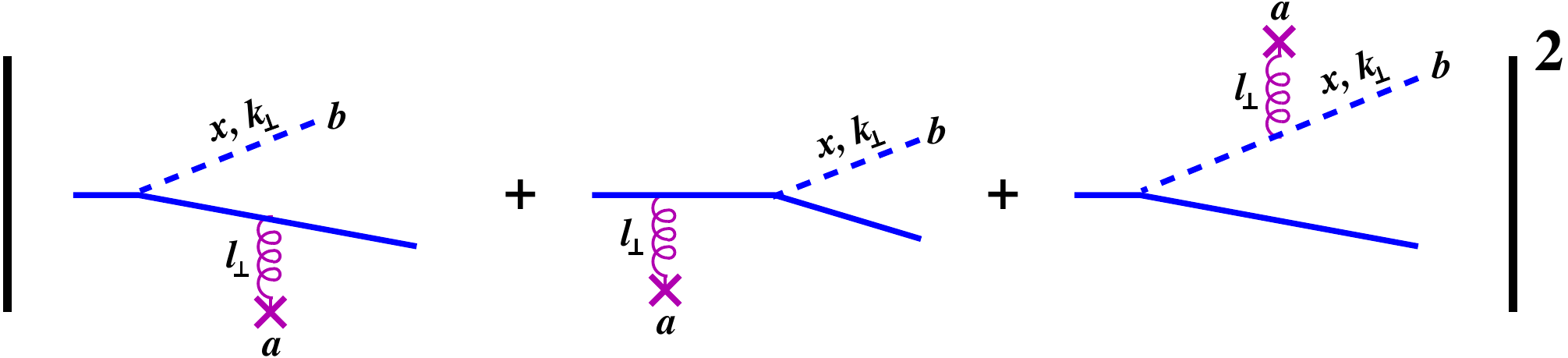}
  \caption{
     \label{fig:GB}
     An analogy to Fig.\ \ref{fig:GB1}: radiation during a high energy,
     small angle, 2-body
     collision, where the collision is mediated by exchanging a
     virtual gauge boson
     with transverse momentum $l_\perp$.  The crosses indicate the
     interaction with the other particle in the 2-body collision, the
     details of which will not be important here.
  }
\end {center}
\end {figure}

If we carry further the model we used in drawing
fig.\ \ref{fig:DGLAP}, we might expect the logarithmic corrections
to the {\it total}\/ rate to be similarly related to fig.\ \ref{fig:GB1}.
If we now abstract the $y$ emission process and its attendant medium
interaction as a net injection of momentum, then Fig.\ \ref{fig:GB1}
looks perhaps analogous to the process of fig.\ \ref{fig:GB}.
The latter figure shows the type of non-Abelian radiation
process long ago considered by Gunion and Bertsch \cite{GB} in
the context of 2-particle collisions in vacuum.
In the high-energy, nearly-collinear limit, the amplitude takes the
form%
\footnote{
   Our $\k_\perp$ is the $\q_\perp$ of Ref.\ \cite{GB}.  We have not
   bothered to show the matrix element representing the cross in
   our fig.\ \ref{fig:GB}, which can be factorized out.
}
\begin {equation}
  - g T^b T^a
  \frac{\k_\perp\cdot {\bm\varepsilon}_\perp (1-x)}
       {\k_\perp^2}
  + g T^a T^b \,
  \frac{(\k_\perp - x \l_\perp)\cdot {\bm\varepsilon}_\perp (1-x)}
       {| \k_\perp  - x \l_\perp|^2}
  - g [T^a,T^b] \,
  \frac{(\k_\perp - \l_\perp)\cdot {\bm\varepsilon}_\perp (1-x)}
       {| \k_\perp - \l_\perp |^2} ,
\label {eq:GB}
\end {equation}
where $\perp$ is defined relative to the initial particle direction,
$\l_\perp$ and $a$ characterize the transverse momentum transfer
and adjoint color index associated with the collision, and $\k_\perp$
and $b$ characterize the transverse momentum and adjoint color index
of the final bremsstrahlung gluon.  In the $k_\perp \gg l_\perp$ limit,
if we keep only the leading behavior of each individual diagram,
the corresponding rate is proportional to
\begin {equation}
  \int d^2k_\perp
  \sum_{a,b} \tr \left|
  - g T^b T^a
  \frac{\k_\perp \cdot {\bm\varepsilon}_\perp (1-x)}{\k_\perp^2}
  + g T^a T^b \,
  \frac{\k_\perp \cdot {\bm\varepsilon}_\perp (1-x)}{\k_\perp^2}
  - g [T^a,T^b] \,
  \frac{\k_\perp\cdot {\bm\varepsilon}_\perp (1-x)}
       {\k_\perp^2}
  \right|^2 .
\label {eq:GBapprox}
\end {equation}
Any individual term, such as
\begin {equation}
  \int d^2k_\perp
  \sum_{a,b} \tr \left|
  - g T^b T^a \,
  \frac{\k_\perp \cdot {\bm\varepsilon}_\perp (1-x)}{\k_\perp^2}
  \right|^2 ,
\label {eq:GBterm}
\end {equation}
generates a factor proportional to
$g^2 \ln(k_\perp^{\rm max}/k_\perp^{\rm min})$, which
is the same sort of logarithmic factor found in (\ref{eq:DGLAPlog}).
However, all the terms in the amplitude in (\ref{eq:GBapprox})
cancel each other, and so all of these logarithmic factors coming
from $k_\perp \gg l_\perp$ cancel each other in the total rate.
Moreover, if the initial particle is a gluon, so that the generators
$T$ above are adjoint-representation generators, then the squares
of each of the individual three terms in (\ref{eq:GBapprox}) are
respectively proportional to
\begin {equation}
  \tr \bigl( (T^b T^a)^\dagger T^b T^a \bigr), \qquad
  \tr \bigl( (T^a T^b)^\dagger T^a T^b \bigr), \qquad
  \tr \bigl( [T^a,T^b]^\dagger [T^a,T^b] \bigr),
\label {eq:clebsch1}
\end {equation}
which (for the adjoint representation) are all the same!  In our
analogy, this corresponds to the equality of the last three rows
of Table \ref{tab:logs}.  Similarly, the cross-terms are proportional
to
\begin {equation}
  -2 \Re \tr \bigl( (T^a T^b)^\dagger T^b T^a \bigr), \quad
  -2 \Re \tr \bigl( [T^a,T^b]^\dagger T^a T^b \bigr), \quad
  2 \Re \tr \bigl( [T^a,T^b]^\dagger T^b T^a \bigr),
\end {equation}
which all equal the negative of (\ref{eq:clebsch1}),
analogous to the first three rows of Table \ref{tab:logs}.
So the cancellation of the logarithms in Table \ref{tab:logs}
appears to be of the Gunion-Bertsch type.

What about the condition $k_\perp \gg l_\perp$ for Gunion-Bertsch
cancellation?  First consider that for $k_\perp \gg l_\perp$ and
small $x$, the off-shellness $\Delta E$ of energy
associated with the intermediate lines in fig.\ \ref{fig:GB1} are
all of order $k_\perp^2/x E$, which corresponds to a time scale
$\Delta t_x \sim x E/k_\perp^2$.  The condition
$k_\perp \gg l_\perp$ is therefore equivalent to the condition
$\Delta t_x \ll x E/l_\perp^2$.  In the context of fig.\ \ref{fig:GB1},
$l_\perp$ consists of the combination of the transverse momentum carried
away by the $y$ bremsstrahlung and the transverse momentum
$\sim \sqrt{\hat q t_{{\rm form}, y}}$ injected by the medium during the
$y$ formation, which are parametrically the same and give
$l_\perp \sim \sqrt{\hat q t_{{\rm form},y}}$.  So the condition
$k_\perp \gg l_\perp$ is equivalent to
\begin {equation}
  \Delta t_x \ll \frac{x E}{\hat q t_{{\rm form},y}} .
\end {equation}
Using the formation time (\ref{eq:tform}), this may be recast as
\begin {equation}
  \Delta t_x \ll t_{{\rm form}, x} \sqrt{\frac{x}{y}} .
\end {equation}
Given our original assumption that $y \ll x$,
the above condition is automatically satisfied for the range (\ref{eq:dtxrange})
of values that generated the logarithm in our case.
The moral of this story is that the logarithms characterized by
Table\ \ref{tab:logs} arise from a kinematic regime that is
analogous to the cancellation of $k_\perp \gg l_\perp$ logarithms
in Gunion-Bertsch.


\subsection {Independent emission model}

There is an alternative picture for understanding and interpreting
two of the entries in Table\ \ref{tab:logs}, which we offer here as
a complement to the previous discussion.


\subsubsection {The approximation}

Imagine that we tried to
estimate double bremsstrahlung in the $y \ll x \ll 1$ limit by assuming that
the $x$ and $y$ bremsstrahlung amplitudes were completely independent
from each other and both given simply by {\it single}-bremsstrahlung
formulas.  This approximation would be somewhat similar to the
small $x$ and $y$ limit of the idealized Monte Carlo approximation
(\ref{eq:MC1}) except that we don't care about the ordering of
$\tau_x$ and $\tau_y$:
\begin {align}
   \left[\frac{d\Gamma}{dx\,dy}\right]^{\rm independent}
   \simeq \int_{-\infty}^{\infty} d(\tau_y{-}\tau_x) &
     \int_0^\infty d(\Delta t_\xx)
     \int_0^\infty d(\Delta t_\yx)
\nonumber\\& \times
     \Re \left[\frac{d\Gamma}{dx\,d(\Delta t_\xx)}\right]_E \,
     \Re \left[\frac{d\Gamma}{dy\,d(\Delta t_\yx)}\right]_E .
\label {eq:indep0}
\end {align}
We will find that this is a useful approximation for
the processes
indicated in the first two rows of Table \ref{tab:logs}, for which
the time integrations above must be correspondingly restricted.
(We'll discuss what happens with the third row of the table later.)
In those processes,
the $y$ emissions (one in the amplitude and one in the conjugate
amplitude) are restricted to occur between the $x$ emissions.
This time ordering modifies (\ref{eq:indep0}) to
\begin {align}
   \left[\frac{d\Gamma}{dx\,dy}
   \right]^{\rm independent}_{2\Re(\bullet y \bar y \bar \bullet + \bullet \bar y y \bar \bullet)}
   \simeq &
     \int_0^\infty d(\Delta t_x)
     \int_0^{\Delta t_x} d(\Delta t_y)
     \int_{-(\Delta t_x-\Delta t_y)/2}^{+(\Delta t_x-\Delta t_y)/2}
          d(\tau_y{-}\tau_x)
\nonumber\\& \qquad\qquad \times
     \Re \left[\frac{d\Gamma}{dx\,d(\Delta t_x)}\right]_E \,
     \Re \left[\frac{d\Gamma}{dy\,d(\Delta t_y)}\right]_E
\nonumber\\
   = &
     \int_0^\infty d(\Delta t_x)
     \int_0^{\Delta t_x} d(\Delta t_y) \>
     (\Delta t_x-\Delta t_y)
\nonumber\\& \qquad\qquad \times
     \Re \left[\frac{d\Gamma}{dx\,d(\Delta t_x)}\right]_E \,
     \Re \left[\frac{d\Gamma}{dy\,d(\Delta t_y)}\right]_E .
\label {eq:indep1}
\end {align}

\begin {figure}[t]
\begin {center}
  \includegraphics[scale=0.5]{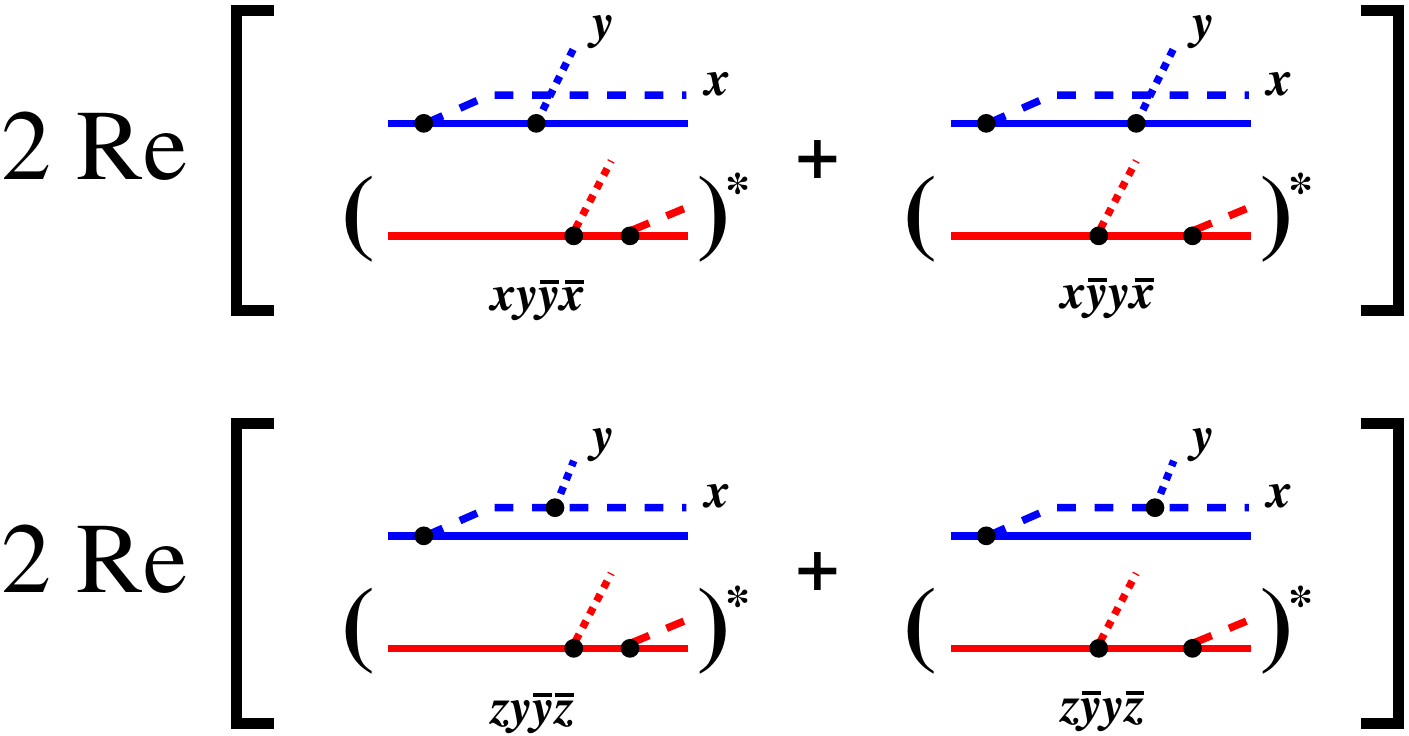}
  \caption{
     \label{fig:indep}
     The processes in the first two rows of
     Table \ref{tab:logs}, drawn in a way that sets us up for
     fig.\ \ref{fig:indep2}.  Above, the black dots denote vertices;
     there is no vertex where lines cross without a dot.
  }
\end {center}
\end {figure}

\begin {figure}[t]
\begin {center}
  \includegraphics[scale=0.5]{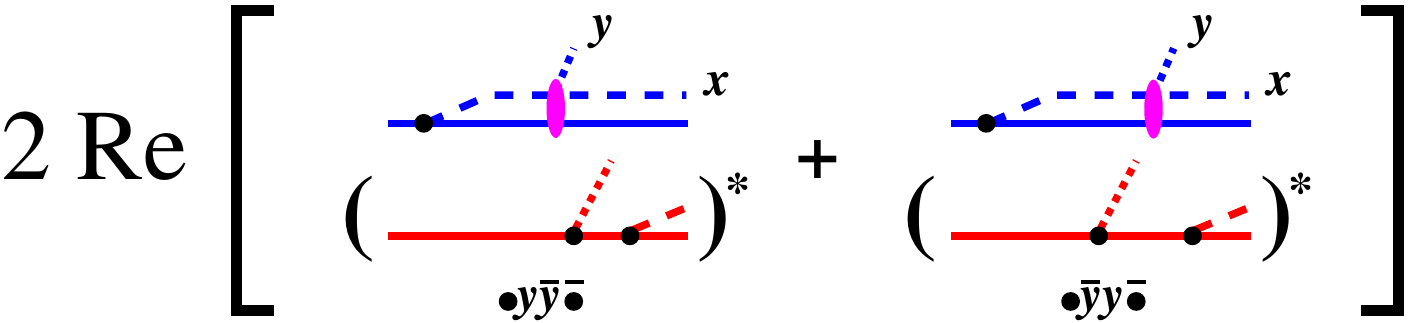}
  \caption{
     \label{fig:indep2}
     Equivalent to fig.\ \ref{fig:indep} in the limit $y \ll x \ll 1$
     where the $y$ emission cannot resolve the $x$ and $1{-}x$
     daughters of the $x$ emission process and couples to them
     (magenta ovals) as to a single adjoint-representation particle.
  }
\end {center}
\end {figure}

Why have we written $\bullet y \bar y \bar \bullet + \bullet \bar y y
\bar \bullet$ in the subscript above instead of, say, $x y \bar y \bar x + x
\bar y y \bar x$?  In our limit $y \ll x \ll 1$, we have $z \equiv
1{-}x{-}y \simeq 1{-}x$.  For {\it single} bremsstrahlung, there is no
difference between referring to the corresponding diagram
(fig.\ \ref{fig:xxrate}) as $x\bar x$ or $z\bar z$
(with $z \equiv 1{-}x$ in the single bremsstrahlung context).
We could use either notation to
describe the $x$ emission in the independent bremsstrahlung
approximation we have
outlined above.  When we talk about actual double bremsstrahlung
diagrams, however, there is a difference.  Let's look at those
diagrams: The interference processes listed in the first two rows of
Table \ref{tab:logs} can be drawn in the form of
fig.\ \ref{fig:indep}.  However, for the reasons given in section
\ref{sec:conclusion}, we expect that the $y$ emission cannot resolve
the two daughters ($x$ and $1{-}x$) of the $x$ emission process when
$y \ll x \ll 1$.  So $y$ emission from those two daughters is
approximately equivalent to $y$ emission from a single particle of energy
$E$ (and so is similar to the cases where $y$ emission occurs instead from the
initial particle of energy $E$).  That is, we can redraw the sum of
diagrams in fig.\ \ref{fig:indep} as fig.\ \ref{fig:indep2}, and it is
this sum that should correspond to the independent approximation given
by the right-hand side of (\ref{eq:indep1}).  The subscript notation
on the left-hand side of (\ref{eq:indep1})
means that this approximation corresponds to the
diagrams of fig.\ \ref{fig:indep} with ``$\bullet$'' summed over the cases
$\bullet = x$ and $\bullet = z$.

\subsubsection {Evaluation and Comparison}

Comparison of (\ref{eq:1bremB}) and (\ref{eq:Gamma1}) identifies
\begin {equation}
   \left[\frac{d\Gamma}{dx\,d(\Delta t)}\right]_E =
   - \frac{\alpha P(x)}{\pi} \, \Omega_{E,x}^2 \csc^2(\Omega_{E,x}\,\Delta t) .
\label {eq:dGdx}
\end {equation}
Using (\ref{eq:Pggg}), the independent emission contribution
(\ref{eq:indep1}) for $y \ll x \ll 1$ is then
\begin {align}
   \left[\frac{d\Gamma}{dx\,dy}
     \right]^{\rm independent}_{2\Re(\bullet y \bar y \bar \bullet + \bullet \bar y y \bar \bullet)}
   &\simeq
   \frac{4 \CA^2 \alphas^2}{\pi^2 x y}
   \int_0^\infty d(\Delta t_y) \>
   \Re\bigl[ \Omega_y^2 \csc^2(\Omega_y \,\Delta t_y) \bigr]
\nonumber\\ & \qquad\qquad\qquad \times
   \int_{\Delta t_y}^\infty d(\Delta t_x) \>
   \Re\bigl[ \Omega_x^2 \csc^2(\Omega_x\,\Delta t_x) \bigr]
   \,
   (\Delta t_x-\Delta t_y)
\nonumber\\
   &=
   \frac{4 \CA^2 \alphas^2}{\pi^2 x y}
   \int_0^\infty d(\Delta t_y) \>
   \Re\bigl[ \Omega_y^2 \csc^2(\Omega_y \,\Delta t_y) \bigr] \,
   \Re\ln\left(\frac{1}{1-e^{-2 i \Omega_x \Delta t_y}}\right) ,
\end {align}
where we define the small $x,y$ expressions
\begin {equation}
   \Omega_x \equiv \sqrt{-\frac{i \hat q_{\rm A}}{2 x E}} ,
   \qquad
   \Omega_y \equiv \sqrt{-\frac{i\hat q_{\rm A}}{2 y E}} .
\label {eq:Omegaxy}
\end {equation}
The integrand falls exponentially for $\Delta t_y \gg \Omega_y^{-1}$,
and so $\Omega_x \,\Delta t_y \ll 1$ in the $y \ll x \ll 1$ limit, giving
\begin {align}
   \left[\frac{d\Gamma}{dx\,dy}
     \right]^{\rm independent}_{2\Re(\bullet y \bar y \bar \bullet + \bullet \bar y y \bar \bullet)}
   &\simeq
   \frac{4 \CA^2 \alphas^2}{\pi^2 x y}
   \int_0^\infty d(\Delta t_y) \>
   \Re\bigl[ \Omega_y^2 \csc^2(\Omega_y \,\Delta t_y) \bigr] \,
   \Re\ln\left(\frac{1}{2 i \Omega_x \Delta t_y}\right)
\nonumber\\
   &=
   \frac{4 \CA^2 \alphas^2}{\pi^2 x y}
   \Re \int_0^\infty d(\Delta t_y) \>
   \Omega_y^2 \csc^2(\Omega_y \,\Delta t_y) \,
   \ln\left(\frac{1}{2 |\Omega_x| \Delta t_y}\right) .
\label {eq:indep1b}
\end {align}
This integral has a $\Delta t {\to} 0$ divergence.
In order to regulate divergences of individual diagrams, we have
routinely subtracted out the vacuum contribution to those diagrams.
Here, we will subtract out the vacuum contribution to the
independent $y$ emission, replacing (\ref{eq:indep1b}) by
\begin {align}
   \left[\frac{d\Gamma}{dx\,dy}
     \right]^{\rm independent}_{2\Re(\bullet y \bar y \bar \bullet + \bullet \bar y y \bar \bullet)}
   &\simeq
   \frac{4 \CA^2 \alphas^2}{\pi^2 x y}
   \Re \int_0^\infty d(\Delta t_y) \>
   \left[
     \Omega_y^2 \csc^2(\Omega_y \,\Delta t_y) - \frac{1}{(\Delta t)^2}
   \right]
   \ln\left(\frac{1}{2 |\Omega_x| \Delta t_y}\right)
\nonumber\\
   &=
   \frac{4 \CA^2 \alphas^2}{\pi^2 x y}
   \Re \left\{
      -i \Omega_y \left[
         \ln \Bigl( \frac{i\Omega_y}{2\pi|\Omega_x|} \Bigr) + \gammaE ,
      \right]
   \right\}
\label {eq:indep2}
\end {align}
where $\gammaE$ is the Euler-Mascheroni constant.
(We leave a more careful and convincing treatment of the $\Delta t{\to} 0$
regularization to other work \cite{dimreg}.)
Using (\ref{eq:Omegaxy}), the above result can be recast as
\begin {equation}
   \left[\frac{d\Gamma}{dx\,dy}
     \right]^{\rm independent}_{2\Re(\bullet y \bar y \bar \bullet + \bullet \bar y y \bar \bullet)}
   \simeq
   \frac{\CA^2 \alphas^2}{\pi^2 x y^{3/2}}
   \sqrt{ \frac{\hat q_{\rm A}}{E} }
   \left[
      -\ln \Bigl( \frac{x}{y} \Bigr)
      +2 \ln(2\pi) - 2\gammaE + \frac{\pi}{2}
   \right] .
\label {eq:indep3}
\end {equation}
The coefficient $-1$ of the logarithm indeed matches up with the sum of the
first two rows in Table\ \ref{tab:logs}, as promised.

We should mention that it is also possible to extract the integral
expression in (\ref{eq:indep2}) not just from the simple
independent bremsstrahlung
approximation above but also from the full integral expressions for
the general result for
$2\Re(x y \bar y\bar x + x \bar y y\bar x)$ and
$2\Re(z y \bar y\bar z + z \bar y y\bar z)$ given in the
preceding work \cite{2brem}.  This can be done
by analytically extracting the behavior of those expressions in the
limit $y \ll x \ll 1$ while formally treating the integration variable
$\Delta t_y \sim \Omega_y^{-1} \propto \sqrt{y}$.  We will not give
details here.
Individually,
$2\Re(x y \bar y\bar x + x \bar y y\bar x)$ and
$2\Re(z y \bar y\bar z + z \bar y y\bar z)$ each contribute half
of the limiting result (\ref{eq:indep3}) in this analysis.

In the first row of Table \ref{tab:logs}, we did not explicitly
list the $2 \Re(x\bar y\bar x y)$ contribution to the group
of diagrams represented by $A(x,y)$.%
\footnote{
  $A(x,y)$ represents the three {\it explicit}\/ diagrams shown
  in fig.\ \ref{fig:subset2} plus their conjugates.
}
To understand this omission, note that the origin of the logarithm in
(\ref{eq:indep2}) can be traced back to the
$\Delta t_x - \Delta t_y \simeq \Delta t_x$ factor inside
the last integral in (\ref{eq:indep1}).  This factor arises because,
for
$\bullet y \bar y \bar \bullet + \bullet \bar y y
\bar \bullet$,
the short-duration $y$ emission can happen at any time in the middle
of the longer-duration $x$ emission.
For the $x\bar y\bar x y$ process shown in
fig.\ \ref{fig:subset}, however, (and similarly for $z\bar y\bar z y$)
the short-duration $y$ emission is
forced by the time-ordering of the diagram to occur only at the
very end of the $x$ emission interval, and so there is no comparable
factor and so no logarithmic enhancement.


\subsubsection {Other processes}

\begin {figure}[t]
\begin {center}
  \includegraphics[scale=0.5]{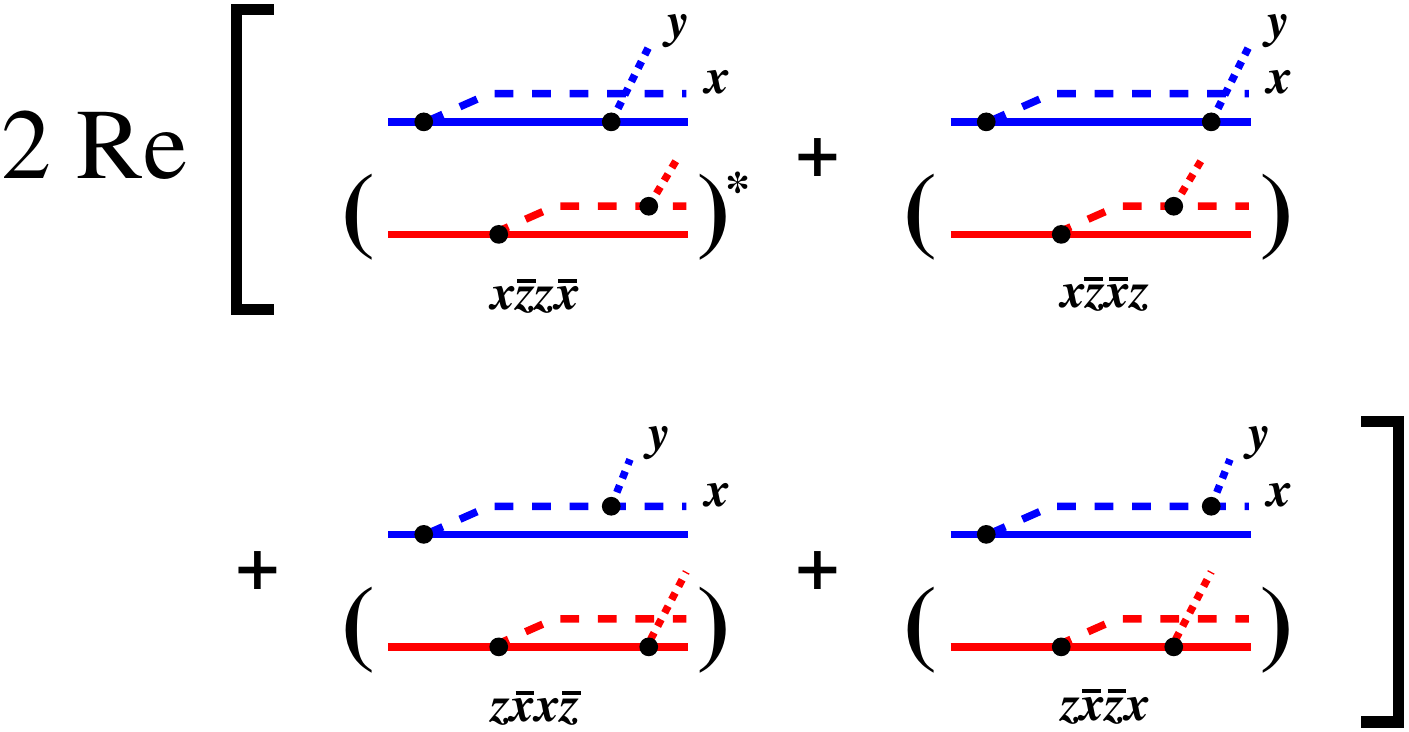}
  \caption{
     \label{fig:indep3}
     The processes in the third row of
     Table \ref{tab:logs}, drawn in a way similar to
     fig.\ \ref{fig:indep}.
  }
\end {center}
\end {figure}

\begin {figure}[t]
\begin {center}
  \includegraphics[scale=0.5]{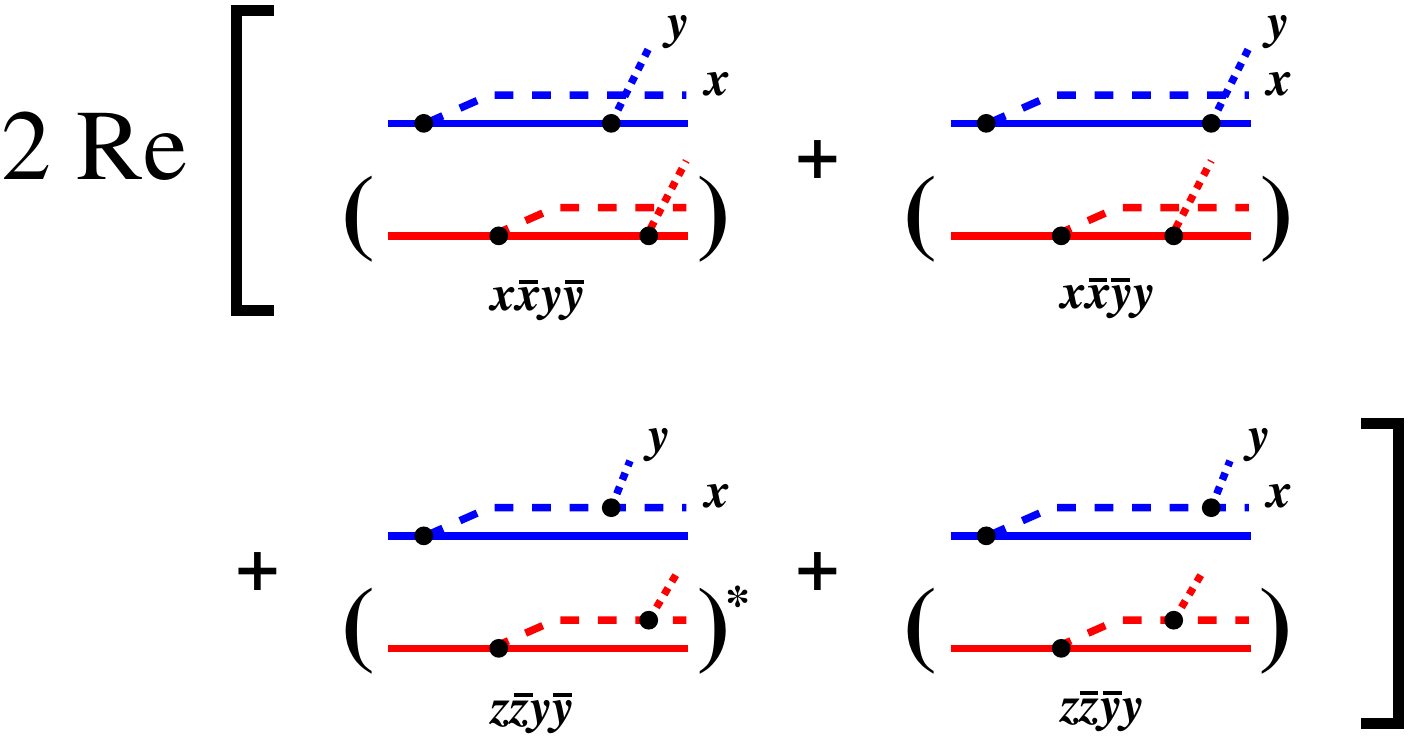}
  \caption{
     \label{fig:indep4}
     The idealized Monte Carlo related processes associated with the
     fourth and fifth rows of
     Table \ref{tab:logs}, drawn in a way for easy comparison with
     fig.\ \ref{fig:indep3}.
  }
\end {center}
\end {figure}

\begin {figure}[t]
\begin {center}
  \includegraphics[scale=0.5]{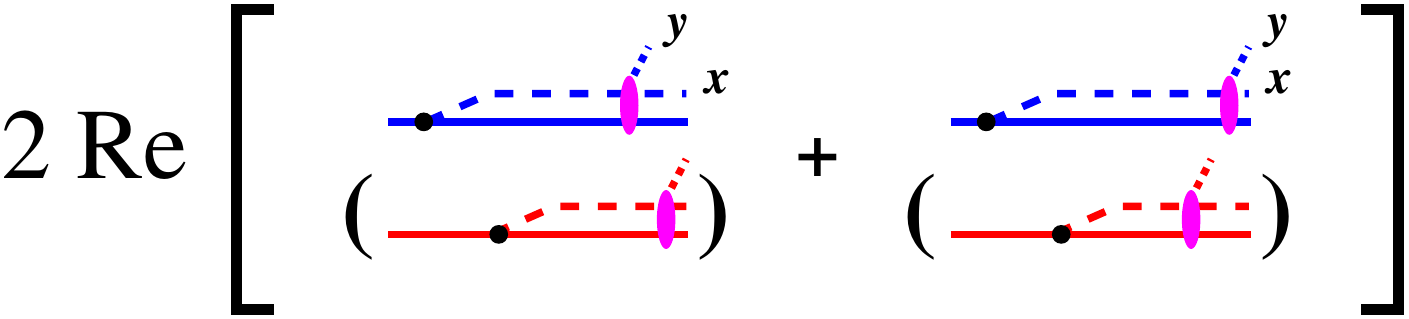}
  \caption{
     \label{fig:indep5}
     The (problematic) independent bremsstrahlung approximation inspired by the
     sum of figs.\ \ref{fig:indep3} and \ref{fig:indep4},
     analogous to fig.\ \ref{fig:indep2}.
  }
\end {center}
\end {figure}

One may investigate similar approximations of other processes in
Table \ref{tab:logs}.  However, we do not know of any sensible way
to apply the independent bremsstrahlung approximation to the third
row of the table, which refers to
$2\Re(x\bar zz\bar x + x\bar z\bar x z
 + z\bar xx\bar z + z\bar x\bar z x)$.
These diagrams are depicted in fig.\ \ref{fig:indep3} in a fashion
similar to fig.\ \ref{fig:indep}.  These are interference diagrams
that combine with the Monte Carlo related diagrams of
fig.\ \ref{fig:indep4} to partly suppress the result along the lines discussed
in section \ref{sec:conclusion}.  Roughly speaking,
the $y$ emission's inability to individually resolve the
$x$ and $1{-}x$ daughters in the $y \ll x$ limit
partly suppresses the rate.  This combined effect is depicted
in fig.\ \ref{fig:indep5}.  The problem with applying an independent
bremsstrahlung approximation to the picture in fig.\ \ref{fig:indep5}, however,
is that there is nothing in that approximation that would constrain
how large could be the separation of the $y$ emission from the $x$
emission.  Consider the two daughters of the $x$ emission.
The issue of how $y$ emission from one daughter decoheres with
$y$ emission from the other, for times $\gg t_{{\rm form},x}$
after the last $x$ emission, is a complicated matter that is simply not
captured by the independent bremsstrahlung approximation that was used so
successfully above for fig.\ \ref{fig:indep3}.

\section {More explicit derivation of  \boldmath$x\bar x y\bar y$}
\label {app:xxyy}

In this appendix, we show how to start with the methods and
diagrammatic rules of the preceding paper \cite{2brem} and
obtain the result for the
$x\bar xy\bar y$ interference given by (\ref{eq:xxyyA}).

Using the notation of
ref.\ \cite{2brem}, the $x\bar x y\bar y$ interference of
fig.\ \ref{fig:xxyy} gives
\begin {multline}
   \left[\frac{dI}{dx\,dy}\right]_{x\bar x y\bar y}
   =
   \left( \frac{E}{2\pi} \right)^2
   \int_{t_\xx < t_\xbx < t_\yx < t_\ybx}
   \sum_{\rm pol.}
   \langle|i\,\overline{\delta H}|\B^\ybx\rangle
   \langle\B^\ybx,t_\ybx|\B^\yx,t_\yx\rangle
   \langle\B^\yx|{-}i\,\delta H|\rangle
\\ \times
   \langle|\rangle^{-1}
   \langle|i\,\overline{\delta H}|\B^\xbx\rangle
   \langle\B^\xbx,t_\xbx|\B^\xx,t_\xx\rangle
   \langle\B^\xx|{-}i\,\delta H|\rangle
\label {eq:xxyy0}
\end {multline}
for the differential probability.
Here, all transverse positions
($\B^\xx$, $\B^\xbx$, $\B^\yx$, $\B^\ybx$) are implicitly integrated over,
and
$\delta H$ is the part of the full Hamiltonian of the theory
that contains the splitting vertices for high-energy particles.

\begin {figure}[t]
\begin {center}
  \begin{picture}(250,150)(0,0)
  \put(0,0){\includegraphics[scale=0.5]{xxyy.pdf}}
  \put(27,44){${-}1,{-}h_{\rm i}$}
  \put(45,100){$x,h_{\rm x}$}
  \put(22,138){$1{-}x,h$}
  \put(170,80){$y,h_{\rm y}$}
  \put(175,138){$1{-}x{-}y,h_{\rm z}$}
  \put(165,44){${-}(1{-}x),{-}\bar h$}
  \end{picture}
  \caption{
     \label{fig:xxyy}
     Our labeling conventions for times $t_i$, longitudinal momenta $x_i$ and
     helicities $h_i$ for the $x\bar x y\bar y$ interference diagram.
  }
\end {center}
\end {figure}

One non-trivial normalization factor above has to do with the
intermediate 2-particle (effectively 0-particle) state $|\rangle$
for $t_\xbx < t < t_\yx$.  This state was normalized
in the preceding paper \cite{2brem} as
\begin {equation}
  \langle|\rangle = x_1^2 = x_2^2 ,
\end {equation}
where, in this context, $(x_1,x_2)=(x_1,-x_1)$ are the longitudinal
momentum fractions of the two particles in the 2-particle state.
This normalization was both natural and convenient in the discussion
of ref.\ \cite{2brem}, and it is similarly convenient here, for the
normalization of the initial and final states in (\ref{eq:xxyy0}).
However, having a non-trivial normalization means that the correct
way to project out the 2-particle (effectively 0-particle) intermediate
state, appropriate to fig.\ \ref{fig:xxyy}, is with
\begin {equation}
   \frac{ |\rangle \langle| }{ \langle|\rangle } \,,
\end {equation}
as in (\ref{eq:xxyy0}).

The effectively 0-particle intermediate state $|\rangle$ does not
have any time dependence in our formalism, and so the integrand in
(\ref{eq:xxyy0}) does not care how far apart in time $t_\xbx$ and
$t_\yx$ are.  That is, we have not bothered to write
a factor of 1 in (\ref{eq:xxyy0}) representing
the time evolution of this intermediate state:
\begin {equation}
   \frac{\langle t_\yx | t_\xbx \rangle}{\langle|\rangle}
   = 1.
\end {equation}
This time independence is consistent with interpreting this diagram as
representing two consecutive splittings that are completely independent
from each other.

Using the rules developed in the preceding paper \cite{2brem},
fig.\ \ref{fig:xxyy} and (\ref{eq:xxyy0}) become
\begin {align}
   \left[\frac{dI}{dx\,dy}\right]_{x\bar xy\bar y}
   &=
   \frac{C_R^2 \alphas^2 }{4 E^4} \,
   (1-x)^{-2}
   \int_{t_\xx < t_\xbx < t_\yx < t_\ybx}
   \sum_{h_\xx,h_\yx,h_\zx}
\nonumber\\ &\times
   \Bigl[
   \sum_{\bar h}
   {\cal P}^{\bar n}_{\bar h \to h_\zx,h_\yx}\bigl(1{-}x \to 1{-}x{-}y,y\bigr) \,
   {\cal P}^{\bar m}_{h_\ix \to \bar h, h_\xx}\bigl(1 \to 1{-}x,x\bigr)
   \Bigr]^*
\nonumber\\ &\times
   \Bigl[
   \sum_h
   {\cal P}^n_{h \to h_\zx,h_\yx}\bigl(1{-}x \to 1{-}x{-}y,y\bigr) \,
   {\cal P}^m_{h_\ix \to h,h_\xx}\bigl(1 \to 1{-}x,x\bigr)
   \Bigr]
\nonumber\\ &\times
   \nabla^{\bar n}_{\B^\ybx}
   \nabla^n_{\B^\yx}
   \langle\B^\ybx,t_\ybx|\B^\yx,t_\yx\rangle
   \Bigr|_{\B^\ybx=0=\B^\yx}
   \nabla^{\bar m}_{\B^\xbx}
   \nabla^m_{\B^\xx}
   \langle\B^\xbx,t_\xbx|\B^\xx,t_\xx\rangle
   \Bigr|_{\B^\xbx=0=\B^\xx} ,
\label {eq:xxyy1}
\end {align}
where ${\cal P}$ is defined in terms of DGLAP splitting functions in
section IV.E of ref.\ \cite{2brem}.
The overall $(1-x)^{-2}$ factor above is the
intermediate-state normalization factor $\langle|\rangle^{-1}$ from
(\ref{eq:xxyy0}).
In the application in this paper, the color representation $R$ of the
solid line in fig.\ \ref{fig:xxyy} will be adjoint,
and so $C_R$ above is $\CA$.

Now turn to the helicity sums, which we will relate to
spin-averaged DGLAP splitting functions.  To this end, it is
useful to first use rotational invariance in the transverse
plane to write
\begin {equation}
   \nabla^{\bar m}_{\B^\xbx}
   \nabla^m_{\B^\xx}
   \langle\B^\xbx,t_\xbx|\B^\xx,t_\xx\rangle
   \Bigr|_{\B^\xbx=0=\B^\xx}
   =
   \tfrac12 \, \delta^{m \bar m}
   \grad_{\B^\xbx} \cdot \grad_{\B^\xx}
   \langle\B^\xbx,t_\xbx|\B^\xx,t_\xx\rangle
   \Bigr|_{\B^\xbx=0=\B^\xx} ,
\label {eq:rotationdot}
\end{equation}
and so
\begin {align}
   \left[\frac{dI}{dx\,dy}\right]_{x\bar x y\bar y}
   &=
   \frac{C_R^2 \alphas^2 }{4 E^4} \,
   (1-x)^{-2}
   \int_{t_\xx < t_\xbx < t_\yx < t_\ybx}
   \tfrac14 \sum_{h_\xx,h_\yx,h_\zx}
\nonumber\\ &\times
   \Bigl[
   \sum_{\bar h}
   {\cal P}^n_{\bar h \to h_\zx,h_\yx}\bigl(1{-}x \to 1{-}x{-}y,y\bigr) \,
   {\cal P}^m_{h_\ix \to \bar h, h_\xx}\bigl(1 \to 1{-}x,x\bigr)
   \Bigr]^*
\nonumber\\ &\times
   \Bigl[
   \sum_h
   {\cal P}^n_{h \to h_\zx,h_\yx}\bigl(1{-}x \to 1{-}x{-}y,y\bigr) \,
   {\cal P}^m_{h_\ix \to h,h_\xx}\bigl(1 \to 1{-}x,x\bigr)
   \Bigr]
\nonumber\\ &\times
   \grad_{\B^\ybx} \cdot \grad_{\B^\yx}
   \langle\B^\ybx,t_\ybx|\B^\yx,t_\yx\rangle
   \Bigr|_{\B^\ybx=0=\B^\yx}
   \grad_{\B^\xbx} \cdot \grad_{\B^\xx}
   \langle\B^\xbx,t_\xbx|\B^\xx,t_\xx\rangle
   \Bigr|_{\B^\xbx=0=\B^\xx} .
\label {eq:xxyy1b}
\end {align}
By reflection invariance of the problem in the transverse plane, we
are free to average over initial helicities $h_\ix$.  The
factor
\begin {equation}
   \sum_{h_\xx,h_\yx,h_\zx}
   \Bigl[
   \sum_{\bar h}
   {\cal P}^n_{\bar h \to h_\zx,h_\yx}
   {\cal P}^m_{h_\ix \to \bar h, h_\xx}
   \Bigr]^*
   \Bigl[
   \sum_h
   {\cal P}^n_{h \to h_\zx,h_\yx}
   {\cal P}^m_{h_\ix \to h,h_\xx}
   \Bigr]
\label {eq:PPPP0}
\end {equation}
(where we've suppressed the other arguments of ${\cal P}$ for brevity)
then becomes
\begin {equation}
     \tfrac12 \sum_{h,\bar h} \Bigl[
     \sum_{h_\yx,h_\zx}
     \bcalP_{\bar h \to h_\zx,h_\yx}^* \cdot \bcalP_{h \to h_\zx,h_\yx}
     \sum_{h_\ix,h_\xx}
     \bcalP_{h_\ix \to \bar h, h_\xx}^* \cdot
     \bcalP_{h_\ix \to h,h_\xx}
     \Bigr] .
\label {eq:PPPP1}
\end {equation}
By rotational invariance,
$\sum_{h_\yx,h_\zx} \bcalP_{\bar h \to h_\zx,h_\yx}^* \cdot \bcalP_{h \to h_\zx,h_\yx}
    = \tfrac12 \delta_{h,\bar h} \sum_{h',h_\yx,h_\zx}
      \bcalP_{h' \to h_\zx,h_\yx}^* \cdot \bcalP_{h' \to h_\zx,h_\yx}$,
and (\ref{eq:PPPP1}) becomes
\begin {equation}
     \tfrac12 \sum_{h',h_\yx,h_\zx}
     \bcalP_{h' \to h_\zx,h_\yx}^* \cdot \bcalP_{h' \to h_\zx,h_\yx}
     \times
     \tfrac12 \sum_{h_\ix,h,h_\xx}
     \bcalP_{h_\ix \to h, h_\xx}^* \cdot
     \bcalP_{h_\ix \to h,h_\xx}
     .
\end {equation}
Up to normalization factors in the definition of $\bcalP$, the
two factors above each represent a spin-averaged DGLAP splitting function.
Taking the precise definition of $\bcalP$ from the preceding 
paper \cite{2brem}, we specifically obtain
\begin {equation}
     \frac{2\,P\bigl(-(1{-}x),1{-}x{-}y,y\bigr)}{C_R (1{-}x)^2y^2(1{-}x{-}y)^2}
     \times
     \frac{2\,P(-1,1{-}x,x)}{C_R x^2(1{-}x)^2}
     =
     \frac{2\,P(\yfrak)}{C_R (1{-}x)y^2(1{-}x{-}y)^2}
     \times
     \frac{2\,P(x)}{C_R x^2(1{-}x)^2} \,.
\label {eq:PPPP2}
\end {equation}
Substituting (\ref{eq:PPPP2}) for (\ref{eq:PPPP0}) in
(\ref{eq:xxyy1b}) then yields%
\footnote{
  An alternative way one could get to this result from
  (\ref{eq:xxyy1b}) is to use (\ref{eq:abcbardef}) to write
  (\ref{eq:PPPP0}) as
  $4(\bar\alpha + \tfrac12 \bar\beta + \tfrac12 \bar\gamma)$.
  Then use (\ref{eq:abcPP2}) to arrive at (\ref{eq:xxyy3}).
}
\begin {align}
   \left[\frac{dI}{dx\,dy}\right]_{x\bar xy\bar y}
   &=
   (1-x)^{-1}
   \int_{t_\xx < t_\xbx < t_\yx < t_\ybx}
\nonumber\\ &\times
   \frac{\alphas P(\yfrak)}{2[(1-x)y(1-x-y)E]^2} \,
   \grad_{\B^\ybx} \cdot
   \grad_{\B^\yx}
   \langle\B^\ybx,t_\ybx|\B^\yx,t_\yx\rangle
   \Bigr|_{\B^\ybx=0=\B^\yx}
\nonumber\\ &\times
   \frac{\alphas P(x)}{2[x(1-x)E]^2} \,
   \grad_{\B^\xbx} \cdot
   \grad_{\B^\xx}
   \langle\B^\xbx,t_\xbx|\B^\xx,t_\xx\rangle
   \Bigr|_{\B^\xbx=0=\B^\xx}
\label {eq:xxyy3}
\end {align}
with $\yfrak$ defined by (\ref{eq:yfrak}).

Now using the definition (\ref{eq:dGdDt}) of $d\Gamma/dx\,d\Delta t$,
(\ref{eq:xxyy3})
will give (once we clarify below some issues about normalization)
\begin {align}
   \left[\frac{dI}{dx\,dy}\right]_{x\bar xy\bar y}
   = (1-x)^{-1} &
     \int_{t_\xx < t_\xbx < t_\yx < t_\ybx} dt_\xx \, dt_\xbx \, dt_\yx \, dt_\ybx
\nonumber\\& \times
     \frac12 \left[\frac{d\Gamma}{d\yfrak\,d(\Delta t_\yx)}\right]_{(1-x)E} \,
     \frac12 \left[\frac{d\Gamma}{dx\,d(\Delta t_\xx)}\right]_E ,
\label {eq:xxyy4}
\end {align}
which is the result claimed in (\ref{eq:xxyyA}) in the main text
(after using time translation invariance to eliminate one of
the time
integrals and so change probability distribution
$dI/dx\,dy$ to rate distribution $d\Gamma/dx\,dy$).

Let's now address the normalization issue, since (\ref{eq:xxyy3}) and
(\ref{eq:xxyy4}) may not look like they exactly match up.
In particular, one might think that
the formula for $[d\Gamma/d\yfrak\,\Delta t]_{(1-x)E}$ is obtained from
the definition (\ref{eq:dGdDt}) of $[d\Gamma/dx\,d\Delta t]_{E}$ by
simply substituting $\bigl((1-x)E,\yfrak\bigr)$ for $(E,x)$.
But this overlooks the fact that the
effective 1-particle quantum mechanics variables $\B$ here and in
the previous paper \cite{2brem} are defined using the longitudinal
momentum fractions defined with respect to the initial parent $E$
of the full double-splitting process.  In the case of
the $t_\yx < t < t_\ybx$ evolution in fig.\ \ref{fig:xxyy} and (\ref{eq:xxyy1}),
\begin {equation}
   \B = \frac{\b_z - \b_y}{z+y} = \frac{\b_z-\b_y}{1{-}x} ,
\label {eq:Bnorm2}
\end {equation}
where $\b_z$ and $\b_y$ are the transverse positions of the $z$ and $y$
daughters.  In contrast, the $\B$ in the corresponding
{\it single}-splitting
calculation would be
\begin {equation}
   \Bfrak \equiv \frac{\b_{1-\yfrak} - \b_\yfrak}{(1{-}\yfrak)+\yfrak}
      = \b_{1-\yfrak}-\b_\yfrak ,
\label {eq:Bnorm1}
\end {equation}
which is (other than the choice of notation) the same thing as
$\b_z-\b_y$.
The correct translation of (\ref{eq:dGdDt}) is then
\begin {equation}
   \left[\frac{d\Gamma}{d\yfrak\,d(\Delta t)}\right]_{(1-x)E} =
   \frac{\alphas P(\yfrak)}{[\yfrak(1-\yfrak)(1-x)E]^2}
   \grad_{\Bfrak^\ybx} \cdot \grad_{\Bfrak^\yx}
   \langle \Bfrak^\ybx,\Delta t | \Bfrak^\yx,0 \rangle_{E,x}
   \Bigr|_{\Bfrak^\ybx = 0 = \Bfrak^\yx} .
\label {eq:dGdDt2}
\end {equation}
Recalling that the states
$|\B\rangle$ are normalized so that
$\langle \B | \B' \rangle = \delta^{(2)}(\B-\B')$
[and so correspondingly 
$\langle \Bfrak | \Bfrak' \rangle = \delta^{(2)}(\Bfrak-\Bfrak')$],
the relationship $\Bfrak = (1-x)\B$
between (\ref{eq:Bnorm2}) and (\ref{eq:Bnorm1})
lets us rewrite (\ref{eq:dGdDt2}) as
\begin {equation}
   \left[\frac{d\Gamma}{d\yfrak\,d(\Delta t)}\right]_{(1-x)E} =
   \frac{\alphas P(\yfrak)}{[(1-x)y(1-x-y)E]^2} \,
   \grad_{\B^\ybx} \cdot
   \grad_{\B^\yx}
   \langle\B^\ybx,t_\ybx|\B^\yx,t_\yx\rangle
   \Bigr|_{\B^\ybx=0=\B^\yx} .
\end {equation}
This indeed matches up with the corresponding factor in
(\ref{eq:xxyy3}) to give (\ref{eq:xxyy4}).


\section {Justification of the prescription (\ref{eq:tauxy})}
\label {app:justify}

We now outline how to justify the midpoint prescription (\ref{eq:tauxy}).
As discussed in the main text, the starting point is to imagine a medium
with, for example,
\begin {equation}
   \hat q(t) = \hat q_0 \, e^{-t^2/\Time^2} .
\label {eq:qhatt}
\end {equation}
In the limit of large $\Time$, $\hat q$ is approximately constant
over any formation time, and so we may still analyze the effects of
overlapping formation times in the constant-$\hat q$ approximation
used for detailed calculations in this paper.  But, at the same time,
(\ref{eq:qhatt}) helps by providing a consistent large-time
regulator for independent splittings that are far separated in time.

There is a subtlety, however, to approximating $\hat q$ as
constant over formation times.  Consider, formally, the
contribution of $x\bar x y\bar y$ to the total probability $I$
for double splitting:
\begin {align}
   \left[\frac{dI}{dx\,dy}\right]_{x\bar xy\bar y}
   = \frac{1}{1-x} &
     \int_{t_\xx < t_\xbx < t_\yx < t_\ybx} \! dt_\xx \, dt_\xbx \, dt_\yx \, dt_\ybx
\nonumber\\& \times
     \tfrac12 \left[\frac{d\Gamma}{dx\,d(\Delta t_\xx)}\right]_E \,
     \tfrac12 \left[\frac{d\Gamma}{d\yfrak\,d(\Delta t_\yx)}\right]_{(1-x)E}
   .
\label {eq:Ixxyy}
\end {align}
When evaluating $d\Gamma/d\yfrak\,d\Delta t_\yx$ in the integrand,
should we take $\hat q$ to be
$\hat q(t_\yx)$ or $\hat q(t_\ybx)$ or pick some
other time in between?
[A similar question arises for $d\Gamma/dx\,d\Delta t_\xx$.]
This may at first sound unimportant
because the integrand falls off exponentially when
$\Delta t_\yx$ exceeds the
formation time, and so (except for negligible corrections)
$t_\yx$ and $t_\ybx$ can be at most roughly
a formation time apart.  That means $\hat q(t_\yx) \to \hat q(t_\ybx)$
as $\Time\to\infty$.  The difference between $\hat q(t_\yx)$ and
$\hat q(t_\ybx)$, and so the different values we might get for
$d\Gamma/d\yfrak\,d\Delta t_\yx$,
scale at worst like $\Time^{-1}$.  The problem is
that this correction can accumulate additively when we integrate
over time.  For simplicity of argument, let's focus on the integral
over $t_\yx$ while holding $t_\xx$, $t_\xbx$, and
$\Delta t_\yx \equiv t_\ybx - t_\yx$ fixed.
If we choose to evaluate $\hat q$ at $t_\yx$, then the $dt_\yx$
integral has
the functional form
\begin {equation}
   \int_{t_\xbx}^\infty dt_\yx \> f\bigl(\hat q(t_\yx)\bigr)
\end {equation}
for some function $f(\hat q)$.  If we instead take $\hat q(t_\ybx)$,
it is
\begin {equation}
   \int_{t_\xbx}^\infty dt_\yx \> f\bigl(\hat q(t_\yx+\Delta t_\yx)\bigr)
   = \int_{t_\xbx+\Delta t_\yx}^\infty dt_\ybx f\bigl(\hat q(t_\ybx)\bigr) .
\end {equation}
The difference between these two prescriptions is
\begin {equation}
   \int_{t_\xbx}^{t_\xbx+\Delta t_\yx} dt \> f\bigl(\hat q(t)\bigr) ,
\end {equation}
which does {\it not}\/ vanish as $\Time \to \infty$, and so the
difference of the prescriptions for $\hat q$ cannot be ignored.

So which choice is correct?
We could avoid this question by
computing $d\Gamma/d\yfrak\,d\Delta t_\yx$
[and similarly $d\Gamma/dx\,d\Delta t_\xx$] {\it without} making the
approximation that $\hat q$ is constant over $\Delta t$, but that makes
the calculation unnecessarily difficult.  Instead, consider the following.
For simplicity of argument, focus on a region of time where
$\hat q(t)$ is monotonically decreasing, i.e.\ $t>0$ in (\ref{eq:qhatt}).
The problem with the choice $\hat q(t_\yx)$
for the time interval $(t_\yx,t_\ybx)$ is then that it overestimates
$\hat q(t)$ for the entire interval.  The problem with the choice
$\hat q(t_\ybx)$ is that it underestimates $\hat q(t)$ for the
entire interval.  However, if we choose the mid-point, $\hat q(\tau_\yx)$ with
$\tau_\yx \equiv (t_\yx + t_\ybx)/2$, we will in almost equal measure overestimate
$\hat q$ in the first half of the interval and underestimate it in the
second half.  That is, the error we make in $d\Gamma/d\yfrak\,d\Delta t_\yx$
will be $O(\Time^{-2})$ instead of $O(\Time^{-1})$.%
\footnote{
  For a more explicit discussion of evaluating single-splitting rates in the
  case of generic, time-varying $\hat q$, see ref.\ \cite{simple}.
  In terms of explicit formulas, the argument about error sizes used
  above can be reproduced from eq.\ (3.8) of ref.\ \cite{simple} by
  expanding
  $\omega_0^2(t) = \omega_0^2(t_{\rm mid}) + \epsilon (t-t_{\rm mid})$,
  where $\epsilon \sim \Time^{-1}$ is small and $t_{\rm mid} \equiv (t_1+t_2)/2$.
  Note that the integral and
  integrand of eq.\ (3.8) of ref.\ \cite{simple} are symmetric under
  $t_1 \leftrightarrow t_2$ by eq. (3.24) of that reference.
  By the corresponding anti-symmetry of
  the $\epsilon (t-t_{\rm mid})$ term in the expansion
  of $\omega_0^2(t)$ under reflection about $t_{\rm mid}$,
  there can be no $O(\epsilon)$ contribution
  to the probability.  The first correction will instead be $O(\epsilon^2)$.
}
This is small enough that the accumulated error after integrating
$t_\yx$ over time will not affect our final results in the
$\Time\to\infty$ limit.

The upshot is that, in the $x\bar xy\bar y$ calculation, $\hat q$ should
be evaluated at the midpoint times $\tau_\xx$ and $\tau_\yx$.
In the single-splitting calculation, it should also be evaluated at the
midpoint time,
and so the formula used for the idealized
Monte Carlo calculation also corresponds
to using $\hat q(\tau_\xx)$ and $\hat q(\tau_\yx)$, where $\tau_\xx$ and
$\tau_\yx$ are the emission times in the idealized Monte Carlo.
We could now go through all the steps from (\ref{eq:MC3}) to
(\ref{eq:Dxxyyetc})%
\footnote{
   We are glossing over a few details here, having conflated calculations
   of contributions to the double-splitting probability (\ref{eq:Ixxyy})
   with calculations of contributions to the double-splitting {\it rate}
   and thence $\Delta\,d\Gamma/dx\,dy$ (\ref{eq:Dxxyyetc}).
   Giving the time dependence
   (\ref{eq:qhatt}), one should initially discuss computing probabilities
   rather than rates.  However, once one computes the correction
   $\Delta\,dI/dx\,dy$ to the probability due to overlapping formation
   times, one may then define the corresponding correction
   $\Delta\,d\Gamma/dx\,dy$ by casting the probability into the form
   $\Delta\,dI/dx\,dy = \int dt \> \Delta\,d\Gamma/dx\,dy$ in the case
   of generic, arbitrarily slowly varying choices of $\hat q(t)$
   (that fall off to zero as $t \to\pm\infty$).
   When thinking about the size of effects that must
   be correctly accounted for, note that $d\Gamma/dx\,dy$ is $O(\Time)$
   and the desired $\Delta\,d\Gamma/dx\,dy$ is $O(\Time^0)$,
   but $dI/dx\,dy$ is $O(\Time^2)$ and the corresponding $\Delta\,dI/dx\,dy$
   is $O(\Time)$.  In particular, one does not need to know
   $\Delta\,dI/dx\,dy$ to order $\Time^0$.
}
explicitly specifying the use of
$\hat q(\tau_\xx)$ and $\hat q(\tau_\yx)$.  However, once we get
to the final result (\ref{eq:Dxxyyetc}), which is finite as
$\Time \to \infty$, we no longer have to worry about this
distinction.  We can then simply take $\Time \to \infty$ and
call $\hat q$ a constant.


\section {Derivation of \boldmath$xy\bar x\bar y_2$}
\label {app:xyxy}


\subsection {Basics}

\begin {figure}[t]
\begin {center}
  \begin{picture}(250,150)(0,0)
  \put(0,0){\includegraphics[scale=0.5]{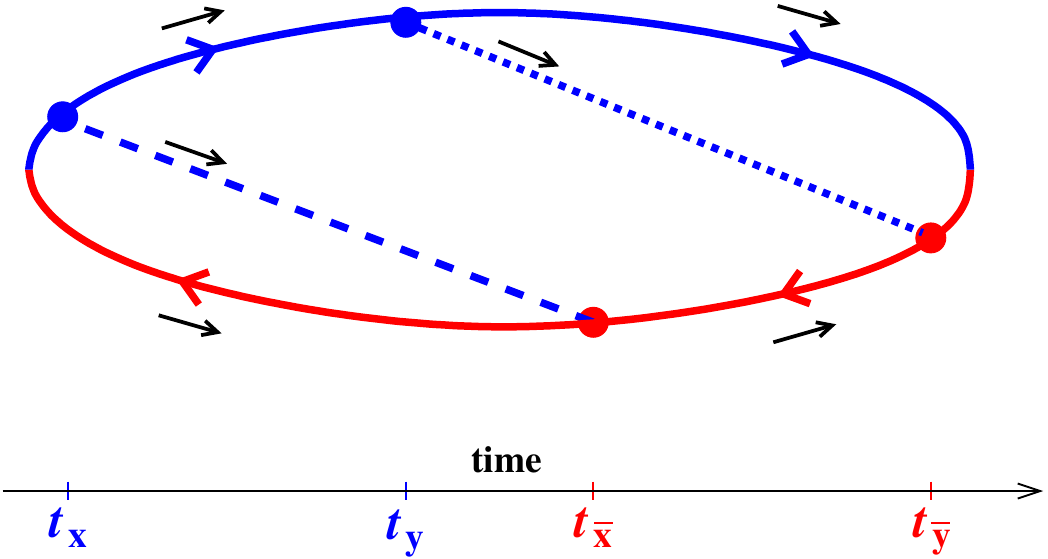}}
  \put(27,44){$\hat x_1,{-}h_{\rm i}$}
  \put(30,83){$\hat x_4,h_{\rm x}$}
  \put(22,138){${-}\hat x_1{-}\hat x_4,h$}
  \put(110,106){$\hat x_2,h_{\rm y}$}
  \put(183,138){$\hat x_3,h_{\rm z}$}
  \put(170,44){$\hat x_1{+}\hat x_4,{-}\bar h$}
  \end{picture}
  \caption{
     \label{fig:xyxy}
     Our labeling conventions for the $xy\bar x\bar y_2$ interference diagram.
     (The only difference between $xy\bar x\bar y_2$ and $xy\bar x\bar y_1$
     is the order $(x_1,x_2,x_3,x_4)$ we have chosen for labeling the
     particles during the 4-particle evolution $t_\yx < t < t_\xbx$.)
  }
\end {center}
\end {figure}
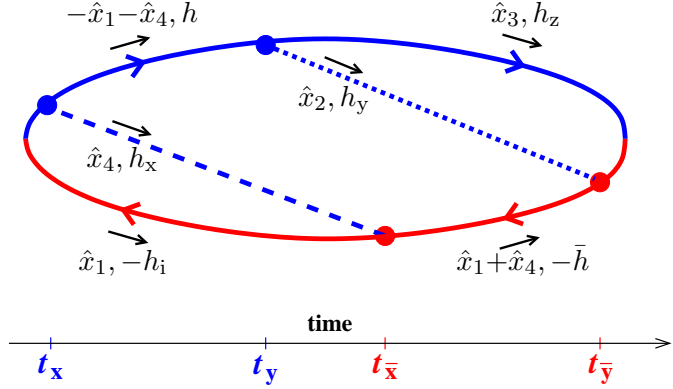

In this appendix, we step through the derivation of the result
(\ref{eq:xyxyresult}) for the $xy\bar x\bar y_2$ interference contribution
of fig.\ \ref{fig:cylinder2B}.
Our notation is given in fig.\ \ref{fig:xyxy}.

A direct application of the methods of the preceding paper \cite{2brem}
gives (remembering that $\hat x_1+\hat x_4 = -\hat x_2 - \hat x_3$)
\begin {align}
   \left[\frac{dI}{dx\,dy}\right]_{xy\bar x\bar y_2}
   &=
   \tfrac12 \CA^2 \,
   \frac{\alphas^2 }{4 E^4} \,
   (\hat x_1+\hat x_4)^{-2}
   \int_{t_\xx < t_\yx < t_\xbx < t_\ybx}
   \int_{\B^\yx,\B^\xbx}
    \sum_{h_\xx,h_\yx,h_\zx}
\nonumber\\ &\times
   \Bigl[
   \sum_{\bar h}
   {\cal P}^{\bar n}_{\bar h \to h_\zx,h_\yx}\bigl(1{-}x \to 1{-}x{-}y,y\bigr) \,
   {\cal P}^{\bar m}_{h_\ix \to \bar h, h_\xx}\bigl(1 \to 1{-}x,x\bigr)
   \Bigr]^*
\nonumber\\ &\times
   \Bigl[
   \sum_h
   {\cal P}^n_{h \to h_\zx,h_\yx}\bigl(1{-}x \to 1{-}x{-}y,y\bigr) \,
   {\cal P}^m_{h_\ix \to h,h_\xx}\bigl(1 \to 1{-}x,x\bigr)
   \Bigr]
\nonumber\\ &\times
   \nabla^{\bar n}_{\B^\ybx}
   \langle\B^\ybx,t_\ybx|\B^\xbx,t_\xbx\rangle
   \Bigr|_{\B^\ybx=0}
\nonumber\\ &\times
   \nabla^{\bar m}_{\C_{41}^\xbx}
   \nabla^n_{\C_{23}^\yx}
   \langle\C_{41}^\xbx,\C_{23}^\xbx,t_\xbx|\C_{41}^\yx,\C_{23}^\yx,t_\yx\rangle
   \Bigr|_{\C_{41}^\xbx=0=\C_{23}^\yx; ~ \C_{23}^\xbx=\B^\xbx; ~ \C_{41}^\yx=\B^\yx}
\nonumber\\ &\times
   \nabla^m_{\B^\xx}
   \langle\B^\yx,t_\yx|\B^\xx,t_\xx\rangle
   \Bigr|_{\B^\xx=0} .
\label {eq:xyxy20}
\end {align}
The color factor $\frac12 \CA^2$ in front is {\it half}\/
of the naive color
factor $d_{\rm A}^{-1} \tr(T^a_{\rm A} T^a_{\rm A} T^b_{\rm A} T^b_{\rm A}) = \CA^2$
from the drawing of fig.\ \ref{fig:xyxy} because
$xy\bar x\bar y_2$ given by (\ref{eq:xyxy20})
represents the
contribution from only {\it one}\/
of the two large-$\Nc$ color routings of $xy\bar x\bar y$.

The helicity sums are not as simple as in the $x\bar xy\bar y$ case of
appendix \ref{app:xxyy} because we do not have any useful analog of
(\ref{eq:rotationdot}) to start the simplification.
However, we may again average over the initial helicity $h_\ix$, and
rotational symmetry in the transverse plane implies that the
initial helicity average of
\begin {align}
   \sum_{h_\xx,h_\yx,h_\zx}
   &\Bigl[
   \sum_{\bar h}
   {\cal P}^{\bar n}_{\bar h \to h_\zx,h_\yx}\bigl(1{-}x \to 1{-}x{-}y,y\bigr) \,
   {\cal P}^{\bar m}_{h_\ix \to \bar h, h_\xx}\bigl(1 \to 1{-}x,x\bigr)
   \Bigr]^*
\nonumber\\ \times
   &\Bigl[
   \sum_h
   {\cal P}^n_{h \to h_\zx,h_\yx}\bigl(1{-}x \to 1{-}x{-}y,y\bigr) \,
   {\cal P}^m_{h_\ix \to h,h_\xx}\bigl(1 \to 1{-}x,x\bigr)
   \Bigr]
\label {eq:calPsum2bar}
\end {align}
above must have the form
\begin {equation}
  \bar\alpha(x,y) \, \delta^{\bar n n} \delta^{\bar m m}
  + \bar\beta(x,y) \, \delta^{\bar n \bar m} \delta^{nm}
  + \bar\gamma(x,y) \, \delta^{\bar n m} \delta^{n \bar m} ,
\label {eq:abcbardef}
\end {equation}
for some functions $\bar\alpha$, $\bar\beta$, and $\bar\gamma$,
analogous to eq.\ (4.38) of ref.\ \cite{2brem}.
Using the explicit formulas for the ${\cal P}$'s for the case
where all high-energy particles are gluons,%
\footnote{
   eqs.\ (4.35) of ref.\ \cite{2brem}.
}
we find%
\footnote{
  Note that, unlike the case of $(\alpha,\beta,\gamma)$ defined in
  ref.\ \cite{2brem}, the $(\bar\alpha,\bar\beta,\bar\gamma)$ here
  are not symmetric under $x \leftrightarrow y$.  However,
  they are symmetric under
  $y \leftrightarrow z \equiv 1{-}x{-}y$.
}
\begin {align}
   \begin{pmatrix}
      \bar\alpha \\ \bar\beta \\ \bar\gamma
   \end{pmatrix}
   =
   \phantom{+}
   & \begin{pmatrix} - \\ + \\ + \end{pmatrix}
       \frac{4}{xy(1{-}x)^6(1{-}x{-}y)}
\nonumber\\
   + & \begin{pmatrix} + \\ - \\ + \end{pmatrix} \Biggl[
       \frac{1}{x^3y^3(1{-}x)^2(1{-}x{-}y)^3}
       + \frac{1{-}x{-}y}{x^3y^3(1{-}x)^2}
\nonumber\\ & \qquad\qquad
       + \frac{x}{y^3(1{-}x)^2(1{-}x{-}y)^3}
       + \frac{y}{x^3(1{-}x)^2(1{-}x{-}y)^3}
   \Biggr]
\nonumber\\
   + & \begin{pmatrix} + \\ + \\ - \end{pmatrix} \Biggl[
       \frac{(1{-}x)^2}{x^3y^3(1{-}x{-}y)^3}
       + \frac{(1{-}x{-}y)}{x^3y^3(1{-}x)^6}
       + \frac{x(1{-}x{-}y)}{y^3(1{-}x)^6}
\nonumber\\ & \qquad\qquad
       + \frac{y}{x^3(1{-}x)^6(1{-}x{-}y)^3}
       + \frac{xy}{(1{-}x)^6(1{-}x{-}y)^3}
   \Biggr]
   .
\label {eq:abcbar}
\end {align}
It will be useful to note for future reference that a certain combination
of $(\bar\alpha,\bar\beta,\bar\gamma)$ can be written in terms of
spin-averaged DGLAP splitting functions:
\begin {align}
   \bar\alpha + \tfrac12 \bar\beta + \tfrac12 \bar\gamma
   &=
   \frac{[1 + x^4 + (1{-}x)^4]}{x^3 (1{-}x)^3} \,
   \frac{[(1{-}x)^4+y^4+(1{-}x{-}y)^4]}{(1{-}x)^3 y^3 (1{-}x{-}y)^3}
\nonumber\\
   &=
   \frac{P(x)}{\CA x^2(1{-}x)^2} \,\,
   \frac{P\bigl(\frac{y}{1{-}x}\bigr)}
        {\CA (1{-}x) y^2 (1{-}x{-}y)^2}
   \, .
\label {eq:abcPP2}
\end {align}
Replacing (\ref{eq:calPsum2bar}) by (\ref{eq:abcbardef}) in
(\ref{eq:xyxy20}) gives
\begin {align}
   \left[\frac{dI}{dx\,dy}\right]_{xy\bar x\bar y_2}
   &=
   \frac{\CA^2 \alphas^2 }{8 E^4} \,
   \frac{
     ( \bar\alpha \delta^{\bar n n} \delta^{\bar m m}
     {+} \bar\beta \delta^{\bar n \bar m} \delta^{nm}
     {+} \bar\gamma \delta^{\bar n m} \delta^{n \bar m} )
   }{
     |\hat x_1 + \hat x_4|^2
   }
   \int_{t_\xx < t_\yx < t_\xbx < t_\ybx}
   \int_{\B^\yx,\B^\xbx}
\nonumber\\ &\times
   \nabla^{\bar n}_{\B^\ybx}
   \langle\B^\ybx,t_\ybx|\B^\xbx,t_\xbx\rangle
   \Bigr|_{\B^\ybx=0}
\nonumber\\ &\times
   \nabla^{\bar m}_{\C_{41}^\xbx}
   \nabla^n_{\C_{23}^\yx}
   \langle\C_{41}^\xbx,\C_{23}^\xbx,t_\xbx|\C_{41}^\yx,\C_{23}^\yx,t_\yx\rangle
   \Bigr|_{\C_{41}^\xbx=0=\C_{23}^\yx; ~ \C_{23}^\xbx=\B^\xbx; ~ \C_{41}^\yx=\B^\yx}
\nonumber\\ &\times
   \nabla^m_{\B^\xx}
   \langle\B^\yx,t_\yx|\B^\xx,t_\xx\rangle
   \Bigr|_{\B^\xx=0} .
\end {align}


\subsection {Multiple scattering (\boldmath$\hat q$) approximation}

In the harmonic oscillator approximation, we may do the first and
last time integrals ($t_\x$ and $t_\ybx$) as in section V.A of
ref.\ \cite{2brem}, with result
\begin {align}
   \left[\frac{d\Gamma}{dx\,dy}\right]_{xy\bar x\bar y_2}
   &=
   - \frac{\CA^2 \alphas^2 M_\ix M_\fx^\seq}{8\pi^2 E^4} \,
   \frac{
     ( \bar\alpha \delta^{\bar n n} \delta^{\bar m m}
     {+} \bar\beta \delta^{\bar n \bar m} \delta^{nm}
     {+} \bar\gamma \delta^{\bar n m} \delta^{n \bar m} )
   }{
     |\hat x_1 + \hat x_4|^2
   }
\nonumber\\ &\quad\times
   \int_0^{\infty} d(\Delta t)
   \int_{\B^\yx,\B^\xbx}
   \frac{B^{\xbx}_{\bar n}}{(B^\xbx)^2} \,
   \frac{B^{\yx}_m}{(B^\yx)^2} \,
   \exp\bigl(
      - \tfrac12 |M_\fx^\seq| \Omega_\fx^\seq (B^\xbx)^2
      - \tfrac12 |M_\ix| \Omega_\ix (B^\yx)^2
   \bigr)
\nonumber\\ &\qquad\times
   \nabla^{\bar m}_{\C_{41}^\xbx}
   \nabla^n_{\C_{23}^\yx}
   \langle\C_{41}^\xbx,\C_{23}^\xbx,t_\xbx|\C_{41}^\yx,\C_{23}^\yx,t_\yx\rangle
   \Bigr|_{\C_{41}^\xbx=0=\C_{23}^\yx; ~ \C_{23}^\xbx=\B^\xbx; ~ \C_{41}^\yx=\B^\yx}
   ,
\label {eq:xyyx21}
\end {align}
where $\Delta t \equiv t_\xbx-t_\yx$ and
\begin {equation}
   M_\ix = \hat x_1 \hat x_4 (\hat x_1{+}\hat x_4) E,
   \qquad
   M_\fx^{\rm seq} = \hat x_2 \hat x_3 (\hat x_2{+}\hat x_3) E .
\label {eq:Mdefs}
\end {equation}
$\Omega_\ix$ and $\Omega_\fx^\seq$ are respectively the
$\Omega_{E,x}$ and $\Omega_{(1-x)E,\yfrak}$ of
(\ref{eq:Omegai}) and (\ref{eq:Omegafseq}).
We put the superscript ``seq'' on $M_\fx^\seq$ and $\Omega_\fx^\seq$
to indicate that they are different
than the corresponding $M_\fx$ and $\Omega_\fx$ in the calculation of
the canonical diagram $xy\bar y\bar x$ in ref.\ \cite{2brem}.

Unlike in the $xy\bar y\bar x$ analysis of the preceding paper \cite{2brem},
the relevant variables $(\C_{41},\C_{23})$ are the same on both sides
of $\langle\C_{41}^\xbx,\C_{23}^\xbx,t_\xbx|\C_{41}^\yx,\C_{23}^\yx,t_\yx\rangle$
above.  In order to keep the notation similar to that of the
preceding paper, it is useful to consider the variables
ordered on the two ends as
\begin {equation}
  \begin{pmatrix} \C^\xbx_{23} \\ \C^\xbx_{41} \end{pmatrix}
  \qquad \mbox{and} \qquad
  \begin{pmatrix} \C^\yx_{41} \\ \C^\yx_{23} \end{pmatrix}
\label {eq:flip}
\end {equation}
so that the derivatives in (\ref{eq:xyyx21}) hit the lower element
of each pair (as in that paper).
The appropriate transformation matrix between
these variables and the normal modes for the 4-particle evolution
is
\begin {equation}
   a_\yx \equiv
   \begin{pmatrix} C^+_{41} & C^-_{41} \\ C^+_{23} & C^-_{23} \end{pmatrix}
\label {eq:ay}
\end {equation}
at $t_\yx$, in the notation of the preceding paper.
The explicit formula is given by eqs.\ (5.24,5.28,5.33) of
that paper.
Given the choice (\ref{eq:flip}),
the corresponding transformation at $t_\xbx$ is simply
\begin {equation}
   a_\xbx^\seq \equiv 
   \begin{pmatrix} 0 & 1 \\ 1 & 0 \end{pmatrix} a_\yx .
\end {equation}
The rest of the analysis proceeds as in section V.C of ref.\ \cite{2brem}
but using
\begin {subequations}
\label {eq:XYZseqdef}
\begin {align}
   \begin{pmatrix} X_\yx^\seq & Y_\yx^\seq \\ Y_\yx^\seq & Z_\yx^\seq \end{pmatrix}
   &\equiv
   \begin{pmatrix} |M_\ix|\Omega_\ix & 0 \\ 0 & 0 \end{pmatrix}
     - i a_\yx^{-1\top} \uOmega \cot(\uOmega\,\Delta t)\, a_\yx^{-1}
   =
   \begin{pmatrix} X_\yx & Y_\yx \\ Y_\yx & Z_\yx \end{pmatrix} ,
\\
   \begin{pmatrix} X_\xbx^\seq & Y_\xbx^\seq \\ Y_\xbx^\seq & Z_\xbx^\seq \end{pmatrix}
   &\equiv
   \begin{pmatrix} |M_\fx^\seq|\Omega_\fx^\seq & 0 \\ 0 & 0 \end{pmatrix}
     - i (a_\xbx^\seq)^{-1\top} \uOmega \cot(\uOmega\,\Delta t)\, (a_\xbx^\seq)^{-1} ,
\\
   \begin{pmatrix} X_{\yx\xbx}^\seq & Y_{\yx\xbx}^\seq \\
                   \Ybar_{\yx\xbx}^\seq & Z_{\yx\xbx}^\seq \end{pmatrix}
   &\equiv
   - i a_\yx^{-1\top} \uOmega \csc(\uOmega\,\Delta t) \, (a_\xbx^\seq)^{-1}
\end {align}
\end {subequations}
in result (\ref{eq:xyxyresult}) of this paper.%
\footnote{
   Compare (\ref{eq:xyxyresult}) here to eq.\ (5.45) of ref.\ \cite{2brem}.
}
The $I^{\rm seq}$'s in (\ref{eq:xyxyresult}) are defined the same way as
the $I$'s of ref.\ \cite{2brem} but using the appropriate
$(X^\seq,Y^\seq,Z^\seq)$ here instead of $(X,Y,Z)$ there:

\begin {subequations}
\label {eq:Iseq}
\begin {align}
   I_0^\seq &=
   \frac{4\pi^2}{[X_\yx^\seq X_\xbx^\seq - (X_{\yx\xbx}^\seq)^2]} \,,
\displaybreak[0]\\
   I_1^\seq &=
   - \frac{2\pi^2}{X_{\yx\xbx}^\seq}
   \ln\left( 1 - \frac{(X_{\yx\xbx}^\seq)^2}{X_\yx^\seq X_\xbx^\seq} \right) ,
\displaybreak[0]\\
   I_2^\seq &=
   \frac{2\pi^2}{(X_{\yx\xbx}^\seq)^2}
     \ln\left( 1 - \frac{(X_{\yx\xbx}^\seq)^2}{X_\yx^\seq X_\xbx^\seq} \right)
   + \frac{4\pi^2}{[X_\yx^\seq X_\xbx^\seq - (X_{\yx\xbx}^\seq)^2]} \,,
\displaybreak[0]\\
   I_3^\seq &=
   \frac{4\pi^2 X_{\yx\xbx}^\seq}
        {X_\xbx^\seq[X_\yx^\seq X_\xbx^\seq - (X_{\yx\xbx}^\seq)^2]} \,,
\displaybreak[0]\\
   I_4^\seq &=
   \frac{4\pi^2 X_{\yx\xbx}^\seq}
        {X_\yx^\seq[X_\yx^\seq X_\xbx^\seq - (X_{\yx\xbx}^\seq)^2]} \,.
\end {align}
\end {subequations}
The 4-particle evolution frequencies $\Omega_\pm$
appearing in (\ref{eq:xyxyresult})
are given by eq.\ (5.21) of ref.\ \cite{2brem}.


\subsection {Small $\Delta t$ limit}

The extraction of the small-$\Delta t$ limit (\ref{eq:xyxy2smalldtA})
follows the procedure of appendix D1 of ref.\ \cite{2brem}.
In our case here,
\begin {subequations}
\begin {align}
   \begin{pmatrix} X_\yx^\seq & Y_\yx^\seq \\ Y_\yx^\seq & Z_\yx^\seq \end{pmatrix}
   &=
     -\frac{i}{\Delta t}\, a_\yx^{-1\top} a_\yx^{-1}
     +\begin{pmatrix} |M_\ix|\Omega_\ix & 0 \\ 0 & 0 \end{pmatrix}
     +O(\Delta t) ,
\\
   \begin{pmatrix} X_\xbx^\seq & Y_\xbx^\seq \\ Y_\xbx^\seq & Z_\xbx^\seq \end{pmatrix}
   &=
     -\frac{i}{\Delta t}\, (a_\xbx^\seq)^{-1\top} (a_\xbx^\seq)^{-1}
     +\begin{pmatrix} |M_\fx^\seq|\Omega_\fx^\seq & 0 \\ 0 & 0 \end{pmatrix}
     +O(\Delta t) ,
\\
   \begin{pmatrix} X_{\yx\xbx}^\seq & Y_{\yx\xbx}^\seq \\
                   \Ybar_{\yx\xbx}^\seq & Z_{\yx\xbx}^\seq \end{pmatrix}
   &=
     -\frac{i}{\Delta t}\, a_\yx^{-1\top} (a_\xbx^\seq)^{-1}
     +O(\Delta t) ,
\end {align}
\end {subequations}
which gives
\begin {subequations}
\label {eq:XYZseqsim}
\begin {align}
   \begin{pmatrix} X_\yx^\seq & Y_\yx^\seq \\ Y_\yx^\seq & Z_\yx^\seq \end{pmatrix}
   &=
     -\frac{i}{\Delta t}
       \begin{pmatrix} M_\ix & 0 \\ 0 & M_\fx^\seq \end{pmatrix}
     +\begin{pmatrix} |M_\ix|\Omega_\ix & 0 \\ 0 & 0 \end{pmatrix}
     + O(\Delta t) ,
\\
   \begin{pmatrix} X_\xbx^\seq & Y_\xbx^\seq \\ Y_\xbx^\seq & Z_\xbx^\seq \end{pmatrix}
   &=
     -\frac{i}{\Delta t}
       \begin{pmatrix} M_\fx^\seq & 0 \\ 0 & M_\ix \end{pmatrix}
     +\begin{pmatrix} |M_\fx^\seq|\Omega_\fx^\seq & 0 \\ 0 & 0 \end{pmatrix}
     + O(\Delta t) ,
\\
   \begin{pmatrix} X_{\yx\xbx}^\seq & Y_{\yx\xbx}^\seq \\
                   \Ybar_{\yx\xbx}^\seq & Z_{\yx\xbx}^\seq \end{pmatrix}
   &=
     -\frac{i}{\Delta t}
     \begin{pmatrix} 0 &  M_\ix \\
                     M_\fx^\seq &  0 \end{pmatrix}
     + O(\Delta t) .
\end {align}
\end {subequations}
Then
\begin {equation}
   X_\yx^\seq X_\xbx^\seq - (X_{\yx\xbx}^\seq)^2
   = - \frac{M_\ix M_\fx^\seq}{(\Delta t)^2}
     \left[
         1 + i(\Omega_\ix\sgn M_\ix + \Omega_\fx^\seq\sgn M_\fx^\seq) \Delta t
     \right]
     + O\bigl((\Delta t)^0\bigr) .
\label{eq:detXseqsim}
\end {equation}
In the current context, $\sgn M_\ix = +1$ and $\sgn M_\fx^\seq = +1$.
We will not need to make use of this result in any other context,
and so we will drop these signs.  Using the above expansions in
(\ref{eq:xyxyresult}) and (\ref{eq:Iseq}) then gives
\begin {align}
   \left[\frac{d\bar\Gamma}{dx\,dy}\right]_{xy\bar x\bar y_2}
   &=
   \frac{\CA^2 \alphas^2 M_\ix M_\fx^\seq}{8\pi^2 E^2}
   (-\hat x_1 \hat x_2 \hat x_3 \hat x_4)
   \bigl(\bar\alpha + \tfrac12\bar\beta + \tfrac12\bar\gamma\bigr)
   \int_0 d(\Delta t)
     \left( \frac{1}{(\Delta t)^2} -
             \frac{i( \Omega_\ix + \Omega_\fx^\seq)}{\Delta t}
     \right)
\nonumber\\ & \quad\qquad
         + {\rm UV~convergent} ,
\end {align}
which may be recast into the form (\ref{eq:xyxy2smalldtA})
with the aid of (\ref{eq:order2}), (\ref{eq:abcPP2}) and (\ref{eq:Mdefs}).


\section {Transverse momentum integration vs.\ Ref.\ \cite{Blaizot0}}
\label {app:Blaizot0}

Consider the case of independent single-splitting processes---what we've
called the idealized Monte Carlo (IMC) picture.
For democratic branchings (no daughter soft), there is a simple
qualitative argument for the IMC picture: The chance of bremsstrahlung
is roughly $\alphas$ per formation time $t_{\rm form}$, and so the typical
time between (non-soft) splittings is
$\tau_{\rm rad} \sim t_{\rm form}/\alphas$,
which is parametrically
large compared to the formation time $t_{\rm form}$ when
$\alphas$ at the relevant energy scale is small.
For the LPM effect, the formation time represents the time over
which the splitting process is quantum mechanically coherent,
and so events which are separated by more than that time should
be quantum mechanically independent.
The effects we calculate
in this paper, in contrast, are suppressed by a factor of
$t_{\rm form}/\tau_{\rm rad} \sim \alphas$, which is the chance for
two consecutive splittings to overlap.

In the context of QCD, the basic scales described above have been
known since the pioneering work of Baier {\it et al.}\/
\cite{BDMPS}
and Zakharov \cite{Zakharov} (BDMPS-Z).
There have been many works on extending (in various limits)
the BDMPS-Z analysis to include transverse momentum distributions,
but the issue we want to mention here arises in the work of
Blaizot {\it et al.}\/ \cite{Blaizot0}.  One of the
issues that
concerned them was how long color coherence between the
daughters might survive after a splitting, and how formally
to show that it does not interfere with the quantum-mechanical
independence of splittings.
They found that color decoherence occurs over a time of order
$t_{\rm form}$ after the time that we call $t_\xbx$ in our
fig.\ \ref{fig:xxrate}, and so all is well qualitatively concerning
the separation of scales justifying the leading-order approximation that
splittings are independent.
The analysis of this decoherence was somewhat complicated,
even in large $\Nc$.

However, as we'll now discuss,
this late-time color decoherence issue (and so our need
to calculate it) disappears completely
if one integrates the
splitting rate over the transverse momenta of the daughters
after the splitting, which means that we can avoid calculating
the details of late-time color decoherence.

Let's start by discussing
the issue in the language used by ref.\ \cite{2brem}, which
developed many of the methods used in our paper here.
The issue is treated in section IV.A of ref.\ \cite{2brem}
in the context of double splitting, where it is shown using
a unitarity argument that, provided one integrates over
the final transverse momenta of the daughters, then one may
ignore what happens to a daughter after it has been emitted
in {\it both}\/ the amplitude and the conjugate amplitude.
So, for instance, consider the $x y\bar y\bar x$ diagram of
figs.\ \ref{fig:subset} and \ref{fig:subset2}.  For this
diagram, one may ignore
what happens to the $y$ daughter after the corresponding
splitting time ($t_\ybx$) in the conjugate amplitude, and
one may additionally ignore what happens to the other two daughters
after the corresponding splitting time ($t_\xbx$) in the
conjugate amplitude.  The same arguments apply to single
splitting, and one may ignore what happens to the two daughters
after the time $t_\xbx$ in the $x\bar x$ diagram of fig.\ \ref{fig:xxrate}.

\begin {figure}[t]
\begin {center}
  \includegraphics[scale=0.7]{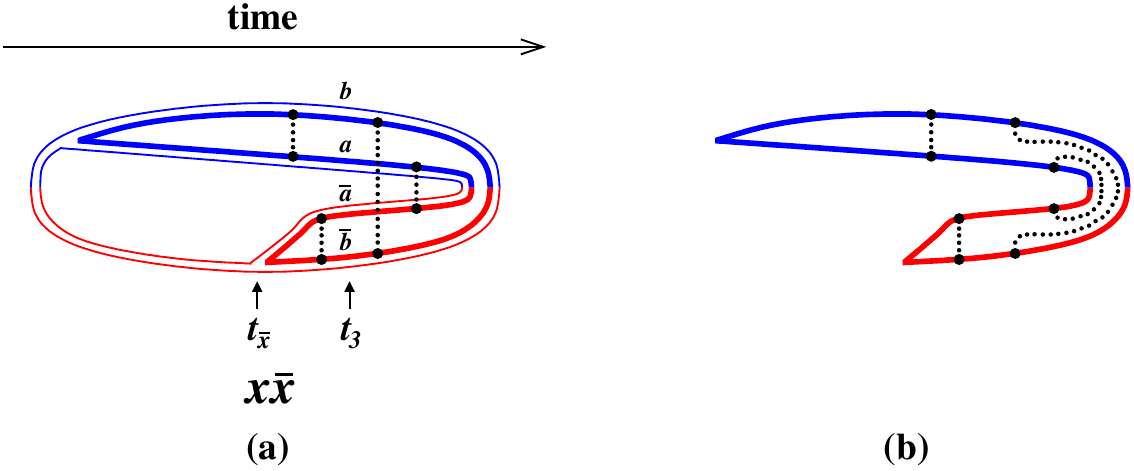}
  \caption{
     \label{fig:Blaizot0}
     (a) An example of non-trivial color interaction in single
     splitting {\it after} the last emission time $t_\xbx$.
     This figure uses a combination of our notation and that
     of Blaizot {\it et al.}\/ \cite{Blaizot0}.
     The high-energy gluons are drawn
     with the double line notation of large $\Nc$.  Correlations
     representing interactions with the medium are show as black
     dotted lines and correspond to the same sorts of correlations
     as depicted by black double lines in fig.\ \ref{fig:cylinder2A}.
     The thick-curve fundamental-representation loop plus the particular
     example of dotted correlation lines shown is equivalent to
     fig.\ 9 of ref.\ \cite{Blaizot0}.  The time $t_3$ indicated above
     is the same as the time $t_3$ in their figure, and our labels
     $a,b,\bar a,\bar b$ here correspond to theirs.
     (b) The same fundamental-representation loop, with the correlations
     drawn in a topologically equivalent way just to make it clear that this
     example represents a planar diagram bounded by the loop and so
     leading order in large $\Nc$.
  }
\end {center}
\end {figure}

If we did not integrate over transverse momenta, then we would
have to worry about later interactions, such as shown in
fig.\ \ref{fig:Blaizot0} for the case of single splitting,
which represents the same interactions as fig.\ 9
of Blaizot {\it et al.}\/ \cite{Blaizot0}.  The development of
the system for times later than $t_\xbx$ in our figure is what
Blaizot {\it et al.}\/ refer to as ``Region III'' in their
discussion, and the tricky color coherence issues are associated
with the time interval $t_\xbx < t < t_3$, which they refer to as
the ``non-factorizable'' piece of the calculation.

In the formalism of Blaizot {\it et al.}, what happens when we do integrate
over final transverse momenta?
Integrating over final transverse momenta $\k_a$ and $\k_b$ in
their differential cross-section (3.13) is equivalent, in the
transverse position space used in their (4.1), to equating
$(\bar\y_a,\bar\y_b)=(\y_a,\y_b)$ and then integrating over
$(\y_a,\y_b)$.  In their formula (4.15) for the effects
of late-time correlations such as shown in our fig.\ \ref{fig:Blaizot0},
the factor $(\y_a;\bar\y_a|S^{(2)}(t_L,t_3)|\z_a;\bar\z_a)$ then
forces $\z_a = \bar\z_a$.  (Similarly, $\z_b = \bar\z_b$.)
This makes the $T(Z)$ of their (4.16) vanish, which means that
their (4.15) vanishes:
what they call the ``non-factorizable''
contributions vanish when integrated over
final transverse momenta.


\end {document}